%% file: paper.tex
\documentclass[useAMS,usenatbib]{mn2e}

\usepackage{lscape}
\usepackage{graphicx}
\usepackage{subfigure}

\usepackage{multirow}
\usepackage{lscape}
\usepackage{longtable}
\usepackage{amssymb}
\usepackage{url}
\usepackage{rotating}
\urldef\mysite\path{http://bd-server.astro.ex.ac.uk/}
\urldef\mysitecalib\path{http://bd-server.astro.ex.ac.uk/calibration}
\urldef\mysitefit\path{http://bd-server.astro.ex.ac.uk/fit}
\urldef\mysiteplots\path{http://bd-server.astro.ex.ac.uk/plots}
\urldef\mysiteiso\path{http://bd-server.astro.ex.ac.uk/isochrones}
\urldef\sdss\path{http://www.sdss.org/dr1/algorithms/fluxcal.html}
\urldef\twomass\path{http://www.ipac.caltech.edu/2mass/releases/allsky/doc/sec6_4a.html}
\urldef\ukirt\path{http://www.jach.hawaii.edu/UKIRT/instruments/wfcam/user_guide/description.html#Tab1.1}
\urldef\irasresp\path{http://irsa.ipac.caltech.edu/IRASdocs/archives/spectral_resp.html}
\urldef\iraszero\path{http://irsa.ipac.caltech.edu/IRASdocs/exp.sup/ch6/C2a.html}
\urldef\jcmt\path{http://www.jach.hawaii.edu/JCMT/continuum/background/background.html}
\urldef\irac\path{http://irsa.ipac.caltech.edu/data/SPITZER/docs/irac/}
\urldef\mips\path{http://irsa.ipac.caltech.edu/data/SPITZER/docs/mips/}
\urldef\pacs\path{http://herschel.esac.esa.int/Docs/PACS/html/pacs_om.html}
\urldef\spire\path{http://herschel.esac.esa.int/Docs/SPIRE/html/spire_om.html#x1-880005.2.3}
\urldef\cfht\path{http://cfht.hawaii.edu/Instruments/Filters/}
\newcommand\logmdot {$\log \dot{M}$}
\title[SED fitting]{Bayesian fitting of Taurus brown dwarf spectral energy distributions} \author[N.J.
Mayne et al.]{N.J. Mayne\thanks{E-mail: nathan@astro.ex.ac.uk
    (NJM)}, Tim  J. Harries, John Rowe and David M. Acreman\\
  School of Physics, University of Exeter, Stocker Road, Exeter, EX4 4QL.\\
 }

\begin{document}

\include{body}

\include{tables}
\include{appendix}

\end{document}

%% file: body.tex
\date{Accepted ?. Received ?; in
  original form ?}

\pagerange{\pageref{firstpage}--\pageref{lastpage}} \pubyear{2009}

\maketitle

\label{firstpage}

\begin{abstract}  

  We present derived stellar and disc parameters for a sample of
  Taurus brown dwarfs both with and without evidence of an associated
  disc. These parameters have been derived using an online fitting
  tool (\mysite{}), which includes a statistically robust derivation
  of uncertainties, an indication of parameter degeneracies, and a
  complete treatment of the input photometric and spectroscopic
  observations.

  The observations of the Taurus members with indications of disc
  presence have been fitted using a grid of theoretical models
  including detailed treatments of physical processes accepted for
  higher mass stars, such as dust sublimation, and a simple treatment
  of the accretion flux. This grid of models has been designed to test
  the validity of the adopted physical mechanisms, but we have also
  constructed models using parameterisation, for example
  semi-empirical dust sublimation radii, for users solely interested
  in parameter derivation and the quality of the fit.

  The parameters derived for the naked and disc brown dwarf systems
  are largely consistent with literature observations. However, our
  inner disc edge locations are consistently closer to the star than
  previous results and we also derive elevated accretion rates over
  non-SED based accretion rate derivations. For inner edge locations
  we attribute these differences to the detailed modelling we have
  performed of the disc structure, particularly at the crucial inner
  edge where departures in geometry from the often adopted vertical
  wall due to dust sublimation (and therefore accretion flux) can
  compensate for temperature (and therefore distance) changes to the
  inner edge of the dust disc. In the case of the elevated derived
  accretion rates, in some cases, this may be caused by the intrinsic
  stellar luminosities of the targets exceeding that predicted by the
  isochrones we have adopted.

\end{abstract}


\section{Introduction}
\label{intro}

In studies of star formation and evolution spectral energy
distribution (SED) fitting is often used to derive parameters for
stars and their circumstellar discs. Fitting tools, such as
\cite{robitaille_2006}, and the increasing availability of
observational data have led to a rapid increase in the number of
studies using SED fitting. As observations of fainter lower-mass stars
(LMS) and brown dwarfs (BDs) have become feasible \citep[see for
example][]{jayawardhana_2003,luhman_2005b,luhman_2008,guieu_2007,monin_2010,olofsson_2010}
such SED fitting techniques have been extended into the lower mass
regime. These studies have generally been limited to small numbers of
members of the Taurus star forming region (SFR) due to its proximity.

For higher mass stars extensive studies have been performed using
detailed treatments of the important physical processes, for example
radiative transfer, dust sublimation and vertical hydrostatic
equilibrium \citep[see for
example][]{tannirkulam_2008,walker_2004}. Such models are
computationally expensive and as such are usually used to model
individual systems. The online fitting tool of \cite{robitaille_2006}
has allowed much greater access to such models via pre-computed
`grids' of radiative transfer simulations covering a selected
parameter space, for example a range of masses and ages. However, the
young-stellar objects (YSOs) modelled by \cite{robitaille_2006} do not
extend into the BD mass regime.

In the low mass regime, studies of populations of objects are
generally performed using parameterisations of physical mechanisms
rendering the problem computationally tractable. $\chi^2$ fitting is
used to derive parameters from a grid of model SEDs computed over
ranges of parameters selected by the user
\citep[e.g.][]{bouy_2008,monin_2010,olofsson_2010}. For example, the
inner boundary of the associated dust disc can be placed at a range of
positions, the resulting SEDs derived, and the best fitting model
selected, leading to an estimate of the inner edge radius.  However,
it is known that the flux from the inner edge is a complicated
function of its geometry, composition and temperature, all of which
can change as a function of the stellar properties \citep[see
discussion in][]{mayne_2010}. A clear correlation between the observed
inner edge location and the dust sublimation radius has been found for
higher mass stars \citep{monnier_2002}. Our approach in this work is
to constrain our modelling of BD and disc (BDD, hereafter) systems by
adopting the physical mechanisms that have proved successful on their
higher mass counterparts such as Classical T Tauri stars and Herbig Ae
stars: (i) we constrain our disc structures so that they are in
vertical hydrostatic equilibrium. (ii) The inner edge of the gas disc
is truncated at the co-rotation radius \citep[as such it is a function
of stellar mass and rotational period,
][]{camenzind_1990,koenigl_1991,muzerolle_2003,mohanty_2008}. (iii)
The location of the dust disc inner radius is defined dust sublimation
\citep{tannirkulam_2008}. (iv) The photosphere luminosity is enhanced
by accretion luminosity from material impacting on the BD surface
\citep{bouvier_1995,herbst_2007}. Note that although we are apparently
increasing the physical complexity of the model, we are in fact {\rm
  reducing} the number of free parameters. As a consequence of this
adequate fits to the SEDs act to corroborate the physical assumptions
that underly our models, whilst inadequate fits lead us to question
them. In contrast allowing (for example) the the disc structure to
vary arbitrarily will necessarily result in an improved spectral fit,
but perhaps at the cost of physical insight. Applying such models to
the studies of populations then requires sufficient computing
resource, or an available repository of pre-computed models.

Comparison of data and models is further complicated by temporal
variability. Simulations of the dynamics of circumstellar discs have
suggested that several mechanisms can lead to structural variations
over a range of timescales, some as short as a few days
\citep{min_2009, acreman_2010}. Combined with variations in emissions
from the stellar photosphere from processes such as accretion
\citep{bouvier_1995,herbst_2007}, this leads to significant variations
in the observable properties. However, non-contemporaneous photometric
data sets are often combined and fitted to derived SEDs, without
consideration of such temporal variability. Additionally, previous
studies have found that the optical photometry of BDs and LMS, which
is vital for constraining the stellar parameters, are poorly
represented by simulated stellar atmospheres. Furthermore, as the
slope of the SED across the optical regime is much steeper than for
higher mass systems, these measurements dominate the $\chi^2$ value
for a given fit and so are often omitted from the fit \citep[see for
example][]{guieu_2007}. Discrepancies in the optical photometry,
however, are often the result of converting magnitudes to a standard
photometric system (to match the models), or indeed converting
magnitudes to monochromatic fluxes, using simple transformations or
calibrations. Such transformations are usually only accurate for
main-sequence (MS) stars, and their use will introduce additional
uncertainty in the observed BD fluxes.

We have used the {\sc torus} radiative transfer code to simulate a
large number of BDD systems covering a range of input parameters and
adopting mechanisms accepted for higher mass systems. These
simulations include a representation of dust sublimation, vertical
hydrostatic equilibrium, accretion flux and radiative transfer. These
models have been incorporated into an online fitting tool freely
accessible to the community at \mysite{}. This fitting tool includes
an accurate treatment of the observational photometry so that it can
be compared consistently with the simulated observations. As a proof
of concept, and to aid development of this tool, we have fitted a
sample of Taurus BD and BDD observations collected from the
literature, the results of which we present in this paper. We have
explored differences in our derived parameters from those in the
literature and discuss some specific examples.

This paper is structured as follows. Section \ref{data} introduces the
adopted data-set for the sample of Taurus BDs and BDDs. The detail of
the transformations and generation of uncertainties (the latter to be
applied to the model during fitting) are mentioned in this section and
explained in full in Appendix \ref{data_problems}. The models used,
including the physics represented, are detailed in Section
\ref{models_fitting} alongside the fitting statistics and fitting
process. Details of the calibration of the models and simulated
observables are explained in full in Appendix \ref{calibration_grid}
where the online interface and tools are also described. The results
of the fitting are presented in Section \ref{results} where we also
discuss individual objects and comparison with other works. Finally we
conclude in Section \ref{conclusions}.

\section{The Data}
\label{data}

We have selected and fitted a sample of LMS and BDs from Taurus
presented in \cite{guieu_2007}. \cite{guieu_2007} present photometry
for these objects and split them into those stars with infrared
excesses (and by inference discs) and those without. For the purpose
of this study we have used the same division. We have also performed a
literature search for photometry and derived parameters of these
objects. This has allowed us to develop an understanding of how to
model the uncertainties of our models during fitting and to explore
temporal variations in the observations.

Table \ref{id_table} shows the short name used in this work and the
2MASS identification (from which the coordinates are easily found) to
facilitate cross matching. Table \ref{target_phot} shows the adopted
photometry for each of our target objects and their sources.

\begin{table}
\begin{tabular}{ll}
\hline
Short Name&2MASS ID\\
\hline
\multicolumn{2}{c}{Sources without infrared excess}\\
\hline
KPNO-Tau 4&J04272799+2612052\\
CFHT-Tau 15&J04274538+2357243\\
KPNO-Tau 5&J04294568+2630468\\
CFHT-Tau 16&J04302365+2359129\\
CFHT-Tau 13&J04312669+2703188\\
CFHT-Tau 7&J04321786+2422149\\
CFHT-Tau 5&J04325026+2422115\\
CFHT-Tau 11&J04350850+2311398\\
KPNO-Tau 9&J04355143+2249119\\
CFHT-Tau 2&J04361038+2259560\\
CFHT-Tau 3&J04363893+2258119\\
ITG 2&J04380083+2558572\\
\hline
\multicolumn{2}{c}{Sources with infrared excess}\\
\hline
CFHT-Tau 9&J04242646+2649503\\
KPNO-Tau 6&J04300724+2608207\\
KPNO-Tau 7&J04305718+2556394\\
CFHT-Tau 12&J04330945+2246487\\
BDD399&J04381486+2611399\\
GM Tau&J04382134+2609137\\
CFHT-Tau 6&J04390396+2544264\\
CFHT-Tau 4&J04394748+2601407\\
CFHT-Tau 8&J04411078+2555116\\
BDD304&J04414825+2534304\\
BDD164&J04442713+2512164\\
\hline
\end{tabular}
\caption{Table showing the short name used in this paper and the 2MASS ID from which the coordinates can be found. The
division of sources with and without infrared excess is from
\citet{guieu_2007}. \label{id_table}}
\end{table}

Table \ref{target_phot} also lists our estimates of the uncertainties
for each observation. The uncertainties are not simply the values
presented alongside the observations but also include components which
attempt to account for the dynamical changes, (i.e. the fact the
observations are non-contemporaneous) and inaccuracies in photometric
transformations. We note that although we have listed the
uncertainties in Table \ref{target_phot} we adopt a Bayesian approach
whereby the observations are precise and the uncertainties are a
property of the model. This is equivalent to asking which underlying
model, when combined with uncertainties representing our incomplete
understanding of this model, will reproduce the precise
observations. The construction of the uncertainties and how they are
applied is described in detail in Appendix \ref{data_problems}.

\section{The Models and Fitting}
\label{models_fitting}

In order to model a combined star and disc system one requires a model
of the stellar properties (luminosity, radius etc), the SED of the
stellar photosphere and an initial model of the disc (structure,
composition etc). The disc model is described as an initial model as
some properties such as the vertical scale height or inner edge
location may change (due to modelled physical processes) to some
equilibrium state. Most existing studies of BDD systems involve
parameterisation in order to remove the requirement for detailed
physical modelling. For instance commonly made assumptions are that the
central star emits as a black body, or as a naked photosphere without
additional accretion flux. More subtly, one can simply pick a range of
atmospheres at different effective ($T_{\rm eff}$) temperatures,
surface gravities ($\log (g)$) and bolometric luminosities ($L_{\rm
  bol}$) to create a fitting grid, as opposed to using stellar
interior models to derive these properties. Additionally, the inner
edge and vertical structure of the disc can simply be prescribed
analytically without inclusion of the detailed physics of magnetic
truncation, dust sublimation and vertical hydrostatic equilibrium.

\subsection{Dust Distribution}
\label{dust}

We have adopted a similar dust model to \cite{wood_2002} with the size
distribution of dust particle given by
\begin{equation}
\label{particle_dist}
n(a)da=C_ia^{-q}\times exp^{[-(a/a_c)^p]}da
\end{equation}
where $n(a)da$ is the number of particles of size $a$ (within the
interval $da$), $a_{\rm c}$ is the characteristic particle size, with
$p$ and $q$ simply used to control the shape of the
distribution. $C_i$ controls the relative abundance of each
constituent species ($i$) in the dust. We have adopted the following
values for each parameter $q=$3.5, $p=$0.6 with $a_{\rm
  c}=$50\,$\mu$m, also with an associated maximum and minimum grain
size of $a_{\rm min}=$5 nm and $a_{\rm max}=$1mm. We set $C_i=$1 and
adjusted the species using a grain fractional abundance \cite[an
equivalent process to that of][]{wood_2002}, using the solar
abundances of \cite{asplund_2006}. The subsequent opacities are
constructed from tabulated opacities for each species. More detail of
the dust model can be found in \cite{mayne_2010}.

\subsection{The Grids}
\label{our_grids}

In order to construct the stellar photosphere we interpolate the
stellar interior models for a given age and mass and subsequently
interpolate the stellar atmosphere models to the required $T_{\rm
  eff}$, $L_{\rm bol}$ and $\log (g)$. This method is in contrast to
the traditional method of selecting spectra over a range of $T_{\rm
  eff}$, $\log (g)$ values and scaling the $L_{\rm bol}$ to match the
observed fluxes at wavelengths shortward of the peak in disc emission
\citep[$\sim$3$\mu$m,][]{dullemond_2001}\footnote{In effect this is a
  `by-eye' pre-fitting to select an input stellar photosphere.}. The
accretion flux is then added, as prescribed in \cite{mayne_2010}, upon
adoption of a rotation period (to derive the co-rotation radius). A
disc component is then added and the radiative transfer through the
system modelled to produce a simulated SED.

The modelling of the BDD systems was performed using the {\sc torus}
radiative transfer code. {\sc torus} is detailed in
\cite{harries_2000} and was subsequently updated by
\cite{harries_2004} and \cite{kurosawa_2004}. {\sc torus} uses the
method of \cite{lucy_1999} to solve radiative equilibrium on an AMR
mesh. The code can also self-consistently solve the equation of
vertical hydrostatic equilibrium and dust sublimation for the disc
\citep{tannirkulam_2007}. The {\sc torus} code has been extensively
benchmarked \citep[for radiative transfer in discs
against][]{pinte_2009}. The addition of variable dust sublimation is
described in \cite{mayne_2010}. We have constructed two grids using
different assumptions about the properties and physics within the star
disc system.

\subsubsection{Ad-hoc Grid}
\label{ad_hoc}

The first model grid includes the effects of accretion. We place the
inner edge of the gaseous and dust disc at the co-rotation radius and
prescribed an initial analytical vertical disc structure. The initial
disc structure is constructed using the analytical relationships (for
the disc density and scaleheight) from \cite{mayne_2010}, $\rho
=\rho_0\frac{R_*}{r^{\alpha}}exp(-\frac{1}{2}[z/h(r)]^2)\propto
r^{-\alpha}$ and $h=h_0(r/R_*)^{\beta}$ where $r$ and $z$ are the
radial and vertical cylindrical polar coordinates. The surface density
($\Sigma$) profile from these equations is $\Sigma(r) \propto
r^{(\beta-\alpha)}$, and $\beta$ and $\alpha$ are set such that
$\Sigma(r) \propto r^{-1}$. This initial disc configuration follows
the analytically prescribed disc structures of previous works, where
SED fitting has been used to derive, for instance the $\beta$
parameters or inner edge locations. These initial models are
subsequently treated in one of two ways. Either, the vertical
structure is fixed in the initial configuration and dust sublimation
is modelled explicitly alongside a calculation of radiative
equilibrium. Alternatively, hydrostatic equilibrium is enforced (via
adjustment of the vertical structure and therefore $\beta$) in concert
with the radiative equlibrium and dust sublimation calculations. This
grid is termed the ad-hoc grid and is an extension of the grid
featured in \cite{mayne_2010}. It covers only BD mass objects for the
ages of 1 and 10\,Myrs.

\subsubsection{Semi-empirical Grid}
\label{semi_emp}

The second grid is designed for non-accreting systems. In this case
the inner edge is placed at the semi-empirically derived dust
sublimation radius \citep[only applicable for negligible accretion
rates, see][]{whitney_2004} and the vertical structure analytically
prescribed, as for the ad-hoc grid. In this grid, the vertical
structure and inner edge location do not change as dust sublimation
and vertical hydrostatic equilibrium are not included. This grid is
termed the semi-empirical grid and has an expanded mass and age range
covering masses from 0.01 to 1.40 M$_{\odot}$ and ages from 1 to
10\,Myrs.

Note that our construction of the initial disc structures assumes that
the dust is evenly distributed through the disc and follows the gas
structure. We therefore implicitly exclude the possibility of dust
settling in the disc for those models within the ad-hoc grid which
enforce a hydrostatic structure on the gas distributions. However, the
remaining models (i.e. a subset of the ad-hoc grid and all of the
semi-empirical models) where the disc scaleheight structure is not
required to fulfil hydrostatic equilibrium, can mimic the settling of
the dusty component of the disc towards the midplane. Further
complexity may be introduced by allowing the dust scaleheight to vary
as a function of grain size
\citep[e.g.][]{tannirkulam_2007,tannirkulam_2008} although we believe
that additional observable constraints \citep[such as scattered light
imaging see][for example]{wood_2002,watson_2007} over and above SED
fitting would be necessary to determine the degree of
sedimentation. However, if dust settling or sedimentation is likely to
be an important factor for a particular object then the semi-empirical
grid or non-hydrostatic ad-hoc models are to be preferred. The details
of which physical process are included in each grid are shown in Table
\ref{grid_tab}\footnote{This information is also available online at
  \mysitecalib{}.}.

\setcounter{table}{2}
\begin{table*}
\begin{tabular}{|l|l|l|l|}
\hline
Grid Reference&ad-hoc Grid&semi-empirical Grid\\
\hline
\multicolumn{3}{|c|}{Physical Processes}\\
\hline
\multirow{2}{*}{Stellar Interior}&\multirow{2}{*}{Dusty-`00$^{(1)}$ }&$M_*>=0.02M_{\odot}$:BCAH98$^{(2)}$\\
&&$M_*<0.02M_{\odot}$: Dusty-`00$^{(1)}$\\
\multirow{3}{*}{Stellar Atmosphere}&\multirow{3}{*}{AMES-Dusty$^{(1)}$}&$T_{\rm eff}<$1400\,K:AMES-Cond$^{(3)}$\\
&&1400\,K$<T_{\rm eff}<$2500\,K:AMES-Dusty$^{(1)}$\\
&&$T_{\rm eff}>$2500\,K:NextGen$^{(4)}$\\
Accretion Luminosity&Yes&Yes-negligible\\
\multirow{2}{*}{Vertical Structure}&VHE&\multirow{2}{*}{Static}\\
&Static&\\
Inner Dust Boundary&Variable Sublimation&Empirical Sublimation Radius\\
Inner Gas Boundary&Co-rotation Radius&Co-rotation Radius\\
Dust Scattering&Isotropic&Full Mie Phase Matrix\\
\hline
\multicolumn{3}{|c|}{Parameters Covered}\\
\hline
Age (Gyrs)&0.001 \& 0.01&0.001--0.01 ($\Delta$=0.001)\\
Mass (M$_{\odot}$)&0.01--0.08 ($\Delta$=0.01)&0.01--0.10 ($\Delta$=0.01), 0.10--1.40 ($\Delta$=0.1)\\
$T_{\rm eff}$&Derived&Derived\\
$\log (g)$&Derived&Derived\\
Luminosity&Derived&Derived\\
Accretion Rate \logmdot&$-$12-- $-$6 ($\Delta$=$-$1)&$-$12\\
Areal Coverage (\%)&1, 10&10\\
Disc Mass (M$_*$)&\multicolumn{2}{|c|}{0.0, 0.01, 0.001}\\
Disc Outer Radius (AU)&100, 300&50, 100\\
$\alpha$ ($\beta^*$)&Derived (VHE), 2.00 (1.00), 2.10 (1.10), 2.25 (1.25)&2.00 (1.00), 2.10 (1.10), 2.25 (1.25)\\\
Scaleheight at 100AU (AU)$^{(2)}$&25&10\\
Inclination ($^{\circ}$)&\multicolumn{2}{|c|}{0, 27, 39, 48, 56, 64, 71, 77, 84, 90}\\
\hline
\multicolumn{3}{|l|}{References. (1) \citet{chabrier_2000}, (2) \citet{baraffe_1998}, (3) \citet{allard_2000} and (4) \citet{hauschildt_1999}}\\
\hline
\end{tabular}
\caption{Table describing the physical processes modelled in the two grids and the range of parameters the grids cover. (1) See explanation in Section \ref{models_fitting}. (2) this initial scaleheight is allowed to adjust if vertical hydrostatic equilibrium is enforced. \label{grid_tab}}
\end{table*}

\subsubsection{Photometry}
\label{photometry}

In most cases parameters are derived from photometric observations,
either magnitudes or monochromatic fluxes, or conversions of the
former to the latter. Therefore our SEDs need to be converted to
apparent SEDs (via the application of a distance and extinction) and
photometric observations simulated. We have simulated such photometric
observations for a range of commonly used filter systems and the
conversions, derivations of photometric measurements and definitions
of the photometric systems are detailed in Appendix
\ref{calibration_grid}. This information is important for observers
who intend to use the fitting tool, as errors when adopting the
incorrect photometric systems can be large (as mentioned in Section
\ref{data} and explained in Appendix \ref{data_problems}).

\subsection{Fitting Statistic}
\label{fitting_statistic}

As shown in Table \ref{target_phot} we have collected a large range of
photometric observations in both magnitude and monochromatic flux
systems. In order to perform a fit we must use consistent units and
therefore we convert magnitudes to monochromatic fluxes. This
conversion is performed in a rigorous manner by reversing the process
described in Appendix \ref{calibration_grid} used to derive a
simulated magnitude and then applying the process required to
construct monochromatic fluxes (again described in Appendix
\ref{calibration_grid}). This allows us to be confident that the
models and data are being compared in a consistent manner.

Practically, fitting is performed by maximising the likelihood of the
data being drawn from a given model. As the probability of the data
being drawn from the $i$th model, $P_m$ is
\begin{equation}
P_m=\prod_i \!e^{-\frac{(x_i-m_i)^2}{2\sigma_i^2}} \, \times \prod_j \! Q(\frac{x_j^u-m_J}{\sigma_J}),
\label{prob}
\end{equation}
where $x_i$ and $m_i$ are the $i$th data and model points (these are
fluxes from the SED, monochromatic fluxes or magnitudes converted to
fluxes), and $\sigma_i$ the uncertainty of the $i$th data point. The
second part of the equation derives the likelihood from the observed
upper limits $x_j^u$ is the $j$th upper limit with uncertainties
$\sigma_j$, $M_j$ the corresponding model flux and where $Q$ is given
by,
\begin{equation}
Q(x)=\frac{1}{2}-\frac{1}{2}{\rm erf}(\frac{x}{\sqrt{2}}),
\label{Q}
\end{equation}
where ${\rm erf}(x)$ is the error function. This in effect assumes
that the $\sigma$ is taken from a measurement of zero, but allows
negative measured fluxes. Ideally, one would provide the measured flux
and uncertainty as oppose to an upper limit, as this ensures the
fitting routine does not have to make further assumptions about the
data. In practice we choose to minimise the statistic $-2{\rm
  ln}(P_m)$, which as described in \cite{naylor_2006} reduces to
$\chi^2$ for the case of data points with Gaussian uncertainties in
one dimension \citep[see also][]{cash_1979}. In practice therefore we
minimise the following statistic,
\begin{eqnarray}
\chi^2_{\rm pseudo}=-2{\rm ln}(\prod_i \!e^{-\frac{(x_i-m_i)^2}{2\sigma_i^2}} \, \times \prod_j \! Q(\frac{x_j^u-m_J}{\sigma_J}))\\
=\chi^2+-2{\rm ln}(Q(\frac{x_j^u-m_J}{\sigma_J})).
\label{min_stat}
\end{eqnarray}
The online fitting tool then returns the $N$ best fitting models to
the user, with associated plots of $\chi^2_{\rm pseudo}$ for each
parameter. Uncertainties are derived using the 68\% confidence
interval by converting the given $\chi^2$ values to a probability
($e^{-(\frac{\chi^2_{\rm pseudo}}{2})}$) and integrating from the most
likely value until 68\% of the total probability has been
reached. Additionally, a Spearman's rank correlation coefficient is
derived for the best fitting 10\% of models for each pair of free
parameters, giving an indication of degeneracy. If two parameters are
found to have evidence of degeneracy 2D $\chi^2_{\rm pseudo}$ maps are
created and presented alongside the fits, although this is only an
indication of degeneracy.

The structure of the online fitting tools and comments on its use are
detailed in Appendix \ref{calibration_grid}. Appendix
\ref{calibration_grid} also includes details of the other tools
available at \mysite{} such as an SED browsing tool and isochrone
server.

\section{Results}
\label{results}

In traditional SED fitting values of extinction and distance are set
to single values (i.e without uncertainties) for a given fit, allowing
the number of free parameters to be reduced by two. However, this
discards useful information as the distance and extinction
uncertainties are well founded constraints. The depth (i.e. range of
distances in a given population) of a given group of stars contributes
little to shifts within the HR-Diagram
\citep{mayne_2007,mayne_2008}. However, when fitting monochromatic
fluxes, or SED sections, to model SEDs variations in distance to
individual objects can affect the flux significantly. For instance
when fitting objects within Taurus the distance is usually set to
140\,pc, or in some cases a modest distance uncertainty of $\pm$10\,pc
is used \citep{bertout_1999}. However, the distance range, from the
most precise derivations of distances to three member stars, is
161$\pm$0.9, 147 and 130\,pc, derived from interferometric
observations \citep{torres_2009}. Therefore a more realistic distance
range is 130--165\,pc, which can lead to flux variations of up to a
factor of 1.6. Assuming a distance of 140\,pc for all objects in the
Taurus region allows us to reduce the number of free parameters, but
is less desirable than using a realistic range of
distances. Uncertainties should also be adopted to apply to the
extinction to each target object. The extinctions shown in Table
\ref{target_phot} have all been sourced from \cite{guieu_2007} (for
consistency), where they are adopted as single values (without
uncertainty). These values, in turn, are derived in \cite{guieu_2006}
where an uncertainty of $\pm$0.8 mags is quoted, which we have
adopted. If we do not include such strongly physically motivated
external constraints, like distance ranges, it is likely that, due to
degeneracies of the parameters, the remaining free parameters will be
optimistically constrained.

\begin{figure}
  \includegraphics[scale=0.3,angle=90]{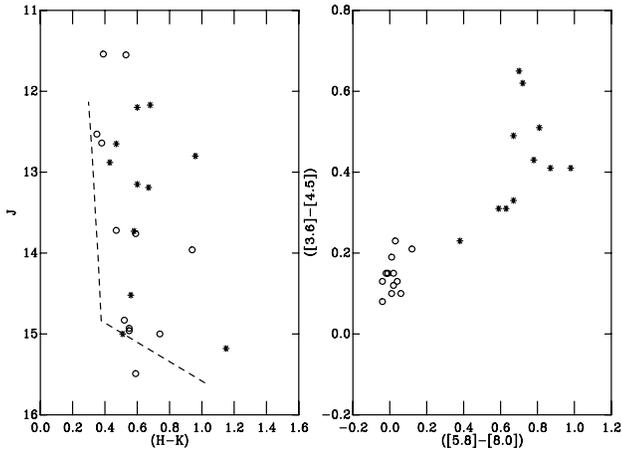}
  \caption{Figure showing the stars with (asterisks) and without
    (empty circles)
    infrared excesses from Table \ref{target_phot}. The left panel
    shows a $J, H-K$ CMD and the right panel a ([3.6]$-$[4.5]),
    ([5.8]$-$[8.0]) CoCoD. The dashed line on the left panel shows an
    isochrone at 1\,Myr for masses between $0.01$ and $0.08M_{\odot}$
    adjusted to a distance of 140\,pc and an extinction of
    $A_V=1$. \label{cmd}}
\end{figure}

Figure \ref{cmd} shows the infrared and near-infrared colour-magnitude
(left panel) and colour-colour (right panel) diagrams (CMD \& CoCoD)
for targets from Table \ref{target_phot}. A 1\,Myr isochrone
constructed using the interior models and atmospheric models of
\cite{chabrier_2000} adjusted to a distance of 140\,pc and an $A_{\rm
  V}=1$, is overlaid on the CMD (left panel).

The left panel of Figure \ref{cmd} shows that the stars with
(asterisks) and without (empty circles) discs lie redward and brighter
than the illustrative isochrone. For the stars without discs several
factors lead to a shift away from the model isochrone. Firstly, some
objects lie behind extinction greater than the illustrative value
selected (i.e. $A_{\rm V} >$1.0). Additionally, some stars without
discs have luminosities above that predicted by the 1\,Myr isochrone at
a given colour or $H-K$ (for instance ITG 2, discussed in more detail
in Section \ref{problem_bd}). Finally, some stars will also have a
mass greater than the maximum mass included in the illustrative
isochrone (0.08$M_{\odot}$, for instance CFHT-Tau 5 also discussed in
more detail in Section \ref{problem_bd}). Of course, for the stars
with discs the same shifts will be present in addition to reddening of
the system caused by the reprocessing of stellar light by the disc. As
expected, the stars with infrared excesses are also clearly
distinguishable in the right hand IRAC CoCoD, and the delineation
agrees with the predicted values of ([3.6]$-$[4.5]) $>+$0.21 and
([5.8]$-$[8.0]) $>+$0.13 \citep{mayne_2010}.

It is clear that the stars in Figure \ref{cmd} do not lie along the
isochrone, and therefore are poorly represented, in a CMD, by a simple
coeval (naked photosphere) population at 1\,Myr at a single distance
and extinction with negligible accretion. The meaning and veracity of
the inferred ``age spreads'' has been discussed at length in the
literature \cite[for a summary see][]{mayne_2008}. Several studies of
classical T-Tauri-stars (CTTS) have shown that active accretion and
variations in distance cannot fully explain the spreads of stellar
populations \citep{burningham_2005,da_rio_2010}. However, accretion,
both past and current, has been identified as a major contributor to
spreads in both luminosity and radii of young stars, and its effects
on the magnitudes and colours of brown dwarfs with discs has been
shown to be significant \citep{mayne_2010}. Indeed \cite{mayne_2010}
showed that the effects of accretion and disc presence on
characterisation are more significant for lower mass systems.

A full discussion of the veracity of age spreads is beyond the scope
of this paper. However, it is clear that fitting for an assumed coeval
population can be performed by adopting a range of ``isochronal
ages'', whether they represent a stars true age or not. Additionally,
a range of accretion rates and inclinations (leading to differing
levels of occulation of the star by the disc) may be sufficient to
provide the spreads observed, with the addition of a small intrinsic
age spread which is less than the disc dissipation time
\citep{littlefair_2011,jeffries_2011}.

\subsection{Literature Stellar Parameters}
\label{literature_parameters}

For each of the objects within our sample we have attempted to find
all the stellar parameters that authors have either derived or adopted
for modelling and analysis of these objects. Table \ref{star_prop}
presents, for each object, each value of the stellar parameter we have
found, with the reference in superscript parenthesis (a legend is
included in the table caption).

Table \ref{star_prop} does not contain information on how each
quantity was derived (or indeed whether it was simply adopted), which
is of course important. However, in cases where parameters were simply
adopted in order to calculate or derive a dependent parameter it is
still important to note the values of the input, which will, of
course, determine the value of the derived property. For example
\cite{muzerolle_2005} adopted approximate values for stellar
parameters such as $T_{\rm eff}$ and mass, in order to model accretion
and derive accretion rates for such objects, whereas, $T_{\rm eff}$
for these objects were derived from fitting to spectral type templates
in \cite{luhman_2004}. Of particular note is that almost all of the
stellar masses derived for these objects rely, in some way, on
isochrones, which are known to be unreliable for low mass stars
\citep[see discussion in][]{mayne_2008}, under the assumption of
coevality at an age of 1\,Myr, and crucially, often at a set distance
of 140\,pc.

\subsection{Naked Stars}
\label{naked_stars}

The naked stars, which we assume are not accreting (supporting
evidence is presented in Table \ref{accrete_table} discussed in
Section \ref{accretion}), were fit using the semi-empirical grid to
allow a range of isochronal ages as observed in the CMD, Figure
\ref{cmd}\footnote{We note that these values should not be adopted as
  the real ages of the system. As discussed in Section \ref{results}
  ages derived from isochrones are currently inaccurate for LMS and
  BDs.}. There is no difference, except the accretion luminosity,
between the naked photospheres used in the ad-hoc and semi-empirical
grids, although the latter includes a wider range of ages and
masses\footnote{Of course, the models of systems at higher masses will
  include stellar atmospheres and interiors appropriate for the higher
  $T_{\rm eff}$ values.}.

\begin{figure}
\includegraphics[scale=0.38,angle=0]{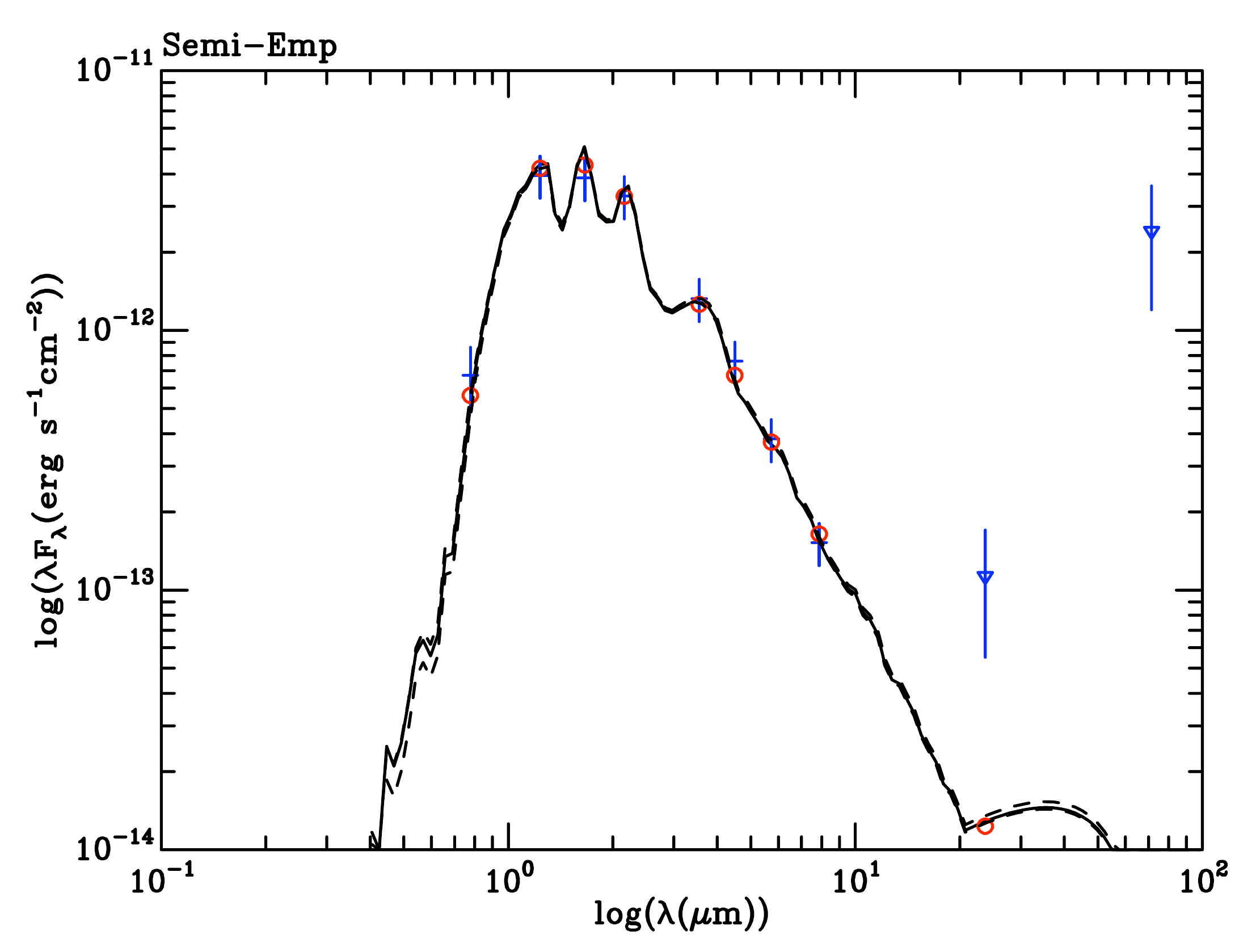}
\caption{SEDs of the best fitting models from semi-empirical grid,
  showing fits to the entire grid (dashed green line), using a set
  distance and extinction (dashed black line) and a restricted grid
  (solid black line). The monochromatic fluxes derived from the best
  fitting model from the restricted grid are shown as red circles and
  the observations as blue crosses (fluxes) or triangles (upper
  limits) with uncertainties shown as bars. Note, the fits for this
  object are closely matched and therefore the resulting SEDs show
  little deviation, meaning the lines for each fit are essentially
  over plotted and some may not be visible.\label{eg_naked}}
\end{figure}

\begin{figure*}
\includegraphics[scale=0.45,angle=90]{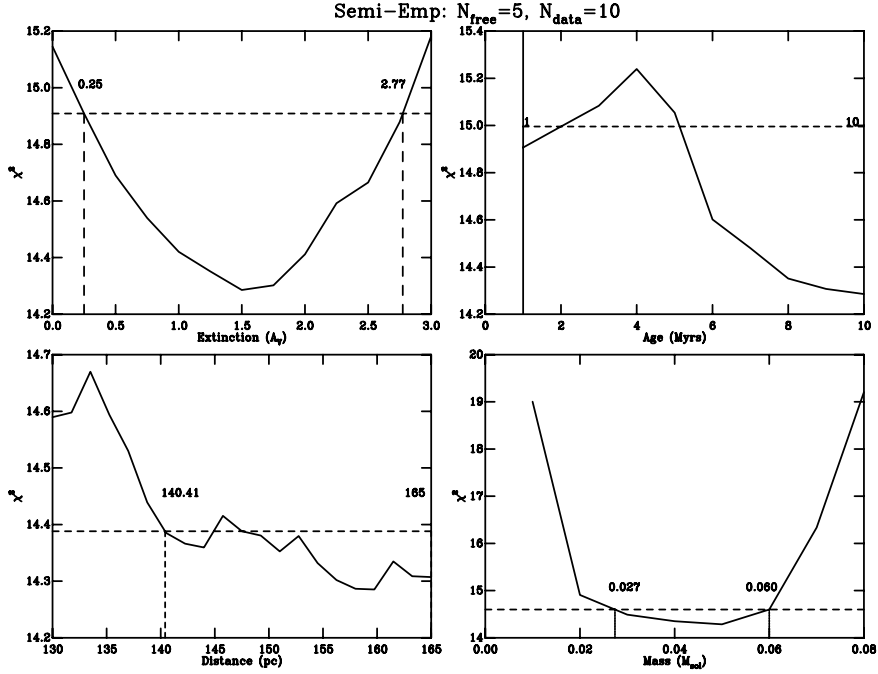}
\caption{Figure showing the $\chi^2$ distributions for the fit to the
  whole semi-empirical grid for CFHT-Tau 16, where the dashed lines
  show the 68\% confidence limits. For each plot the remaining
  parameters have been held at their best fitting
  values.\label{eg_chi_one}}
\end{figure*}

Figure \ref{eg_naked} shows an example of a fit to CFHT-Tau 16 using
the semi-empirical grid of models and the observations presented in
Table \ref{target_phot}. We have fitted this star in three different
ways. Firstly, we have fitted the star allowing all of the parameters
of the model grid to be free. Secondly, we have fitted the star by
setting the distance to 140\,pc and extinction to the single values
from Table \ref{target_phot} \citep[following][]{guieu_2007}. Finally,
we have performed a fit using observations, shown in Table
\ref{star_prop} to constrain the model grid parameters alongside
distance and extinction uncertainties. In the case of CFHT-Tau 16, all
of the fits yielded similar results. The resulting $\chi^2$ parameter
space for the case where the parameters are all free is shown in
Figure \ref{eg_chi_one}.

Figure \ref{eg_naked} shows that in the case of CFHT-Tau 16 the best
fitting SEDs, for each fit type, are well matched. This is reflected
in the consistency of the derived parameters, for this object, shown
in Table \ref{naked_best_par} (described later). However, Figure
\ref{eg_chi_one} shows that even in this excellent case some
parameters remain formally unconstrained when the confidence intervals
are examined. The 68\% confidence intervals for the extinction and
mass parameters are within the fitted parameter space, but the the
distance upper bound is unconstrained. The distance ranges selected,
however, are strongly constrained by the range of distances found to
existing members and should not be extended. It is usually inadvisable
for an object to be fitted to the entire grid, since degeneracies in
the parameter space can lead to derivations of incorrect
parameters. For instance the effects of increasing mass, decreasing
distance and increasing accretion rate all act to increase the
luminosity of the system and therefore can lead to compensating
errors, especially when combined with adjustments in extinctions. This
is especially significant for lower temperature stars as the region of
the SED sensitive to stellar and disc properties overlap
\citep{clarke_2006}. Indeed \cite{mayne_2010} show that a population
of accreting BDD systems can appear both spectroscopically and
photometrically similar to a population of CTTS under heavy extinction
and at a different distance.

\begin{figure*}
\includegraphics[scale=0.65,angle=90]{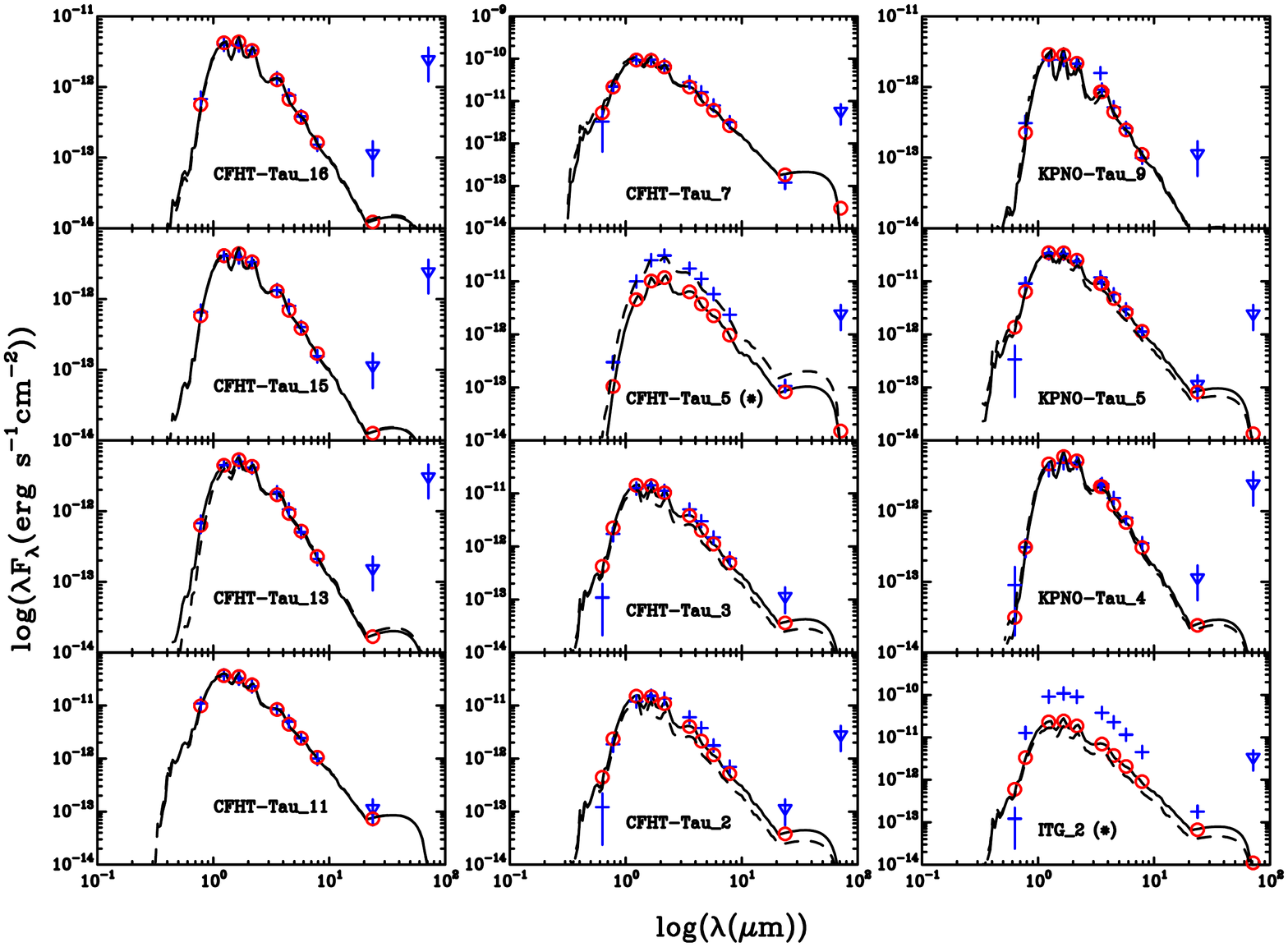}
\caption{The best fitting SEDs for the naked objects, all from the
  semi-empirical grid. The solid black lines are the SEDs from the
  restricted fits and dashed lines show the best fitting SEDs when the
  distance and extinction have been set. The monochromatic fluxes
  derived for the best fitting restricted model are shown as circles
  and the observed fluxes, (from Table \ref{target_phot}) are shown as
  crosses, or triangles for upper limits, with the bars denoting the
  uncertainties. It can be seen that we failed to find a satisfactory
  fit for ITG 2 and CFHT-Tau 5 denoted by (*). \label{naked_best_sed}}
\end{figure*}

For the remaining sample of naked BDs we have fitted using the final
two methods applied to CFHT-Tau 16, i.e. fitting using a set distance
and extinction, and fitting using a restricted grid with all
parameters constrained by observations (in other words we do not fit
using a grid where all the parameters are free). Figure
\ref{naked_best_sed} shows the best fitting SEDs for each of the naked
objects. The solid and dashed lines show the fits allowing distance
and extinction to vary over the prescribed ranges and setting them to
fixed values, respectively. The model fluxes, for the restricted case,
are shown as circles and the observations as crosses, with triangles
denoting upper limits.

The parameters derived for the restricted fitting grid are shown in
Table \ref{naked_best_par}. Table \ref{naked_best_par} shows the best
fitting values for the independent variables with 68\% confidence
intervals and the dependent variables associated with the best fitting
model. Some of the confidence intervals indicate a 0.0 increment in
either direction, this is sometimes caused by the best fitting value
reaching a boundary (for example 165\,pc in the range 130--165\,pc as
is the case for CFHT-Tau 16). This may indicate that the parameter is
poorly constrained, however, as the boundaries for the fitting grid
have been chosen using external literature constraints this is not
necessarily a problem.

Comparison between the values derived in Table \ref{naked_best_par}
and the literature values (Table \ref{star_prop}) show reasonable
agreement. For most objects our derived masses match, within the
uncertainties, those found in the literature and the effective
temperatures match to within $\sim$100\,K, with the exception of ITG
2, CFHT-Tau 5 (which are discussed in Section \ref{problem_bd}) and
KPNO-Tau 4. For KPNO-Tau 4 the discrepancy in the effective
temperatures is large with a difference of $\sim$350\,K. This is most
likely caused by this object existing at the boundary in our grid
between the AMES-Dusty \citep{chabrier_2000} and AMES-Cond
\citep{allard_2000} models, which as shown in \cite{mayne_2010} leads
to a significant disconuity in photometric magnitudes.

For most cases allowing the distance to change over the observed
range, and adding an uncertainty to the extinction, results in
parameters which are consistent (within the uncertainties) with those
derived with a set distance and extinction (excepting ITG 2 and
CFHT-Tau 5). However, the best fitting values are often slightly
different, meaning the derived dependent variables do vary. For
example changes in the $T_{\rm eff}$ are, for the most extreme cases
of CFHT-Tau 11 and KPNO-Tau 9, 134 or 289\,K respectively, however for
the remaining fits they are below 100\,K.

\subsubsection{Problem Fits}
\label{problem_bd}

\begin{figure}
\includegraphics[scale=0.38,angle=0]{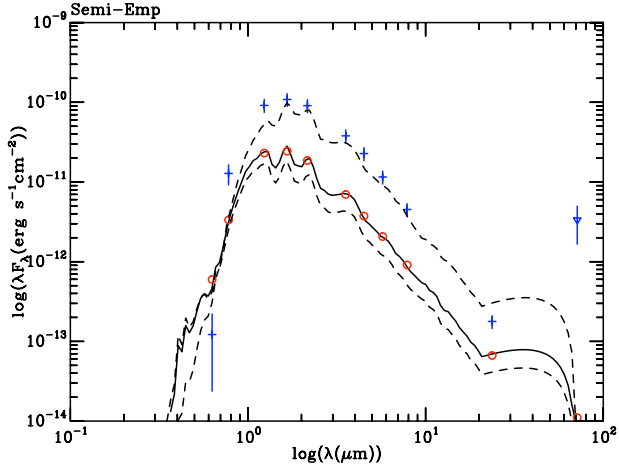}
\caption{SEDs and monochromatic fluxes for ITG 2 where symbols have
  the same meanings as Figure \ref{eg_naked}.\label{itg_2_one}}
\end{figure}

\begin{figure}
\includegraphics[scale=0.38,angle=0]{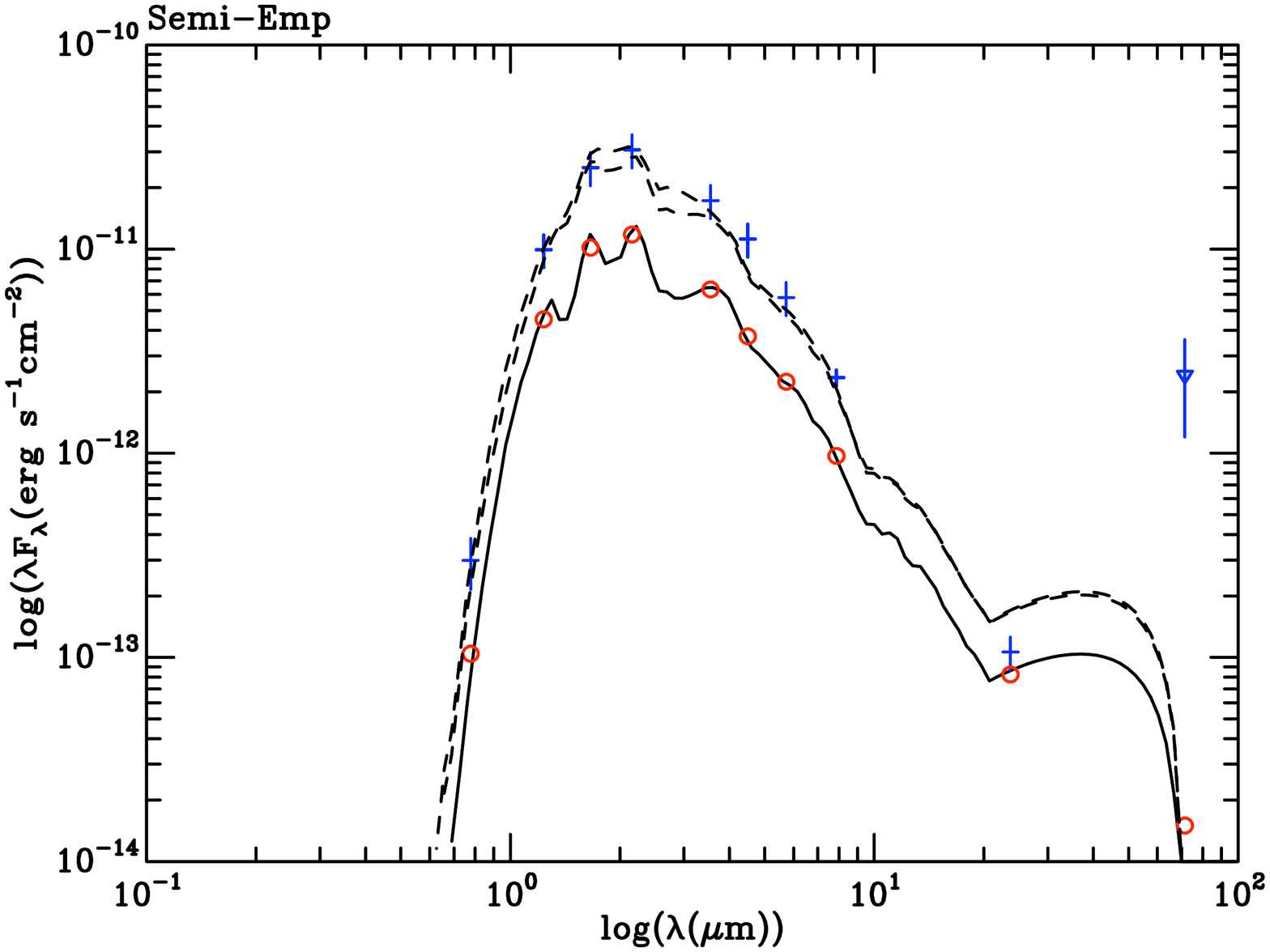}
\caption{SEDs and monochromatic fluxes for CFHT-Tau 5 where symbols
  have the same meanings as Figure \ref{eg_naked}.\label{cfht-tau_5_one}}
\end{figure}

We were unable to achieve acceptable fits using a restricted parameter
space, to the objects ITG 2 and CFHT-Tau 5. For ITG 2 setting the
distance and extinction and allowing the rest of the grid parameters
to run free also resulted in a poor best fit model. For CFHT-Tau 5,
however, an acceptable fit was found using a set distance and
extinction but allowing the stellar mass to increase significantly. To
explore the problems with these objects we have also fitted them using
the entire grid. The resulting best fits are presented in Figures
\ref{itg_2_one} and \ref{cfht-tau_5_one}, alongside those already
presented in Figure \ref{naked_best_sed}. For ITG 2 the only way we
could derive a best fitting SED close to the observed data points was
to allow the mass of the star ($M_*$) to increase to 0.2 M$_{\odot}$
and allow the extinction and distance to increase to $\sim$5 A$_{\rm
  V}$ and 165\,pc respectively. The resulting star has a much higher
effective temperature of 3193\,K, $\log (g)$ of 3.40, radius of 1.45
R$_{\odot}$ and a luminosity of $\log\left( L/L_{\odot}\right)
$=$-$0.700. These values are inconsistent with the literature
properties presented in Table \ref{star_prop}. For CFHT-Tau 5 the best
fitting SED returned values of, M$_*$=0.1 M$_{\odot}$,
distance=140.50\,pc, A$_{\rm V}$=0.50, $\log \left( L/L_{\odot}
\right)$=$-$1.160, T$_{\rm eff}$=3002, $\log (g)$=3.45 and a radius of
0.968R$_{\odot}$. Again, as with ITG 2, these parameters are
inconsistent with the literature values presented in Table
\ref{star_prop}.

These objects are at much higher luminosities, for their given masses,
than the youngest (i.e. 1\,Myr) isochrone. This means that these stars
are poorly fit by both of our grids. Hence, we can only achieve these
luminosities by increasing the stellar mass and compensating for the
SED shape by increasing the extinction. Table \ref{star_prop} shows
that CFHT-Tau 5 has a similar luminosity to CFHT-Tau 7, however, the
latter object has a significantly higher mass and is, therefore, not
affected by this problem. This limitation means that for very,
``isochronally'' young stars we will be unable to reproduce their SEDs
with our current grid. We are currently running a grid of models which
are divorced from isochronal theory and will extend to this region of
parameter space (explained in more detail in Section
\ref{conclusions}). It is interesting to note that the SEDs of such
objects cannot be reproduced if we allow the accretion rates to
increase, even though they are naked stars \citep[some supposed naked
stars have been found with non-negligible accretion rates][these are
discussed in Section \ref{accretion}]{kennedy_2009}. The accretion
diagnostics for the sample have been collected (Table
\ref{accrete_table}) and are discussed later in Section
\ref{accretion}.

\subsection{Stars with Discs}
\label{stars_discs}

For the BDD objects we have fitted to the ad-hoc grid to allow for a
range of accretion rates, again using either a set distance and
extinction or the restricted grid with varying distance and
extinction. One important note is that, as discussed in Section
\ref{ad_hoc}, the observed dust disc stuctures may not appear in
hydrostatic equilibrium due to grain settling. Therefore, we have
included the ad-hoc models with prescribed disc structures in the
fitting process. This may also be a reason to favour the
semi-empirical models for systems with significant grain settling, but
with negligible accretion rates.

\begin{figure*}
\includegraphics[scale=0.65,angle=90]{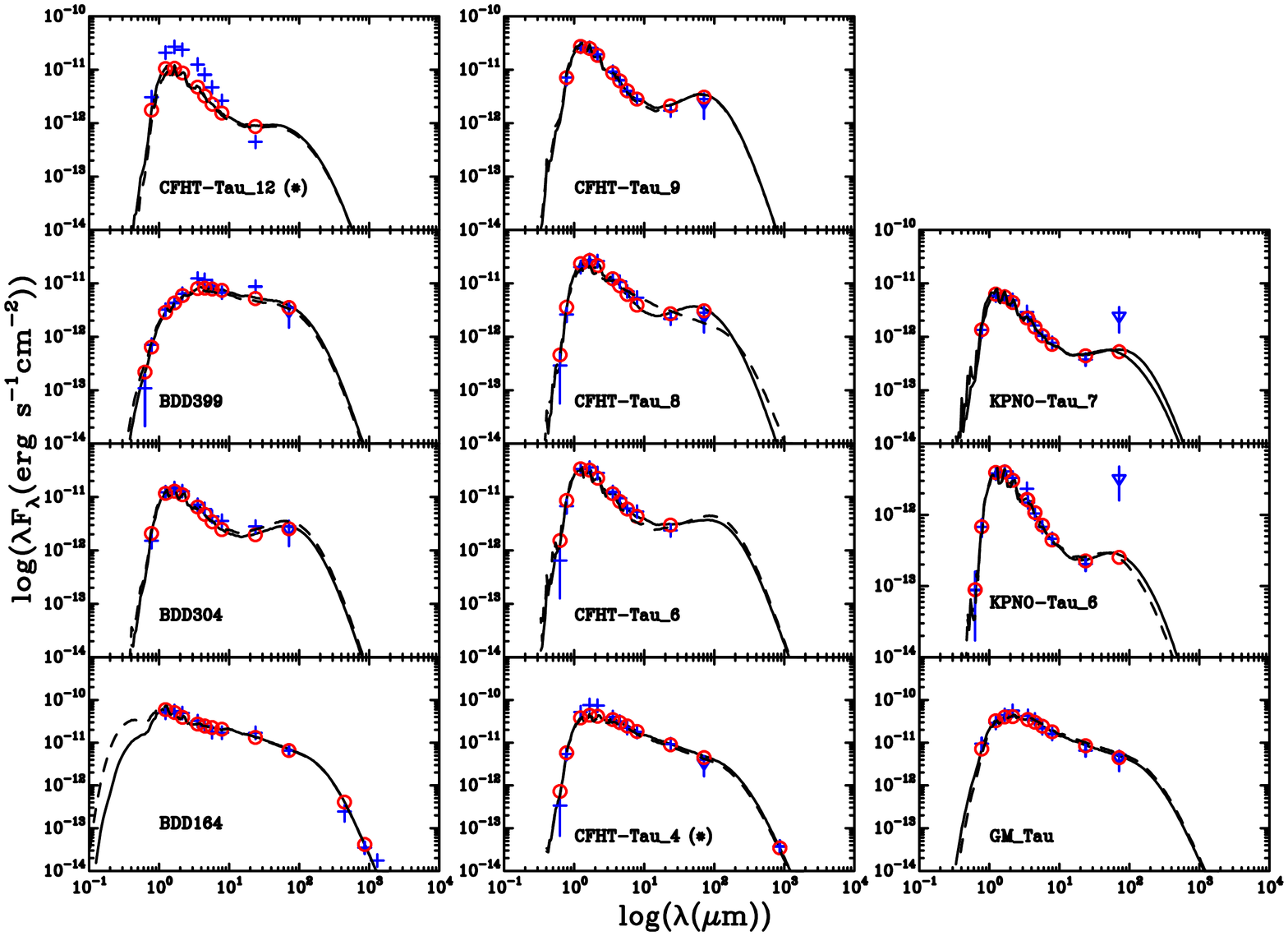}
\caption{The best fitting SEDs for the BDD objects, from the ad-hoc
  grid. The solid black lines are the SEDs from the restricted fits
  and dashed lines show the best fitting SEDs when the distance and
  extinction are fixed. The monochromatic fluxes derived for the best
  fitting restricted model are shown as circles and the observed
  fluxes, from Table \ref{target_phot}, are shown as crosses, or
  triangles for upper limits, with the bars denoting the
  uncertainties. It can be seen that we failed to find a satisfactory
  fit CFHT-Tau 12 and, to a lesser extent, CFHT-Tau 4, which are
  denoted by (*). \label{disc_best_sed_AH}}
\end{figure*}

Figure \ref{disc_best_sed_AH} shows the best fitting SEDs for the BDD
objects from the ad-hoc grid. As for Figure \ref{naked_best_sed} the
solid and dashed lines show the fits allowing distance and extinction
to vary over the prescribed ranges (the restricted fits) and those
with distance and extinction set at fixed values, respectively. The
model fluxes, for the restricted case, are shown as circles and the
observations as crosses, with triangles denoting upper limits.

Figure \ref{disc_best_sed_AH} shows that the fits do not change
significantly upon fixing the distance and extinction, with the
exception of CFHT-Tau 8. It is clear that all observations are well
fit with accreting 1\,Myr disc structures drawn from the ad-hoc grid,
again excepting CFHT-Tau 12. The best fitting parameters derived for
the BDD systems using a restricted grid drawn from the ad-hoc model
grid are presented in Table \ref{disc_best_par_AH}

We found that fixing the distance and extinction only affects the
derived best fit parameters significantly in a few examples when using
the ad-hoc grid. For example, the best fitting mass for BDD304
increases by 0.02M$_{\odot}$ when the distance and extinction are
set. For most objects the derived accretion rates are consistent with
those found in previous studies (see Table
\ref{accrete_table}). Additionally, in all cases except CFHT-Tau 12
and CFHT-Tau 9 the parameters derived for the targets using the
accretion models (i.e. ad-hoc) are consistent with the spread of
values presented in Table \ref{star_prop}. It is clear from Table
\ref{disc_best_par_AH} that we are unable to differentiate between
accretion rates ranging from \logmdot=$-$12 to $-$9. This is due to
unavailability, or poor quality of, optical photometry which covers
the shorter wavelength range where accretion flux is the dominant
contributor to the SED. The derived accretion rates are discussed in
detail in Section \ref{accretion}.

We also fitted the BDD objects to the semi-empirical grid, where
accretion rates are negligible but a larger range of ages and masses
is available. For all BDDs, excepting BDD399, BDD304, BDD164 and
CFHT-Tau 4, acceptable fits could be found using the non-accreting
models. This is as the effects of a disc and accretion on observations
of a BDD system can mimic a star of higher mass or younger age
\citep{mayne_2010}. Therefore observations of systems with discs, can
be fit using SEDs of accreting and non-accreting models. For instance
for CFHT-Tau 8 and KPNO-Tau 6 best fitting masses of 0.1 and 0.05
M$_{\odot}$ (and ages of 5\,Myrs) were found with the semi-empirical
grid, compared to 0.08 and 0.03 found using the ad-hoc grid, and 0.078
and $\sim$0.025 M$_{\odot}$ (for an age of 1\,Myr) found in the
literature (see Table \ref{star_prop}). It is perhaps surprising that
even in these cases, where a large range of photometric measurements
are used, that there can still be some confusion as to whether a
source is accreting, or if it is in fact a more massive and younger
object. This suggests, as detailed in \cite{mayne_2010} that selection
of stars from CMDs and subsequent investigation into their mass and
accretion rate relationship is unreliable. In the case of CFHT-Tau 12
the fit is actually better when using the semi-empirical grid, however
several models cannot be fit without allowing for accretion flux,
i.e. BDD399, BDD304, BDD164, GM Tau and to a lesser extent CFHT-Tau 4.

\subsection{Derived Parameters}
\label{derived_pars}

The main differences between our models of BDD systems and those used
in previous studies is the additional modelling of accretion
luminosity, vertical hydrostatic equilibrium and dust sublimation. The
main derived parameters which indicate the veracity of these
mechanisms, when compared to observations, are the accretion rate,
inner edge location and flaring parameter. Without the inclusion of
these physical mechanisms it is clear from previous works and from the
fits in this paper that reasonable fits can still be achieved and
parameters derived. However, if important physical processes are
neglected it is questionable whether these parameters are reliable and
whether we are learning anything about the physical mechanisms
governing these systems.

\subsubsection{Accretion}
\label{accretion}

Accretion rates of \logmdot $<-$9 contribute little to shifts within
the HR-Diagram \citep{mayne_2010}.  However, when fitting
monochromatic fluxes or SED sections to model SEDs variations in
accretion rate below this threshold can affect the flux significantly
\citep{mayne_2010}. This can apply to objects with and without discs
e.g. \cite{kennedy_2009} find 43 stars in their sample of 1253 are
actively accreting whilst no disc is detected.

The adopted magnitudes (from Table \ref{target_phot}) do not extend
across the optical regime, i.e. magnitudes at U, B and V. This does
mean that a large section of the SED remains unconstrained. Fitting
has been performed by adopting magnitudes or deriving fluxes from
photographic plate observations, such as the NOMAD catalogue
\citep{zacharias_2005} by \cite{bouy_2008}. However, the spectral
responses of the photographic plates are poorly constrained and
therefore subsequent transformation to another photometric system or
to a monochromatic flux are unreliable (see Appendix
\ref{phot_detail}). However, as accretion flux can contribute
significantly in the optical regime allowing this parameter to run
free may lead to best fitting models having elevated or unrealistic
levels of accretion. Therefore, we have collected photometry from
\cite{zacharias_2005} for as many of our objects as possible alongside
H$_{\alpha}$ equivalent widths and derived accretion rates, which are
presented in Table \ref{accrete_table}. As many of our objects cannot
be satisfactorily fit without accretion flux these values can be used
to support their status as active accretors. These values can only be
used as a guide to add support for an accreting or non-accreting
model. The optical photometry is quite unreliable and it is very
difficult to compare measured widths of H$_{\alpha}$ made with
different instruments or over different epochs due to their
variability \citep{jayawardhana_2006}, so one cannot use these values
as strong constraint.

Table \ref{accrete_table} shows accretion diagnostics in the form of
H$_{\alpha}$ equivalent widths for our targets as found by various
authors, and derived accretion rates for some objects. Table
\ref{accrete_table} also shows our derived accretion rates for the BDD
objects.

\setcounter{table}{6}
\begin{table*}
\begin{tabular}{llllll}
\hline
Name&Equivalent Width of H$_{\alpha}$ ($\AA$)&\logmdot&$B_{\rm plate}$&$V_{\rm plate}$&\logmdot$^*$\\
\hline
\multicolumn{6}{c}{Sources without infrared excess}\\
\hline
CFHT-Tau 16&$-$16.76$^{(1)}$&&&&$<$$-$12\\
CFHT-Tau 15&$-$18.90$^{(1)}$&&&&$<$$-$12\\
CFHT-Tau 13&$-$3.60$^{(1)}$&&&&$<$$-$12\\
CFHT-Tau 11&$-$45.07$^{(1)}$&&20.31&17.97&$<$$-$12\\
CFHT-Tau 7$^*$&$-$8.63$^{(1)}$,5.5$^{(8)}$&&18.83&17.96&$<$$-$12\\
CFHT-Tau 5&$-$29.84$^{(1)}$&&&&$<$$-$12\\
CFHT-Tau 3&$-$10.5$^{(4)}$,50$\pm$4 or 65$\pm$1$^{(3)}$, $-$55$\pm$4$^{(7)}$&$<$$-$12$^{(4)}$&&&$<$$-$12\\
CFHT-Tau 2&11$\pm$1$^{(3)}$,$-$13$\pm$4$^{(7)}$&&20.70&&$<$$-$12\\
KPNO-Tau 9&20$^{(2)}$&&&&$<$$-$12\\
KPNO-Tau 5&30$^{(3)}$&&20.56&&$<$$-$12\\
KPNO-Tau 4&150$^{(2)}$,$-$38.4$^{(4)}$$-$158.08$^{(1)}$&$<$$-$12$^{(4)}$&&&$<$$-$12\\
ITG 2\\
\hline
\multicolumn{6}{c}{Sources with infrared excess}\\
\hline
CFHT-Tau 12&$-$79.4$^{(1)}$&&&&$-$12$^{+3.00}_{-0.00}$\\
BDD399&$-$47$^{(4)}$&$-$10.8$^{(4)}$&&&$-$8$^{+0.07}_{-1.00}$\\
BDD304&$-$233.7$^{(4)}$&$-$11.3$^{(4)}$&20.33&&$-$9$^{+1.03}_{-1.00}$\\
BDD164&&$-$11.03$^{(4)}$, $-$9.7$^{(9)}$&19.27&17.89&$-$8$^{+0.55}_{-3.34}$\\
CFHT-Tau 9&$-$9.95$^{(1)}$&&20.59&&$-$9$^{+0.13}_{-3.00}$\\
CFHT-Tau 8&$-$52$^{(1)}$&&&&$-$10$^{+1.03}_{-1.00}$\\
CFHT-Tau 6&$-$102$^{(4)}$,$-$63.7$^{(1)}$&$-$11.3$^{(4)}$&19.99&&$-$9$^{+0.06}_{-3.00}$\\
CFHT-Tau 4&$-$129.3$^{(4)}$, 69$\pm$4$^{(3)}$&$-$11.3$^{(4)}$&&&$-$9$^{+0.14}_{-3.00}$\\
GM Tau&&$-$8.7$^{(6)}$&18.39&17.45&$-$8$^{+0.63}_{-3.11}$\\
KPNO-Tau 6&350$^{(2,3)}$,$-$207.91$^{(1)}$,$-$41.1$^{(3)}$&$-$11.4$^{(4,5)}$&&&$-$10$^{+0.05}_{-2.00}$\\
KPNO-Tau 7&300$^{(2,3)}$,$-$31.1$^{(4)}$&$-$11.4$^{(4)}$&&&$-$10$^{+0.90}_{-2.00}$\\
\hline
\end{tabular}
\caption{Table of accretion indicators for sources with and without infrared excesses as defined in \citet{guieu_2007}. (*) the \logmdot in column 6 is from this work. References: (1) \citet{guieu_2006}, (2) \citet{briceno_2002}, (3) \citet{jayawardhana_2003}, (4) \citet{muzerolle_2005}, (5) \citet{herczeg_2008}, (6) \citet{white_2003}, (7) \citet{martin_2001}, (8) \citet{kraus_2009} and \citet{bouy_2008}. B$_{\rm plate}$ and V$_{\rm plate}$ photometry is from \citet{zacharias_2005}. \label{accrete_table}}
\end{table*}

We have assumed negligible accretion rates for all of the BD systems
and Table \ref{accrete_table} shows that this is a good assumption for
these objects. For the BDD objects, in most cases, our accretion rates
are consistent with those derived in the literature. Alternatively,
where we have derived non-negligible accretion rates the
EW(H$_{\alpha}$) and blue photometry support this. Several individual
cases, however, will be investigated further in Section
\ref{specific_cases}.

\subsubsection{Disc Structure}
\label{disc_structure}

The ad-hoc grid includes a state-of-the-art treatment of dust
sublimation and vertical hydrostatic equilibrium. Therefore, one would
expect to find significant differences between the inner edge
locations, inclinations and flaring parameters derived in this work,
and those derived by studies using an analytical or semi-empirical
approach (for instance placing the inner edge at a predefined
sublimation radius).

Table \ref{disc_prop} shows our best fitting inclinations, inner edge
locations (converted to AU) and $\beta$ ($1-\alpha$, the flaring
parameter) compared against those derived in \cite{guieu_2007} for the
BDD targets, where available\footnote{\cite{guieu_2007} do not
  publish their best fitting inclinations or the fits for CFHT-Tau 12
  and BDD399}. The disc structures and surface density profiles are
varied in \cite{guieu_2007} but the corresponding value of the surface
density exponent of the best fitting model is not published, so
comparisons in Table \ref{disc_prop} are incomplete. Our best fitting
total disc masses are presented in Table \ref{disc_best_par_AH} for
each of the BDD systems. Whereas the models used in \cite{guieu_2007}
adopted a single disc mass of 1 M$_J$.

\setcounter{table}{7}
\begin{table*}
\begin{tabular}{llccc}
\hline
Name&Source&Inclination ($^{\circ}$)&$\beta$ (0=VHE)$^{(1)}$&R$_{\rm sub}$ (AU)$^{(2)}$\\
\hline
\multicolumn{4}{c}{Sources with infrared excess}\\
\hline
\multirow{2}{*}{CFHT-Tau 9}&This paper&64$^{+2.49}_{-37}$&0 (1.25)&0.0081\\
&\citet{guieu_2007}&&1.125&0.032--0.067\\
\hline
\multirow{2}{*}{KPNO-Tau 6}&This paper&64$^{+5.22}_{-37}$&0 (1.25)&0.0036\\
&\citet{guieu_2007}&&1.0--1.125&0.032\\
\hline
\multirow{2}{*}{KPNO-Tau 7}&This paper&71$^{+0.99}_{-71}$&1.25 (0)&0.0043\\
&\citet{guieu_2007}&&1.0--1.125&0.015\\
\hline
\multirow{2}{*}{CFHT-Tau 12(*)}&This paper&64$^{+5.87}_{-16}$&1.25 (0)&0.0058\\
&\citet{guieu_2007}&\multicolumn{3}{c}{Not Published}\\
\hline
\multirow{2}{*}{BDD399}&This paper&64$^{+0.00}_{-0}$&1.10&0.025\\
&\citet{guieu_2007}&\multicolumn{3}{c}{Not Published}\\
\hline
\multirow{2}{*}{GM Tau}&This paper&48$^{+8.01}_{-9}$&1.10&0.025\\
&\citet{guieu_2007}&&1.0&0.015--0.067\\
\hline
\multirow{2}{*}{CFHT-Tau 6}&This paper&48$^{+8.66}_{-21}$&1.25 (0)&0.064\\
&\citet{guieu_2007}&&1.0--1.125&0.067--0.14\\
\hline
\multirow{2}{*}{CFHT-Tau 4}&This paper&39$^{+9.44}_{-12}$&1.10&0.025\\
&\citet{guieu_2007}&&1.0&0.14--0.3\\
\hline
\multirow{2}{*}{CFHT-Tau 8}&This paper&71$^{+0.76}_{-49.43}$&0 (1.25)&0.025\\
&\citet{guieu_2007}&&1.0--1.125&0.015--0.067\\
\hline
\multirow{2}{*}{BDD304}&This paper&71$^{+2.17}_{-71}$&1.25 (0)&0.022\\
&\citet{guieu_2007}&&1.125&0.015--0.032\\
\hline
\multirow{2}{*}{BDD164}&This paper&0$^{+56}_{-0}$&1.10&0.025\\
&\citep{guieu_2007}&&1.125&0.032--0.067\\
\hline
\end{tabular}
\caption{Table of disc structure parameters for sources with infrared excesses as defined in \citet{guieu_2007}. Here our R$_{\rm sub}$ radii have been converted to AU and a $\beta$ parameter constructed using $1-\alpha$ (see Section \ref{models_fitting}). Inclinations from our work are included but inclinations are not published in \citet{guieu_2007}. (1) As the $\beta$ variables can take only four values uncertainties are meaningless we simply quote the best fitting values and if models with the alternate value were found to have $\chi^2$ values within the 68\% confidence interval their values are quoted in brackets. (2) the R$_{\rm sub}$ from our work is positioned according to the balance of stellar flux and disc temperature, the range chosen in \citet{guieu_2007} is 0.015--0.3 AU.\label{disc_prop}}
\end{table*}

Table \ref{disc_prop} shows that, in general, our inner edge
locations, or more precisely the position of our dust sublimation
radii, are often much closer to the star than those found in
\cite{guieu_2007}. This is caused by the inclusion of accretion flux
in our models and realistic dust sublimation as opposed to the
prescription of a vertical wall. As discussed in \cite{mayne_2010} the
resulting inner edge structures are curved, consistent with detailed
studies \citep{tannirkulam_2007} of the inner edges of circumstellar
discs, which leads to significant changes in the way the SED around
1--10$\mu$m changes with inclination. This means we are able to
achieve fits aesthetically similar to those of \cite{guieu_2007} but
with significantly different inner edge locations. It is however
difficult to perform a complete comparison as \cite{guieu_2007} do not
publish their best fitting inclinations. Additionally,
\cite{guieu_2007} only model inner edge positions of 0.015--0.3 AU
from the central star, which does not extend low enough to match some
of our models.

\subsubsection{Specific Cases}
\label{specific_cases}

Further comparison and investigation into the structures of the discs
can be performed in a few cases where either, the fitting process has
revealed interesting results or further studies have been performed on
the object in question.

\begin{figure*}
\includegraphics[scale=0.45,angle=90]{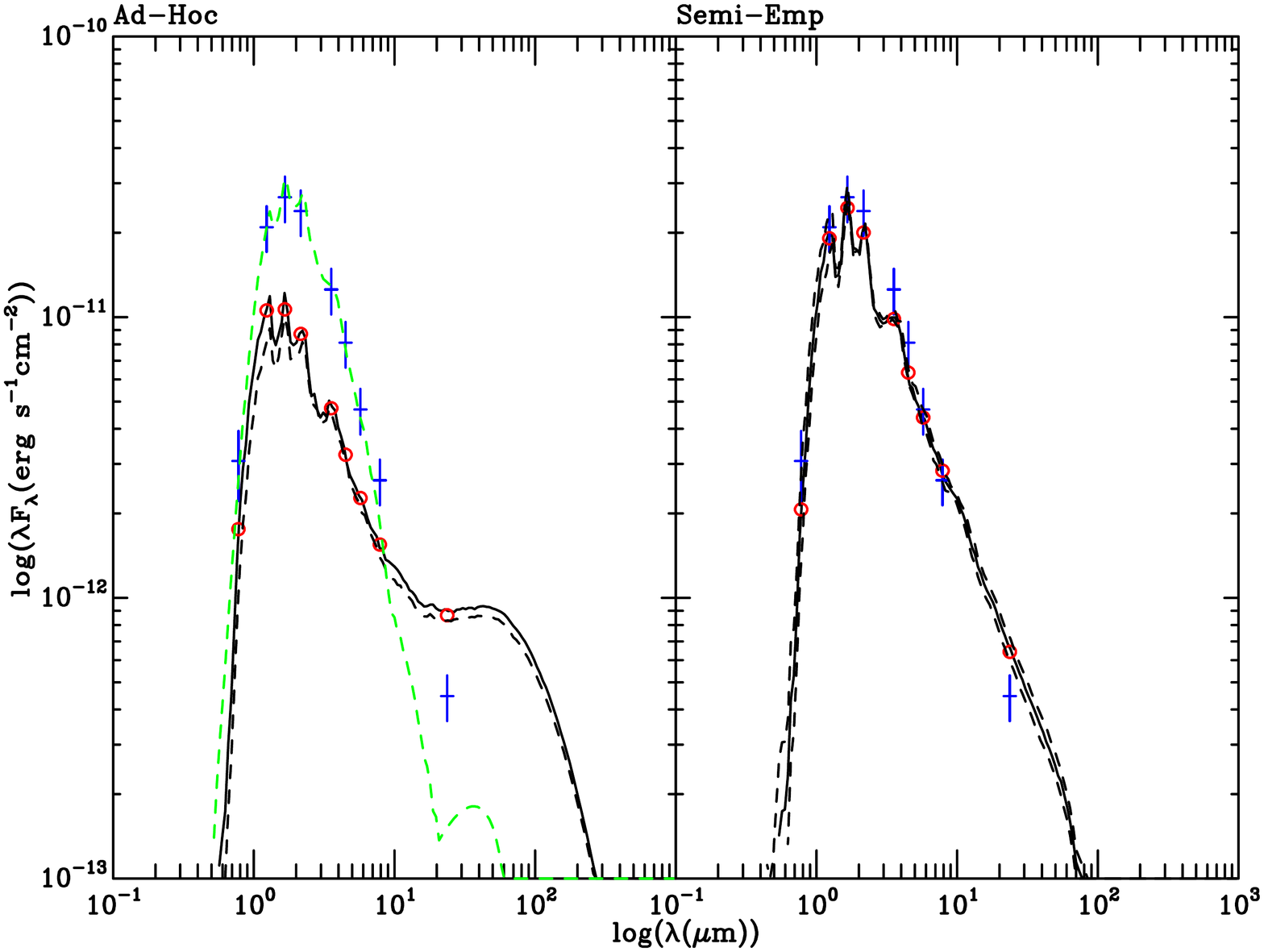}
\caption{SEDs and monochromatic fluxes for CFHT-Tau 12 where symbols have
  the same meanings as Figure \ref{eg_naked}.\label{cfht-tau_12}}
\end{figure*}

CFHT-Tau 12 is better fit using a semi-empirical model with a
negligible accretion rate, rather than an accreting model from the
ad-hoc grid. Figure \ref{cfht-tau_12} shows the fits for CFHT-Tau 12
using both grids. Figure \ref{cfht-tau_12} shows that the
semi-empirical grid yields a much better fitting SED in all
cases. Most parameters, and their uncertainties, are consistent
between the fits in the two grids. However, the best fitting mass is
marginally increased, the distance and inclination decreased, and the
age moves from 1\,Myr to 2\,Myrs. Only by allowing the accretion rate
to move to \logmdot=$-$7 can we get an acceptable fit using the
accreting, ad-hoc models (dashed green line left panel, Figure
\ref{cfht-tau_12}). Such an extreme accretion rate is unlikely but
there is little additional information in Table \ref{accrete_table} to
constrain this, other than a non-contemporaneous H$_{\alpha}$
equivalent width measurement. It is interesting to note that CFHT-Tau
12 has no measurements at wavelengths longer than 24\,$\mu$m (see
Table \ref{target_phot}) leading to poor constraint of the cooler dust
disc at larger radii, however our best fitting disc mass is M$_{\rm
  disc}$=0.001M$_*$, with no acceptable fit to an 0.01$M_*$ disc model
found.

BDD399, BDD164, BDD304, GM Tau and CFHT-Tau 4 are all poorly fit when
using the semi-empirical grid, (i.e. assuming negligible accretion and
allowing age to vary). This is not unsurprising as for each of these
objects there is evidence of non-negligible accretion rates, as shown
in Table \ref{accrete_table}. If such objects are fitted without
accretion luminosity the mass of the central star is increased and its
distance decreased, to increase the short-wavelength luminosity.

\begin{figure*}
\includegraphics[scale=0.45,angle=90]{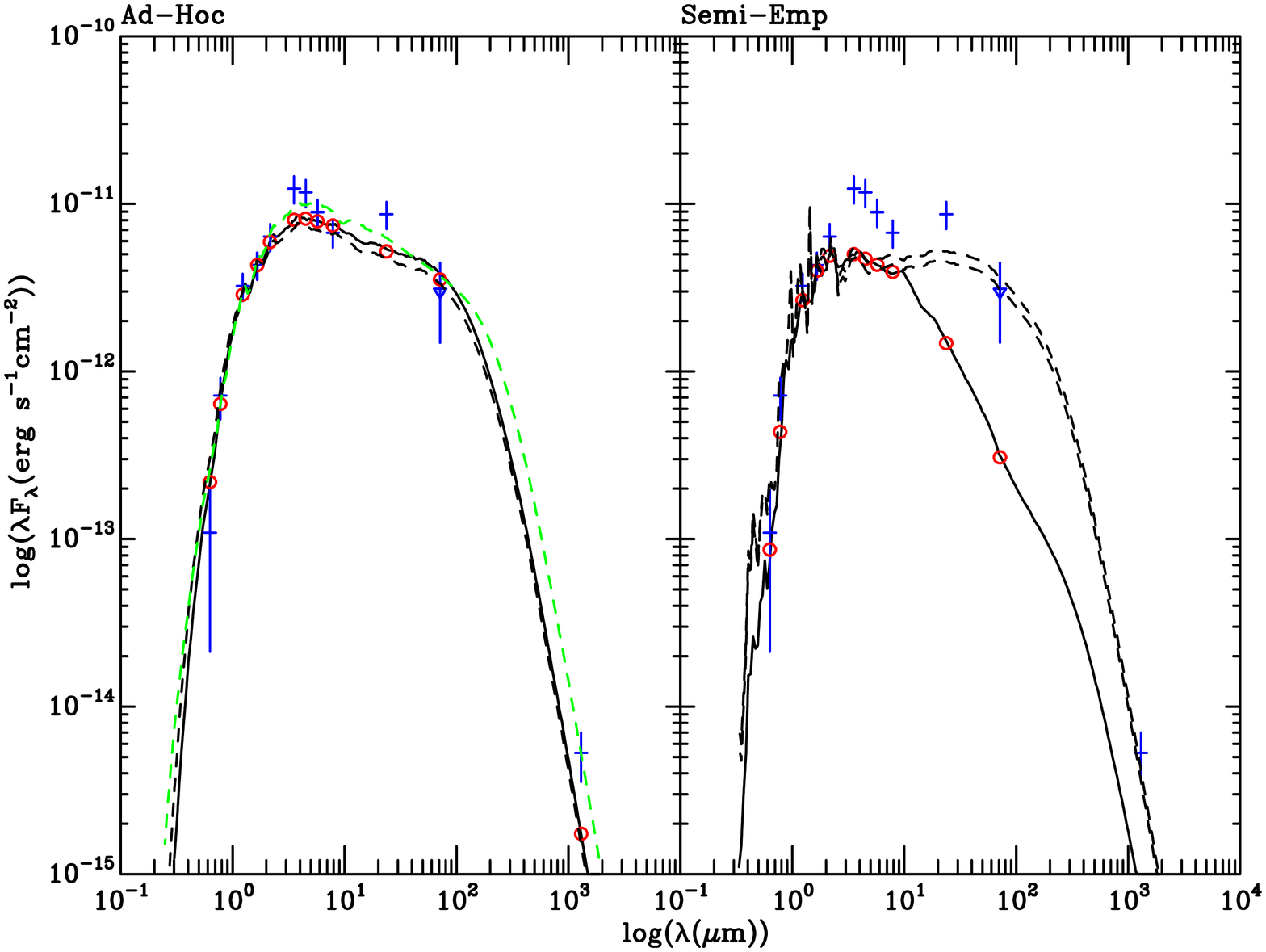}
\caption{SEDs and monochromatic fluxes for BDD399 where symbols have
  the same meanings as Figure \ref{eg_naked}.\label{bdd399}}
\end{figure*}

\begin{figure*}
\includegraphics[scale=0.45,angle=90]{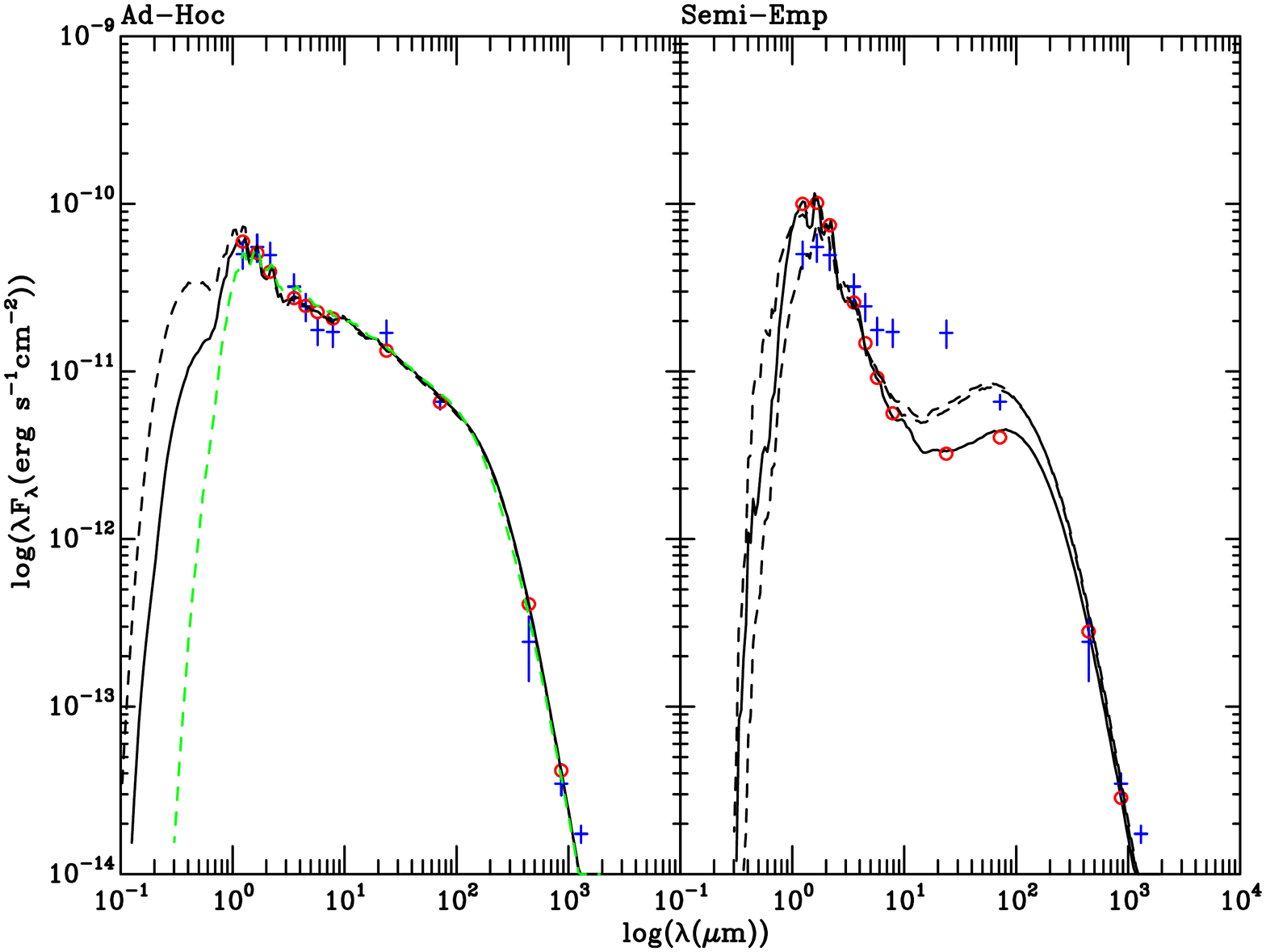}
\caption{SEDs and monochromatic fluxes for BDD164 where symbols have
  the same meanings as Figure \ref{eg_naked}.\label{bdd164}}
\end{figure*}

\begin{figure*}
\includegraphics[scale=0.45,angle=90]{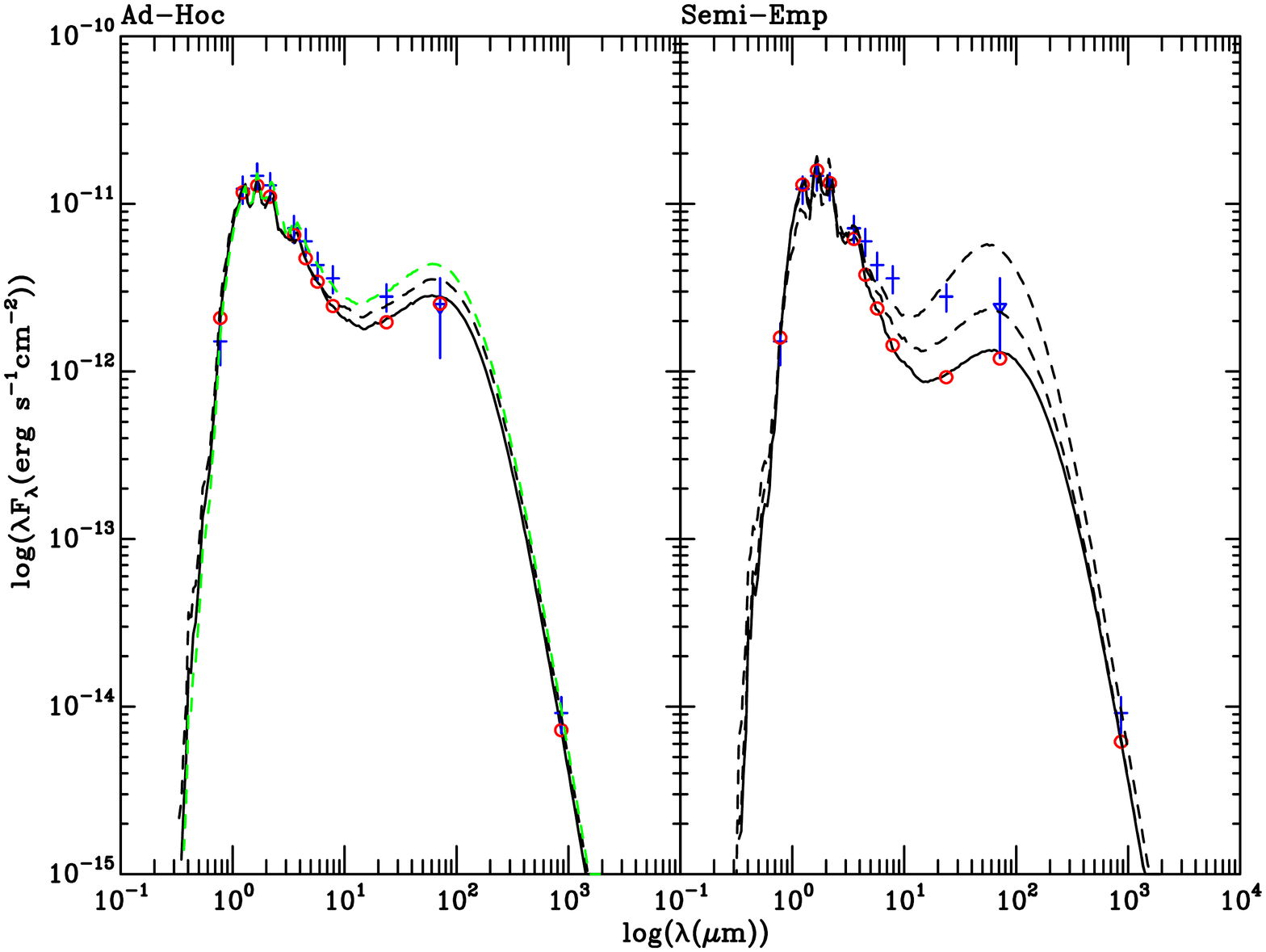}
\caption{SEDs and monochromatic fluxes for BDD304 where symbols have
  the same meanings as Figure \ref{eg_naked}.\label{bdd304}}
\end{figure*}

\begin{figure*}
\includegraphics[scale=0.45,angle=90]{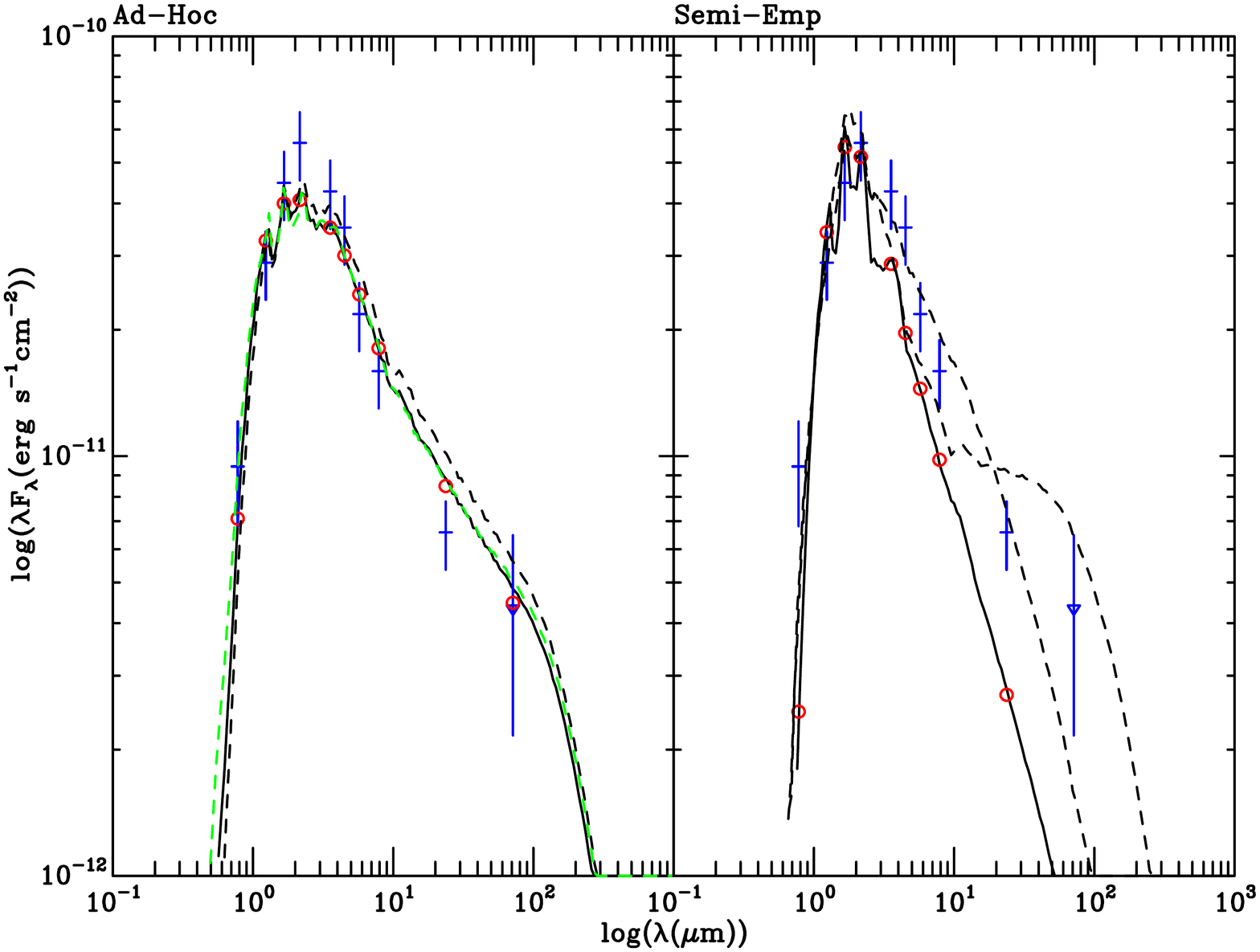}
\caption{SEDs and monochromatic fluxes for GM Tau where symbols have
  the same meanings as Figure \ref{eg_naked}. \label{gm_tau}}
\end{figure*}

\begin{figure*}
\includegraphics[scale=0.45,angle=90]{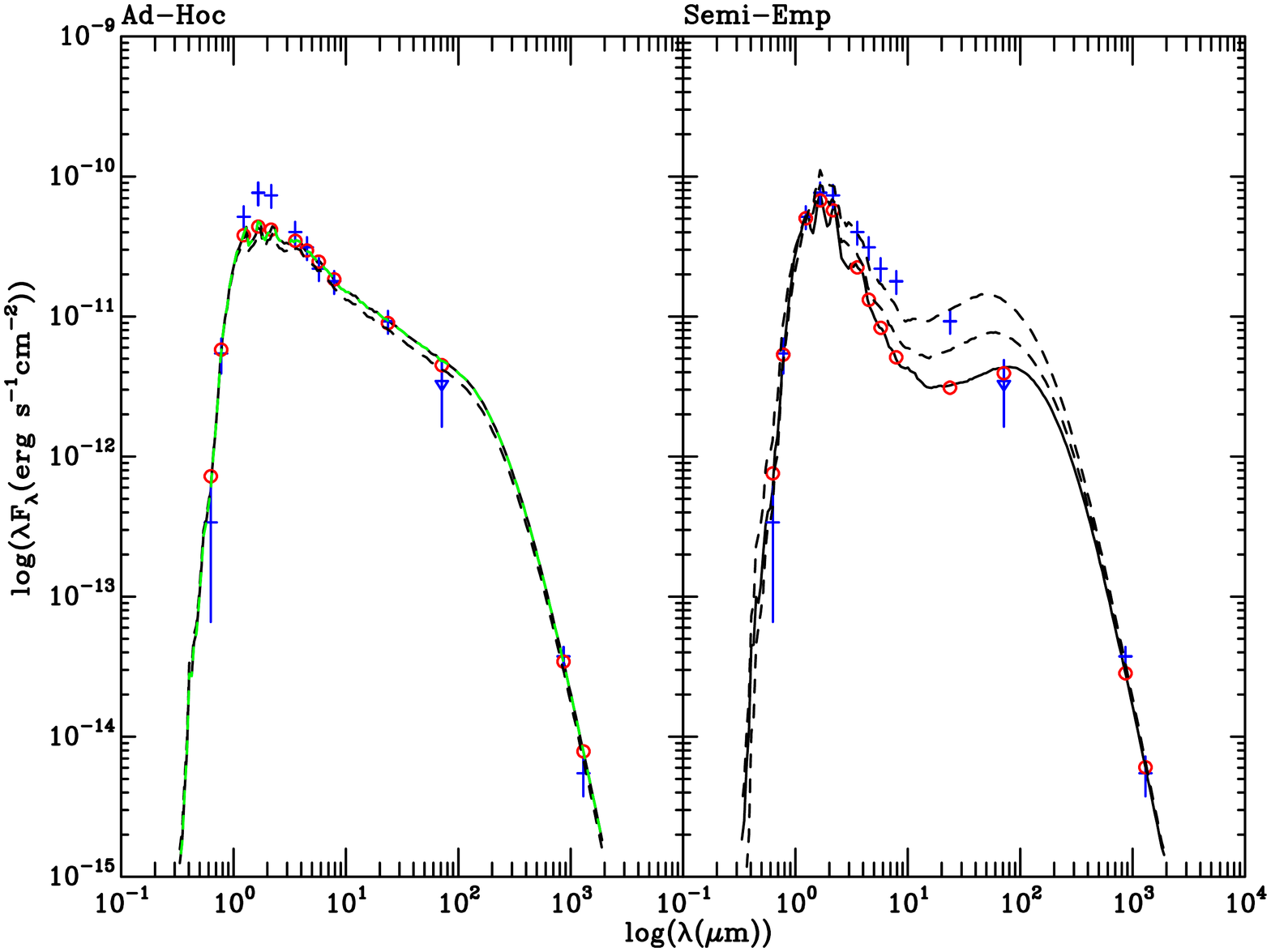}
\caption{SEDs and monochromatic fluxes for CFHT-Tau 4 where symbols
  have the same meanings as Figure \ref{eg_naked}.\label{cfht-tau_4}}
\end{figure*}

Figures \ref{bdd399}, \ref{bdd164}, \ref{bdd304}, \ref{gm_tau} and
\ref{cfht-tau_4} show the fits from both grids (semi-empirical and
ad-hoc) for all fit types (i.e. using the entire grids, refining the
fit and finally setting only the distance and extinction).

For BDD399 fitting to the entire semi-empirical grid, still does not
result in a satisfactory fit. As stated in \cite{luhman_2007} the
photosphere is too faint for any model of a young BD. However, as we
can see in the left panel of Figure \ref{bdd399} the fit is much more
consistent for all fit types in the ad-hoc grid, and indeed yields
similar parameters. The derived accretion rate for this object is
\logmdot=$-$8$^{+0.07}_{-1.00}$ (for the restricted fit) which is
greater than the literature result of $-$10.8
\citep{muzerolle_2005}. This object was studied in detail in
\cite{luhman_2007} where an inclination of 70$^{\circ}$ was found with
an inner disc radius of 58R$_*$ and outer radius of R=20--40 AU. The
central star of \cite{luhman_2007} had a luminosity of $-$1.2218
$\log \left( \frac{L_*}{L_{\odot}} \right)$, a mass of 0.05M$_{\odot}$
and a radius of 1.02R$_{\odot}$. We find an inclination angle of
64$^{\circ}$, an inner disc radius of 6.4R$_*$ and an outer disc
radius of 100 AU, with a corresponding star of luminosity $-$1.3400
$\log \left( \frac{L_*}{L_{\odot}}\right) $ mass of 0.08M$_{\odot}$ and
radius 0.829R$_{\odot}$. We derived a best fitting disc mass for this
object of M$_{\rm disc}$=0.001M$_*$. Whilst our inclinations are
consistent with that of \cite{luhman_2007} our fit is hampered by the
fact that the luminosity of the stellar photosphere of BDD399 is
under-luminous when compared to isochronal models (a grid is in
preparation to improve on this limitation). As \cite{luhman_2007}
obtained images of the disc directly their derived disc parameters
should be viewed as more reliable.

For BDD164 a satisfactory fit can only be achieved with an accreting
model, however in this case our derived accretion rate of
\logmdot=$-$8$^{+0.55}_{-3.34}$ is consistent with the values of
$-$11.03 and $-$9.7 \citep{muzerolle_2005,bouy_2008}, albeit with the
best fitting rate elevated over the literature values
again. \cite{bouy_2008} studied this object in detail using
multi-wavelength photometry (although as noted in Section
\ref{accretion} they used NOMAD photometry in the optical regime which
is derived using photographic plates in this case). They found the
best fit for a disc with inner radius at 0.04 AU and a $\beta$ value
of 1.1 or 1.2 (consistent with an $\alpha$ of 2.10 or 2.20 in this
work) for a central star with mass 0.045M$_{\odot}$. We find an
$\alpha$ of 2.10 and an inner disc radius of 0.025 AU for a star with
mass 0.07--0.08M$_{\odot}$, with a best fitting total disc mass of
M$_{\rm disc}$=0.01M$_*$.

For BDD304, Figure \ref{bdd304} again reveals a requirement of an
accretion flux when this object is fitted and comparison of the
derived parameters from Tables \ref{disc_best_par_AH} and
\ref{accrete_table} shows that we derive an elevated accretion rate,
\logmdot=$-$9$^{+1.03}_{-1.00}$, compared to $-$11.3
\citep{muzerolle_2005}. We find a best fitting disc mass of M$_{\rm
  disc}$=0.001M$_*$ for this object.

For GM Tau, even using the accreting models of the ad-hoc grid we are
unable to achieve a completely satisfactory fit, as shown in Figure
\ref{gm_tau}. This is caused by the fact that the mass of GM Tau is
estimated to be at or above the higher mass boundary of the ad-hoc
grid (M$_*=0.08$M$_{\odot}$, compared to 0.078, see Table
\ref{star_prop}).  Indeed, this can be seen when fitting using the
semi-empirical grid where the mass derived is much higher.  It is
interesting to note that if the monochromatic fluxes were fitted to
the entire grid (only setting the distance and extinction), as shown
in Figure \ref{gm_tau} a satisfactory fit can be obtained whereby the
object looks very similar to a higher mass object at higher
inclinations, M$_*$=0.8M$_{\odot}$ at 64$^{\circ}$ \citep[as
highlighted by][]{mayne_2010}. This highlights the problems of using
fitting tools such as the one featured in this paper and that of
\cite{robitaille_2006} without pre-selecting, or refining the
parameter space \citep[this point is also emphasised
in][]{robitaille_2006}. The best fitting disc mass derived for this
object was M$_{\rm disc}$=0.01M$_*$.

As with GM Tau for CFHT-Tau 4 we cannot achieve a fit comparable to
the other objects in our sample using any of the accreting models (see
Figure \ref{cfht-tau_4}). However, our derived accretion rate,
\logmdot=$-$9$^{+0.14}_{-3.00}$ is consistent with the literature
value of $-$11.3 \cite{muzerolle_2005}, whilst our best fitting model
accretion rate is significantly elevated (\logmdot=$-$9). We again
derive a best fitting disc mass of M$_{\rm disc}$=0.01M$_*$.

Whilst many stars have evidence of lower accretion rates as presented
in Table \ref{accrete_table}, BDD164 was found to have a lower
accretion rate, \logmdot=$-$11.03, and subsequently \cite{bouy_2008}
derived a much higher accretion rate of \logmdot=$-$9.7 using SED
fitting. Therefore, within our uncertainties we are unable to rule out
the higher accretion rates based on the values presented in Table
\ref{accrete_table}, and more work must be done to ascertain whether
SED fitting in general overestimates the accretion rates of BDD
targets, or whether other methods systematically underestimate the
accretion rates.

\section{Adopted Parameters}
\label{adopt_par}

Adopting realistic uncertainties for the distance and extinction, as
oppose to fixing them, does not result in significant variation in the
best fitting parameters (as shown in Section \ref{results}). As there
is no physical reason to assume all objects in Taurus are at 140\,pc,
and that the extinctions derived in the literature are infinitely
precise, we have adopted the fits where these parameters are varied
over a reasonable range.

For the naked objects we have found that for most cases, excepting ITG
2 and CFHT-Tau 5, the SEDs are well represented and the best fitting
parameters consistent with literature values, when fitting using the
semi-empirical grid, (i.e. assuming negligible accretion and allowing
the ``model'' or ``isochronal'' age to vary). Clearly, this variation
in age simply reflects a variation in luminosity and its underlying
causes are beyond the scope of this paper but are discussed in the
literature \citep[e.g.][]{mayne_2008}. For the cases of ITG 2 and
CFHT-Tau 5 it is apparent that these objects are at luminosities
elevated, for their masses, above the 1\,Myr isochrone and therefore
could not be fitted satisfactorily (we are running a grid to address
this, see Section \ref{conclusions}).

All of the targets with discs, BDDs, were best fit using the accreting
models of the ad-hoc grid, with the exception of CFHT-Tau 12. This
suggests that variations in the accretion rate and occultation by the
disc are sufficient to account for the luminosity spreads in this
population. Although again some objects are more luminous than the
1\,Myr isochrone namely, GM Tau and CFHT-Tau 4.

The resulting adopted parameters for each BD and BDD object are
included in Tables \ref{adopt_par_naked}, \ref{adopt_par} and
\ref{adopt_par_disc}, for naked and disc objects respectively.

\section{Conclusions}
\label{conclusions}

We have collected a set of photometric observations for a sample of
Taurus low-mass star and brown dwarf candidates. We then derived best
fitting parameters and uncertainties for the stellar and disc
properties (if required). We have shown that most naked brown dwarfs
in Taurus can be fit using negligibly accreting models over a range of
ages, with the exception of ITG 2 and CFHT-Tau 5 where the stars were
found to be over-luminous when compared to the 1\,Myr isochrone. For
targets with a disc we found satisfactory fits and derived a new list
of accretion rates for all targets using accreting models with an age
of 1\,Myr. The only exception is CFHT-Tau 12 where a different age,
amongst small adjustments to other parameters, yielded a better
fit. Additionally, we have shown that changes in the disc structure
can combine in complex ways to reproduce the SEDs of the brown dwarf
and disc objects but with significantly different geometrical,
e.g. inner disc position, properties to those found in the
literature. As we have allowed the disc structure (in the ad-hoc grid)
to adjust to an equilibrium state using a more realistic, vertical
structure we favour our best fitting models. We have included physical
mechanisms which are largely accepted for CTTS systems, in the ad-hoc
grid. If the BD regime is a mere extension of CTTS physical regime to
lower masses these mechanism will govern the structure of BDD
systems. So, by comparing these models with observations we are
discovering limitations in our theoretical models, which directs
improvements. On the other-hand fitting such systems using models with
parameterisations, although it leads to acceptable fits and predicted
parameters may, obscure problems with our understanding of this
regime. Therefore, we would encourage fitting of observations using
models derived using robust physical ingredients, as opposed to
fitting to large grids constructed by adopting significant
parameterisation, given the degenerate nature of SED fitting for BD
and BDD objects.

The models presented in this paper have been incorporated into an
online fitting tool. This tool currently allows users to select from
two models grids to which observations can be fitted. However, the
fitting tool itself is independent of these grids and simply requires
a structured list of SEDs in order for observations to be fitted to
any new model grid. We are currently running a further model grid to
expand the parameter space covered and address some of the issues with
the over- or under-luminous sources (when compared to the
isochrones). It is our hope that continued development of the models
of LMS and BD stars with discs will be informed by feedback from
fitting tools such as the one presented in this paper.

We have attempted to publish a complete description of our techniques
on \mysite{} allowing users and researchers to recreate, check and
compare to what we have done.

\section[]{ACKNOWLEDGMENTS}
The calculations for this paper were performed on the DiRAC Facility
jointly funded by STFC, the Large Facilities Capital Fund of BIS, and
the University of Exeter. NJM and TJH were supported by STFC grant
ST/F003277/1.

\bibliographystyle{mn2e}
\bibliography{references}

%% file: tables.tex
\setcounter{table}{1}
\begin{landscape}
\begin{table}
\begin{center}
\begin{tabular}{l|lllllllllllllll|lll}
\hline
Data&A$_{\rm V}$&R&I&J&H&K&L&[3.6]&[4.5]&[5.8]&[8.0]&24&70&450&850&1300\\
Units&\multicolumn{12}{c}{Magnitudes (except where stated, i.e. mJy)}&\multicolumn{4}{l}{mJy}\\
1$\sigma$ (default)&0.8&0.8&0.3&\multicolumn{8}{c}{0.2}&\multicolumn{4}{l}{Upper limits ($<$) assumed as 2$\sigma$, otherwise stated.}\\
\hline
\multicolumn{17}{c}{Sources without infrared excess}\\
\hline
KPNO-Tau 4&2.45&20.54&18.75&15.00&14.02&13.28&12.60&12.56&12.35&12.23&12.11&$<$0.89 (mJy)&$<$57.20&&&$-$1.58$\pm$0.90\\
CFHT-Tau 15&1.30&&17.94&14.93&14.24&13.69&&13.20&13.05&12.98&13.00&$<$0.89 (mJy)&$<$57.20\\
KPNO-Tau 5&0.00&19.10&15.08&12.64&11.92&11.54&10.83&11.05&10.92&10.81&10.85&$<$0.89 (mJy)&$<$57.20&&&$-$0.69$\pm$0.75\\
CFHT-Tau 16&1.51&&17.91&14.96&14.25&13.70&&13.21&13.11&13.03&13.02&$<$0.89 (mJy)&$<$57.20\\
CFHT-Tau 13&3.49&&17.90&14.83&13.97&13.45&&12.85&12.75&12.73&12.67&$<$1.20 (mJy)&$<$72.20\\
CFHT-Tau 7&0.00&16.63&14.12&11.54&10.79&10.40&&9.91&9.79&9.74&9.72&9.69&$<$134.00\\
CFHT-Tau 5&9.22&23.37&18.79&13.96&12.22&11.28&&10.42&10.19&10.08&10.05&9.81&$<$57.20\\
CFHT-Tau 11&0.00&&14.88&12.53&11.94&11.59&&11.19&11.06&11.02&10.98&$<$0.89 (mJy)\\
KPNO-Tau 9&0.00&&18.76&15.49&14.76&14.17&13.03&13.61&13.53&13.44&13.48&$<$0.89 (mJy)&&&&$-$2.62$\pm$0.82\\
CFHT-Tau 2&0.00&20.21&16.81&13.76&12.76&12.17&&11.57&11.38&11.37&11.36&$<$0.89 (mJy)&$<$65.90&&$<$5.81&$-$0.60$\pm$0.80\\
CFHT-Tau 3&0.00&20.33&16.88&13.72&12.84&12.37&&11.77&11.62&11.56&11.54&$<$0.89 (mJy)&&&&0.37$\pm$0.77\\
ITG 2&0.64&20.21&14.71&11.55&10.63&10.10&&9.57&9.42&9.33&9.34&9.25&$<$79.00&&&$-$0.46$\pm$0.80\\
\hline
\multicolumn{17}{c}{Sources with infrared excess}\\
\hline
CFHT-Tau 9&0.91&&15.35&12.88&12.19&11.76&&11.11&10.80&10.48&9.85&6.79&$<$57.20\\
KPNO-Tau 6&0.88&20.56&17.90&15.00&14.20&13.69&12.60&13.08&12.77&12.41&11.82&9.10$\pm0.22$&$<$75.80&&&$-$0.66$\pm$0.79\\
KPNO-Tau 7&0.00&&17.16&14.52&13.83&13.27&12.35&12.60&12.27&11.93&11.26&8.44&$<$57.20&&&0.70$\pm$0.88\\
CFHT-Tau 12&3.44&&16.26&13.15&12.15&11.55&&10.77&10.54&10.31&9.93&8.25&\\
BDD399&0.00&20.33&17.84&15.18&14.13&12.98&&10.79&10.14&9.61&8.91&5.03&$<$70.70&&&2.29$\pm$0.75\\
GM Tau&4.34&&15.04&12.80&11.59&10.63&&9.44&8.95&8.64&7.97&5.33&$<$103.00\\
CFHT-Tau 6&0.41&18.39&15.40&12.65&11.84&11.37&&10.78&10.37&10.03&9.16&6.44&&&&2.86$\pm$0.76\\
CFHT-Tau 4&3.00&19.10&15.64&12.17&11.01&10.33&&9.51&9.08&8.63&7.85&4.96&$<$77.80&&10.8$\pm$1.8&2.38$\pm$0.75\\
CFHT-Tau 8&1.80&19.27&16.43&13.19&12.12&11.45&&10.85&10.23&9.86&9.14&6.53&$<$57.20\\
BDD304&1.06&&17.03&13.73&12.80&12.22&&11.38&10.87&10.40&9.59&6.26&$<$57.20&&2.64$\pm0.64$\\
BDD164&0.00&&&12.20&11.36&10.76&&9.75&9.34&8.87&7.89&4.30&157.00$\pm$15.7&36$\pm$15&10$\pm$1.5&7.55$\pm$0.89\\
\hline
\end{tabular}
\caption{The adopted photometry and extinctions \citep[A$_{\rm V}$,][]{guieu_2005,guieu_2006,guieu_2007} for each target. $R$ and $I$ photometry are from CFHT12K and MEGACAM and are sourced from \citet{martin_2001}, \citet{luhman_2004} and \citet{guieu_2006}. The JHK magnitudes are from 2MASS. $L$ band observations are presented in \citet{jayawardhana_2003}. The IRAC photometry has been published in \citet{luhman_2006}, \citet{guieu_2007}, \citet{monin_2010} and \citet{luhman_2010}. MIPs photometric magnitudes are from \citet{luhman_2010} and the monochromatic fluxes from \citet{guieu_2007}. The SCUBA 450 and 850 data are from \citet{klein_2003} and the 1300\,$\mu$m fluxes are from \citet{klein_2003} and \citet{scholz_2006}. For MIPs 24$\mu$m column contains magnitudes unless denoted with (mJy) where a flux in mJy is given. $<$ denotes upper limits (in mJy) which are assumed to be $2\sigma$ upper limits. The uncertainties presented in the third row are adopted unless a specific uncertainty is stated with the observation. The division of sources with and without infrared excess is from \citet{guieu_2007}.}
\label{target_phot}
\end{center}
\end{table}
\end{landscape}

\setcounter{table}{3}
\begin{landscape}
\begin{table}
\begin{center}
\begin{tabular}{llllll}
\hline
Name&Spectral Type (SpT)&A$_V$&$T_{\rm eff}$(K)&Mass (M$_{\odot}$)&$\log \left( \frac{L_{\rm bol}}{L_{\odot}} \right)$\\
&($\pm$0.5 unless stated $^{(1, 3, 7)}$)&($\pm$0.8 unless stated $^{(1, 3, 7)}$)&\\
\hline
\multicolumn{6}{c}{Sources without infrared excess}\\
\hline
KPNO-Tau 4&M9.50$^{(1,2,4,5,6,7,8)}$, M9.75$^{(3)}$&2.45$^{(1)}$,2.5$^{(3,7)}$&2360$^{(3)}$, 2300$^{(4)}$&0.011$^{(4)}$, 0.014$^{(5)}$, 0.012$^{(8)}$&$-$2.138$^{(3)}$,$-$2.638$^{(4)}$\\
CFHT-Tau 15&M8.25$^{(1,7,8)}$&1.30$^{(1,7)}$&&0.027$^{(7)}$\\
KPNO-Tau 5&M7.50$^{(1,2,5,6,8)}$($\pm$0.25$^{(2)}$)&0.0$^{(1)}$,1.30$^{(6)}$**&&0.052$^{(5)}$,0.044$^{(8)}$&\\
CFHT-Tau 16&M8.50$^{(1,7,8)}$&1.51$^{(1,7)}$&&0.022$^{(8)}$&\\
CFHT-Tau 13&M7.25$^{(1,7,8)}$&3.49$^{(1,7)}$&&0.051$^{(8)}$&\\
CFHT-Tau 7$^{\rm B}$&M6.50$^{(1,3,7,8,9)}$,M5.75$^{(10)}$&0.0$^{(1,3,7)}$&2833$^{(3)}$&0.07$^{(9)}$,0.078$^{(8)}$&$-$0.976$^{(3)}$\\
CFHT-Tau 5&M7.50$^{(1,3,7)}$&9.22$^{(1,3,7)}$&2688$^{(3)}$&0.044$^{(8)}$&$-$0.877$^{(3)}$\\
CFHT-Tau 11&M6.75$^{(1,7,8)}$&0.0$^{(1,7)}$&&0.068$^{(8)}$&\\
KPNO-Tau 9&M8.50$^{(1,2,5,6,8)}$($\pm$0.25$^{(2)}$)&0.0$^{(1)}$, 9.07$^{(6)}$**&&0.026$^{(5)}$,0.022$^{(8)}$&\\
CFHT-Tau 2&M7.50$^{(1,2,6,8)}$($\pm$0.25$^{(2)}$),M8$^{(2,5,11)}$&0.0$^{(1,11)}$, 1.99$^{(2)}$*&2795$^{(2)}$&0.036$^{(5)}$,0.044$^{(8)}$&\\
CFHT-Tau 3&M7.75$^{(1,2,4,6,8)}$($\pm$0.25$^{(2)}$),M9$^{(2,5,11)}$&0.0$^{(1,11)}$, 0.99$^{(2)}$*&2752$^{(2)}$&0.038$^{(8)}$,0.035$^{(4)}$,0.016$^{(5)}$&$-$2.056$^{(4)}$\\
ITG 2&M7.25$^{(1,5,12)}$&0.64$^{(1)}$, 0.60$^{(12)}$*&2838$^{(12)}$&0.065$^{(5)}$,0.051$^{(8)}$&$-$1.222$^{(12)}$\\
\hline
\multicolumn{6}{c}{Sources with infrared excess}\\
\hline
CFHT-Tau 9&M6.25$^{(1,7,8)}$,M5.75$\pm$0.25$^{(10)}$&0.91$^{(1,7)}$,0.50$^{(10)}$&2959$^{(1,7)}$,3020$\pm$35$^{(10)}$&0.09$^{(8)}$,0.087$^{+0.012}_{-0.01}$$^{(10)}$&$-$1.644$^{\rm M}$$^{(10)}$\\
KPNO-Tau 6&M9.0$^{(3,7)}$,M8.5$^{(2,4,6,8,10)}$($\pm$0.25$^{(2,10)}$)&0.88$^{(1,7)}$,0.90$^{(10)}$,7.6$^{(6)}$**&2469$^{(3)}$,2571$^{(7)}$,2555$^{(4,8)}$($\pm$78$^{(8)}$)&0.022$^{(8)}$,0.021$^{+0.007}_{-0.007}$$^{(10)}$,0.025$^{(4)}$,0.026$^{(5)}$&$-$2.460$^{\rm M}$$^{(10)}$,$-$2.274$^{(3)}$,$-$2.678$^{(4)}$\\
KPNO-Tau 7&M8.25$^{(1,2,4,5,6,8,19)}$($\pm$0.25$^{(2,10)}$)&0.0$^{(1,10)}$, 4.07$^{(6)}$**&2630$\pm$578$^{(10)}$,2632$^{(4)}$&0.027$^{(8)}$,0.03$^{(4,5,10)}$($^{+0.12}_{-0.14}$$^{(10)}$)&$-$2.482$^{(4)}$, $-$2.364$^M$$^{(10)}$\\
CFHT-Tau 12&M6.50$^{(1,7,8)}$&3.44$^{(1,7)}$&2924$^{(7)}$&0.078$^{(8)}$&\\
BDD399&M7.25$^{(1,4,5,7,8,12)}$&0.0$^{(1,12)}$&2838$^{(4)}$&0.05$^{(13)}$,0.051$^{(7)}$,0.065$^{(5)}$,0.07$^{(4)}$&$-$2.745$^{(4)}$,$-$1.222$^{(13)}$\\
GM Tau&M6.50$^{(1,7,8,10,12)}$($\pm$0.5$^{(10)}$)&4.34$^{(1)}$, 4.08$^{(12)}$*&2935$^{(10,12)}$($\pm$55$^{(10)}$)&0.078$^{(7)}$, 0.074$^{+0.015}_{-0.013}$$^{(10)}$&$-$1.328$^{(12)}$, $-$1.452$^M$$^{(10)}$\\
CFHT-Tau 6&M7.25$^{(1,3,4,5,7,12)}$, M8.25$^{(8)}$&0.4$^{(3)}$,0.41$^{(1, 7)}$,0.25$^{(12)}$*&2838$^{(12)}$,2818$^{(7)}$,2724$^{(3)}$&0.051$^{(8)}$,0.050$^{(4)}$,0.065$^{(5)}$&$-$1.699$^{(12)}$,$-$1.383$^{(3)}$,$-$1.721$^{(4)}$\\
CFHT-Tau 4$^{\rm
  B}$&M7.00$^{(1,3,4,58,10,11,12)}$($\pm$0.25$^{(10)}$)&3.0$^{(1)}$,2.48$^{(12)}$*&2880$^{(12,10)}$($\pm$33$^{(10)}$)&0.058$^{(8)}$,0.06$^{(4)}$,0.068$^{+0.012}_{-0.007}$$^{(10)}$,0.85$^{(5)}$&$-$1.268$^{(12)}$,
$-$1.092$^M$$^{(10)}$\\
CFHT-Tau 8&M5.5$^{(12)}$,M6.50$^{(1,3,7,8)}$&1.8$^{(1,3)}$,1.77$^{(7)}$,2.48$^{(12)}$*&3058$^{(12)}$,2833$^{(3)}$,2924$^{(7)}$&0.078$^{(8)}$&$-$1.620$^{(12)}$,$-$1.455$^{(3)}$\\
BDD304&M7.75$^{(1,4,5,8,12)}$&1.06$^{(1)}$,0.99$^{(12)}$*&2752$^{(4,12)}$&0.035$^{(4)}$,0.038$^{(8)}$,0.04$^{(5)}$&$-$2.036$^{(12)}$,$-$2.066$^{(4)}$\\
BDD164&M7.25$^{(1,5,8,9,10,12)}$($\pm$0.25$^{(10)}$)&0.0$^{(1,12,10)}$$^{(1)}$&2838$^{(12)}$,2840$\pm$43$^{(10)}$&0.051$^{(8)}$, 0.054$^{+0.008}_{-0.008}$$^{(10)}$,0.065$^{(5)}$&$-$1.553$^{(12)}$, $-$1.432$^M$$^{(10)}$\\
\hline
\end{tabular}
\caption{A table of the stellar properties, either derived or adopted, for the sample of BDs with and without infrared excesses \citep[as defined by][]{guieu_2007}. References: (1) \citet{guieu_2007}, (2) \citet{briceno_2002}, (3) \citet{guieu_2005}, (4), \citet{muzerolle_2005}, (5) \citet{scholz_2006}, (6) \citet{jayawardhana_2003}, (7) \citet{guieu_2006}, (8) \citet{kraus_2007}, (9) \citet{konopacky_2007}, (10) \citet{kraus_2009}, (11) \citet{martin_2001}, (12) \citet{luhman_2004} and (13) \citet{luhman_2007}. * denotes A$_V$ values derived from A$_J$ values using the extinction relations of \citet{rieke_1985}. ** denotes A$_V$ values derived from E(K$-$L) using relations of \citet{rieke_1985}. $^{\rm M}$, column four, denotes an approximate L$_{\rm bol}$ derived from M$_{\rm bol}$. $^{\rm B}$, column one, denotes a possible binary system \citep{kraus_2009}. \label{star_prop}}
\end{center}
\end{table}
\end{landscape}

\setcounter{table}{4}
\begin{landscape}
\begin{table}
\begin{center}
\begin{tabular}{lllllllllllll}
\hline
\hline
\multirow{2}{*}{Target Name}&CFHT-&CFHT-&CFHT-&CFHT-&CFHT-&CFHT-&CFHT-&CFHT-&KPNO-&KPNO-&KPNO-&ITG 2\\
&Tau 16&Tau 15&Tau 13&Tau 11&Tau 7&Tau 5 (*)&Tau 3&Tau 2&Tau 9&Tau 5&Tau 4&(*)\\
\hline
\hline
Variable&\multicolumn{11}{c}{Independent Variables}\\
\hline
Mass (M$_{\odot}$)&0.040$^{+0.000}_{-0.010}$&0.050$^{+0.000}_{-0.020}$&0.070$^{+0.000}_{-0.030}$&0.090$^{+0.000}_{-0.030}$&0.100$^{+0.000}_{-0.010}$&0.060$^{+0.000}_{-0.000}$&0.040$^{+0.020}_{-0.009}$&0.040$^{+0.020}_{-0.005}$&0.020$^{+0.010}_{-0.001}$&0.060$^{+0.019}_{-0.010}$&0.030$^{+0.020}_{-0.001}$&0.050$^{+0.010}_{-0.007}$\\
\hline
Age (Myr)&8$^{+2}_{-7}$&10$^{+0}_{-9}$&10$^{+0}_{-9}$&4$^{+0.03}_{-3}$&1$^{+1}_{-0}$&1$^{+0}_{-0}$&4$^{+0.3}_{-3}$&4$^{+0.06}_{-3}$&2$^{+8}_{-0.15}$&1$^{+1}_{-0}$&3$^{+3}_{-1}$&1$^{+2}_{-0}$\\
\hline
Dist (pc)&159.75$^{+3.60}_{-20.36}$&156.25$^{+8.75}_{-17.64}$&159.75$^{+5.25}_{-17.50}$&130$^{+24.91}_{0.00}$&140.50$^{+8.75}_{-10.50}$&130$^{+3.50}_{-0.00}$&133.50$^{+15.75}_{-3.50}$&130$^{+17.50}_{-0.00}$&159.75$^{+5.75}_{-17.54}$&138.75$^{+21.48}_{-8.75}$&131.75$^{+33.25}_{-1.75}$&130$^{+21.00}_{-0.00}$\\
\hline
A$_{\rm V}$&1.27$^{+0.58}_{-0.48}$&1.62$^{+0.48}_{-0.53}$&2.69$^{+0.96}_{-0.00}$&0.48$^{+0.40}_{-0.20}$&0.48$^{+0.29}_{-0.32}$&8.42$^{+0.38}_{-0.00}$&0.80$^{+0.00}_{-0.32}$&0.80$^{+0.00}_{-0.34}$&0.80$^{+0.00}_{-0.48}$&0.80$^{+0.00}_{-0.24}$&3.25$^{+0.00}_{-0.64}$&1.44$^{+0.00}_{-0.14}$\\
\hline
&&\multicolumn{11}{c}{Dependent Variables of Best Fitting Model}\\
\hline
$\log \left( \frac{L_*}{L_{\odot}} \right)$&$-$2.323&$-$2.300&$-$2.090&$-$1.622&$-$1.160&$-$1.570&$-$2.002&$-$2.002&$-$2.580&$-$1.570&$-$2.208&$-$1.730\\
\hline
$T_{\rm eff}$ (K)&2760&2826&2968&3021&3002&2887&2789&2789&2519&2887&2703&2844\\
\hline
$\log (g)$)&4.07&4.18&4.21&3.88&3.45&3.58&3.77&3.77&3.87&3.58&3.80&3.63\\
\hline
R$_*$(R$_{\odot}$)&0.300&0.294&0.340&0.562&0.968&0.653&0.426&0.426&0.268&0.653&0.358&0.560\\
\hline
\hline
\end{tabular}
\caption{Table showing the best fitting independent (with uncertainties at the 68\% confidence interval) and dependent variables for the naked objects using the semi-empirical grid and fitting using a restricted grid (using the input from Table \ref{target_phot} and parameter ranges from Table \ref{star_prop}).The Objects ITG 2 and CFHT-Tau 5 are denoted with (*) as no satisfactory fit was achieved. Note the age is the model age and should not be construed as an exact age of the system. Additionally, the values of the dependent or derived variables are only for the best fitting models, strictly these should span a range of values corresponding to that covered by the independent variables.\label{naked_best_par}}
\end{center}
\end{table}
\end{landscape}
 
\setcounter{table}{5}
\begin{landscape}
\begin{table}
\begin{center}
\begin{tabular}{llllllllllll}
\hline
\hline
\multirow{2}{*}{Target Name}&CFHT-&BDD&BDD&BDD&CFHT-&CFHT-&CFHT-&CFHT-&GM &KPNO-&KPNO-\\
&Tau 12 (*)&399&304&164&Tau 9&Tau 8&Tau 6&Tau 4 (*)&Tau&Tau 6&Tau 7\\
\hline
\hline
Variable&\multicolumn{11}{c}{Independent Variables}\\
\hline
Mass (M$_{\odot}$)&0.05$^{+0.030}_{-0.000}$&0.08$^{+0.000}_{-0.000}$&0.06$^{+0.000}_{-0.020}$&0.08$^{+0.000}_{-0.010}$&0.08$^{+0.000}_{-0.020}$&0.08$^{+0.000}_{-0.010}$&0.06$^{+0.019}_{-0.010}$&0.08$^{+0.000}_{-0.000}$&0.08$^{+0.000}_{-0.010}$&0.03$^{+0.010}_{-0.011}$&0.03$^{+0.010}_{-0.010}$\\
\hline
Age (Myrs) (1)&1 (10)&1&1&1&1&1&1&1&1&10 (1)&1 (10)\\
\hline
\logmdot&$-$12$^{+3.00}_{-0.00}$&$-$8$^{+0.07}_{-1.00}$&$-$9$^{+1.03}_{-1.00}$&$-$8$^{+0.55}_{-3.34}$&$-$9$^{+0.13}_{-3.00}$&$-$10$^{+1.03}_{-1.00}$&$-$9$^{+0.06}_{-3.00}$&$-$9$^{+0.14}_{-3.00}$&$-$8$^{+0.63}_{-3.11}$&$-$10$^{+0.05}_{-2.00}$&$-$10$^{+0.90}_{-2.00}$\\
\hline
Period (days)&0.5 (5.0)&5.0 (0.5)&5.0 (0.5)&5.0&0.5 (5.0)&5.0 (0.5)&0.5 (5.0)&5.0&5.0 (0.5)&0.5 (5.0)&0.5 (5.0)\\
\hline
Coverage (\%)&10 (1)&10 (1)&10 (1)&1 (10)&10 (1)&10 (1)&10 (1)&10 (1)&10 (1)&10 (1)&10 (1)\\
\hline
M$_{\rm disc}$ (M$_*$)$^{(1)}$&0.001&0.001&0.001&0.01&0.001 (0.01)&0.001 (0.01)&0.01&0.01&0.01 (0.001)&0.001&0.001 (0.01)\\
\hline
R$_{\rm outer}$ (AU)$^{(1)}$&300&100&100&100&100 (300)&300 (100)&300 (100)&300&300 (100)&300 (100)&100 (300)\\
\hline
$\alpha$ (0=VHE)$^{(1,2)}$&2.25 (0)&2.10&2.25 (0)&2.10&0 (2.25)&0 (2.25)&2.25 (0)&2.10&2.10&0 (2.25)&2.25 (0)\\
\hline
Inclination ($^{\circ}$)&64$^{+5.87}_{-16}$&64$^{+0}_{-0}$&71$^{+2.17}_{-71}$&0$^{+56}_{-0}$&64$^{+2.59}_{-37}$&71$^{+0.76}_{-49.43}$&48$^{8.66}_{-21}$&39$^{+9.44}_{-12}$&48$^{+8.01}_{-9}$&64$^{+5.22}_{-37}$&71$^{+0.99}_{-71}$\\
\hline
Dist (pc)&161.50$^{+3.50}_{-31.50}$&130.00$^{+14.00}_{-0.00}$&130.00$^{+17.50}_{-0.00}$&149.25$^{+4.16}_{-19.25}$&163.25$^{+1.75}_{-33.25}$&138.75$^{+19.25}_{-8.75}$&140.50$^{+22.75}_{-9.10}$&130.00$^{+14.00}_{-0.00}$&151.00$^{+6.63}_{-17.50}$&131.75$^{+19.25}_{-1.75}$&130.00$^{+31.93}_{-0.00}$\\
\hline
A$_{\rm V}$&2.64$^{+0.64}_{-0.00}$&0.64$^{+0.16}_{-0.32}$&1.86$^{+0.00}_{-0.64}$&0.80$^{+0.00}_{-0.56}$&1.15$^{+0.56}_{-0.58}$&2.60$^{+0.00}_{-0.56}$&1.21$^{+0.00}_{-0.48}$&3.24$^{+0.29}_{-0.48}$&3.54$^{+0.96}_{-0.00}$&1.52$^{+0.16}_{-0.90}$&0.80$^{+0.00}_{-0.48}$\\
\hline
&&\multicolumn{10}{c}{Dependent Variables of Best Fitting Model}\\
\hline
$\log \left( \frac{L_*}{L_{\odot}} \right)$&$-$1.730&$-$1.340&$-$1.570&$-$1.340&$-$1.340&$-$1.340&$-$1.570&$-$1.340&$-$1.340&$-$2.490&$-$2.130\\
\hline
T$_{\rm eff}$ (K)&2846&2941&2893&2941&2941&2941&2893&2941&2941&2667&2703\\
\hline
$\log (g)$&3.63&3.50&3.58&3.50&3.50&3.50&3.58&3.50&3.50&4.05&3.72\\
\hline
R$_*$ (R$_{\odot}$)&0.569&0.829&0.658&0.829&0.829&0.829&0.658&0.829&0.829&0.270&0.398\\
\hline
$\log \left( \frac{\dot{L}}{L_{\odot}} \right)$&$-$5.755&$-$1.782&$-$2.762&$-$1.782&$-$2.782&$-$3.782&$-$2.762&$-$2.782&$-$1.782&$-$3.555&$-$3.780\\
\hline
$\dot{T_{\rm eff}}$ (K)&2496&4044&2582&7192&2274&1279&2582&2274&4044&2554&1847\\
\hline
$\dot{\lambda_{\rm peak}}$ ($\AA$)&58452&7165&11221&4029&12742&22659&11221&12742&7165&11345&15688\\
\hline
R$_{\rm sub}$ (R$_*$)&2.2&6.4&7.3&6.4&2.1&6.4&2.1&6.4&6.4&2.9&2.3\\
\hline
\hline
\end{tabular}
\caption{Table showing the best fitting independent (with uncertainties at the 68\% confidence interval) and dependent variables for the BDD objects using the ad-hoc grid and fitting using either a set distance (140\,pc) and extinction (values from Table \ref{target_phot}) or allowing these parameters to vary over the ranges prescribed by uncertainties, i.e. 130--165\,pc for distance and A$_{\rm V}$$\pm$0.8. The Objects CFHT-Tau 12 is denoted with an (*) as no satisfactory fit was achieved as was found, to a lesser extent for CFHT-Tau 4 which is denoted in the same way. Additionally, the values of the dependent (or derived) variables are only for the best fitting models, strictly these should span a range of values corresponding to that covered by the independent variables. (1) These variables can take only two or three available options, i.e. Age is 1 or 10\,Myrs in this grid, therefore uncertainties are meaningless we simply quote the best fitting values and if models with the alternate value where found to have $\chi^2$ values within the 68\% confidence interval their values are quoted in brackets. For example in the model age column: 1 (10) means a best fit of 1\,Myr was achieved but a 10\,Myr model cannot be ruled out (see Table \ref{grid_tab} for list of available parameters). (2) The $\alpha$ parameter can be used to calculate the $\beta$ parameter ($\alpha$-$\beta$=-1), a nominal value of 0 is used to denote where Vertical Hydrostatic Equilibrium (VHE) is used (see discussion in Section \ref{models_fitting}). \label{disc_best_par_AH}}
\end{center}
\end{table}
\end{landscape}

\setcounter{table}{8}
\begin{landscape}
\begin{table}
\begin{center}
\begin{tabular}{lllllllllllll}
\hline
\hline
\multirow{2}{*}{Target Name}&CFHT-&CFHT-&CFHT-&CFHT-&CFHT-&CFHT-&CFHT-&CFHT-&KPNO-&KPNO-&KPNO-&ITG 2\\
&Tau 16&Tau 15&Tau 13&Tau 11&Tau 7&Tau 5&Tau 3&Tau 2&Tau 9&Tau 5&Tau 4&\\
\hline
\hline
Variable&\multicolumn{12}{c}{Independent Variables}\\
\hline
Mass ($_{\odot}$)&0.040$^{+0.000}_{-0.010}$&0.050$^{+0.000}_{-0.020}$&0.070$^{+0.000}_{-0.030}$&0.090$^{+0.000}_{-0.030}$&0.100$^{+0.000}_{-0.010}$&0.060$^{+0.000}_{-0.000}$&0.040$^{+0.020}_{-0.009}$&0.040$^{+0.020}_{-0.005}$&0.020$^{+0.010}_{-0.001}$&0.060$^{+0.019}_{-0.010}$&0.030$^{+0.020}_{-0.001}$&0.050$^{+0.010}_{-0.007}$\\
Age (Myrs)&8$^{+2}_{-7}$&10$^{+0}_{-9}$&10$^{+0}_{-9}$&4$^{+0.03}_{-3}$&1$^{+1}_{-0}$&1$^{+0}_{-0}$&4$^{+0.3}_{-3}$&4$^{+0.06}_{-3}$&2$^{+8}_{-0.15}$&1$^{+1}_{-0}$&3$^{+3}_{-1}$&1$^{+2}_{-0}$\\
Dist (pc)&159.75$^{+3.60}_{-20.36}$&156.25$^{+8.75}_{-17.64}$&159.75$^{+5.25}_{-17.50}$&130$^{+24.91}_{0.00}$&140.50$^{+8.75}_{-10.50}$&130$^{+3.50}_{-0.00}$&133.50$^{+15.75}_{-3.50}$&130$^{+17.50}_{-0.00}$&159.75$^{+5.75}_{-17.54}$&138.75$^{+21.48}_{-8.75}$&131.75$^{+33.25}_{-1.75}$&130$^{+21.00}_{-0.00}$\\
A$_{\rm V}$&1.27$^{+0.58}_{-0.48}$&1.62$^{+0.48}_{-0.53}$&2.69$^{+0.96}_{-0.00}$&0.48$^{+0.40}_{-0.20}$&0.48$^{+0.29}_{-0.32}$&8.42$^{+0.38}_{-0.00}$&0.80$^{+0.00}_{-0.32}$&0.80$^{+0.00}_{-0.34}$&0.80$^{+0.00}_{-0.48}$&0.80$^{+0.00}_{-0.24}$&3.25$^{+0.00}_{-0.64}$&1.44$^{+0.00}_{-0.14}$\\
\hline
&\multicolumn{12}{c}{Dependent Variables of Best Fitting Model}\\
\hline
$\log \left( \frac{L}{L_{\odot}}\right) $&$-$2.323&$-$2.300&$-$2.090&$-$1.622&$-$1.160&$-$1.570&$-$2.002&$-$2.002&$-$2.580&$-$1.570&$-$2.208&$-$1.730\\
T$_{\rm eff}$ (K)&2760&2826&2968&3021&3002&2887&2789&2789&2519&2887&2703&2844\\
$\log (g)$&4.07&4.18&4.21&3.88&3.45&3.58&3.77&3.77&3.87&3.58&3.80&3.63\\
R$_*$ (R$_{\odot}$)&0.300&0.294&0.340&0.562&0.968&0.653&0.426&0.426&0.268&0.653&0.358&0.560\\
\hline
\hline
\end{tabular}
\caption{Table showing the adopted independent variables (with uncertainties at the 68\% confidence interval) and the dependent variables of the best fitting model for the naked objects. Note the age here is the ``model'' or isochronal age and should not be used as the exact chronological age of the system \citep[see discussion][]{mayne_2008}.\label{adopt_par_naked}}
\end{center}
\end{table}
\end{landscape}

\setcounter{table}{9}
\begin{landscape}
\begin{table}
\begin{center}
\begin{tabular}{llllllllllll}
\hline
\hline
\multirow{2}{*}{Target Name}&CFHT-&BDD&BDD&BDD&CFHT-&CFHT-&CFHT-&CFHT-&GM &KPNO-&KPNO-\\
&Tau 12 (*)&399&304&164&Tau 9&Tau 8&Tau 6&Tau 4&Tau&Tau 6&Tau 7\\
\hline
\hline
Variable&\multicolumn{11}{c}{Independent Variables}\\
\hline
Mass ($_{\odot}$) &0.06$^{+0.030}_{-0.002}$&0.08$^{+0.000}_{-0.000}$&0.06$^{+0.000}_{-0.020}$&0.08$^{+0.000}_{-0.010}$&0.08$^{+0.000}_{-0.020}$&0.08$^{+0.000}_{-0.010}$&0.06$^{+0.019}_{-0.010}$&0.08$^{+0.000}_{-0.000}$&0.08$^{+0.000}_{-0.010}$&0.03$^{+0.010}_{-0.011}$&0.03$^{+0.010}_{-0.010}$\\
Age (Myrs)$^{(1)}$&2$^{+2}_{-1}$&1&1&1&1&1&1&1&1&10 (1)&1 (10)\\
\logmdot&$-$12&$-$8$^{+0.07}_{-1.00}$&$-$9$^{+1.03}_{-1.00}$&$-$8$^{+0.55}_{-3.34}$&$-$9$^{+0.13}_{-3.00}$&$-$10$^{+1.03}_{-1.00}$&$-$9$^{+0.06}_{-3.00}$&$-$9$^{+0.14}_{-3.00}$&$-$8$^{+0.63}_{-3.11}$&$-$10$^{+0.05}_{-2.00}$&$-$10$^{+0.90}_{-2.00}$\\
Period (days)$^{(1)}$&1&5.0 (0.5)&5.0 (0.5)&5.0&0.5 (5.0)&5.0 (0.5)&0.5 (5.0)&5.0&5.0 (0.5)&0.5 (5.0)&0.5 (5.0)\\
Coverage (\%)$^{(1)}$&10&10 (1)&10 (1)&1 (10)&10 (1)&10 (1)&10 (1)&10 (1)&10 (1)&10 (1)&10 (1)\\
M$_{\rm disc}$ (M$_*$)$^{(1)}$&0.001&0.001&0.001&0.01&0.001 (0.01)&0.001 (0.01)&0.01&0.01&0.01 (0.001)&0.001&0.001 (0.01)\\
R$_{\rm outer}$ (AU)$^{(1)}$&100 (50)&100&100&100&100 (300)&300 (100)&300 (100)&300&300 (100)&300 (100)&100 (300)\\
$\alpha$ (0=VHE)$^{(1)}$&2.00 (2.10)&2.10&2.25 (0)&2.10&0 (2.25)&0 (2.25)&2.25 (0)&2.10&2.10&0 (2.25)&2.25 (0)\\
Inclination ($^{\circ}$)&48$^{+12.80}_{-48.00}$&64$^{+0}_{-0}$&71$^{+2.17}_{-71}$&0$^{+56}_{-0}$&64$^{+2.59}_{-37}$&71$^{+0.76}_{-49.43}$&48$^{8.66}_{-21}$&39$^{+9.44}_{-12}$&48$^{+8.01}_{-9}$&64$^{+5.22}_{-37}$&71$^{+0.99}_{-71}$\\
Dist (pc)&140.50$^{+24.50}_{-4.45}$&130.00$^{+14.00}_{-0.00}$&130.00$^{+17.50}_{-0.00}$&149.25$^{+4.16}_{-19.25}$&163.25$^{+1.75}_{-33.25}$&138.75$^{+19.25}_{-8.75}$&140.50$^{+22.75}_{-9.10}$&130.00$^{+14.00}_{-0.00}$&151.00$^{+6.63}_{-17.50}$&131.75$^{+19.25}_{-1.75}$&130.00$^{+31.93}_{-0.00}$\\
A$_{\rm V}$&2.64$^{+0.72}_{-0.00}$&0.64$^{+0.16}_{-0.32}$&1.86$^{+0.00}_{-0.64}$&0.80$^{+0.00}_{-0.56}$&1.15$^{+0.56}_{-0.58}$&2.60$^{+0.00}_{-0.56}$&1.21$^{+0.00}_{-0.48}$&3.24$^{+0.29}_{-0.48}$&3.54$^{+0.96}_{-0.00}$&1.52$^{+0.16}_{-0.90}$&0.80$^{+0.00}_{-0.48}$\\
\hline
&\multicolumn{11}{c}{Dependent Variables of Best Fitting Model}\\
\hline
$\log \left( \frac{L}{L_{\odot}}\right) $&$-$1.610&$-$1.340&$-$1.570&$-$1.340&$-$1.340&$-$1.340&$-$1.570&$-$1.340&$-$1.340&$-$2.490&$-$2.130\\
T$_{\rm eff}$ (K)&2887&2941&2893&2941&2941&2941&2893&2941&2941&2667&2703\\
$\log (g)$&3.62&3.50&3.58&3.50&3.50&3.50&3.58&3.50&3.50&4.05&3.72\\
R$_*$ (R$_{\odot}$)&0.624&0.829&0.658&0.829&0.829&0.829&0.658&0.829&0.829&0.270&0.398\\
$\log \left( \frac{\dot{L}}{L_{\odot}}\right) $&-&$-$1.7818&$-$2.7617&$-$1.7818&$-$2.7818&$-$3.7818&$-$2.7617&$-$2.7818&$-$1.7818&$-$3.5545&3.7804\\
$\dot{T_{\rm eff}}$ (K)&-&4044&2582&7192&2274&1279&2582&2274&4044&2554&1847\\
$\dot{\lambda_{\rm peak}}$ ($\AA$)&-&7165&11221&4029&12742&22659&11221&12742&7165&11345&15688\\
R$_{\rm sub}$ (R$_*$)&3.5&6.4&7.3&6.4&2.1&6.4&2.1&6.4&6.4&2.9&2.3\\
\hline
\hline
\end{tabular}
\caption{
  Table showing the adopted independent (with uncertainties at the 68\%
  confidence interval) and dependent variables for the BDD objects. Note
  the age here is the ``model'' or isochronal age and should not be used
  as the exact chronological age of the system \citep[see
  discussion][]{mayne_2008}. These objects have all be fitted using the 
  ad-hoc grid, except for CFHT-Tau 12. (1) These variables can take only two or
  three available options, i.e. Age is 1 or 10\,Myrs in this grid,
  therefore uncertainties are meaningless we simply quote the best
  fitting values and if models with the alternate value where found to
  have $\chi^2$ values within the 68\% confidence interval their values
  are quoted in brackets. For example in the model age column: 1 (10)
  means a best fit of 1\,Myr was achieved but a 10\,Myr model cannot be
  ruled out (see Table \ref{grid_tab} for list of available
  parameters). The $\alpha$ parameter can be used to 
  calculate the $\beta$ parameter ($\alpha$-$\beta$=-1), a nominal value of 0
  is used to denote where Vertical Hydrostatic Equilibrium (VHE) is used 
  (see discussion in Section \ref{models_fitting}).\label{adopt_par_disc}}
\end{center}
\end{table}
\end{landscape}

%% file: appendix.tex
\appendix

\section{Data Problems}
\label{data_problems}

We attempted initial fitting of the observations using the associated
uncertainties quoted in the source publications for the
photometry. During this fitting, the optical photometry clearly
contributed a disproportionate statistical weight to the fitting
process. This resulted in best fitting SEDs which were systematically
shifted away from the IR and NIR observations. Previous studies have
dealt with this issue by simply neglecting the shortwave, optical
photometry during the fitting process \citep{guieu_2007}. However,
given the powerful constraint such observations can provide on the
stellar photosphere we feel such observations must be retained, where
possible. 

This initial fitting essentially adopted uncertainties based on a
frequentist statistical approach, where we assume that the observation
recorded is a value drawn from an underlying distribution. The shape
of this distribution is determined by the properties of the measuring
instrument and reduction or processing properties. Quoted
uncertainties on observational data points are then the values
encompassing 68\% percent of the integrated (about the observed value)
assumed or modelled Gaussian uncertainty curve. These uncertainties are
by nature underestimates (encompassing 68\% of the probability curve)
and are to be understood as the probability that the observation,
under the same exact conditions (including time), would be repeated.
We adopt a Bayesian perspective whereby the uncertainties are a
property of the model not the data. Such uncertainties must include a
component accounting for poor or incomplete model formulation, and a
component accounting for uncertainties in conversion of the model to
the observational domain. In effect we are asking which model, from
our set, provides the best representation of the data, as oppose to,
from which model distribution are the data most likely drawn. In
practice, it was clear from the initial fits that the quoted
observational uncertainties were insufficient to account for all
sources of discrepancy between the model and observations,
particularly for the optical regime. The resulting best fitting models
were systematically shifted towards the shortwave length observations
leading to a shift away from the longer wavelength points. Therefore,
by adjusting the uncertainties iteratively with the fitting process,
for the entire sample, we constructed a representative model of the
total uncertainty for each photometric regime, short, mid and long
wavelength (these are presented in Table
\ref{target_phot}). Practically, this resulted in a significant
widening of the quoted photometric uncertainties, which was most
extreme for the optical photometry.

In this study we have combined many disparate sources of photometry
using different optical equipment (telescopes, cameras and filters)
spread over several epochs, whilst the models are time independent. We
have also combined several model formulations, for instance models of
the stellar interiors, atmospheres and models of radiative transfer in
discs. Therefore, our combined uncertainties, applied to our models,
will have several contributions. In this section we explore some of
the underlying contributions to observational uncertainties in each
photometric regime.

\subsubsection{IRAC Photometry}
\label{irac_phot}

For the IRAC photometry an uncertainty of 0.2 mags was found to best
represent the required uncertainties. Is this reflected by
investigation of the data? Tables \ref{irac_phot_naked} and
\ref{irac_phot_disc} show IRAC photometry for each target object, as
published in several studies, for objects in our sample both without
and with discs respectively. In almost all cases the magnitudes of the
IRAC photometry vary over at most four observations, much more than
the quoted uncertainties. In some cases the differences could be
attributed to improved reductions
\citep[e.g.][]{luhman_2006,luhman_2010}, whilst in others new
observations have been performed
\citep[e.g.][]{luhman_2006,monin_2010}. In either case it is clear
that the natural variation in such photometry is around a factor ten
larger than the quoted uncertainties. Some of this variation will be
caused by temporal changes in the structure of the associated disc, as
demonstrated by the fact that the dispersion of the photometry is
generally less for naked stars (Table \ref{irac_phot_naked}) than
those with discs (Table \ref{irac_phot_disc}). However, there is still
variation in naked stars in excess to the quoted uncertainties, which
could be due to variations in the stellar emission. Even without
natural variation in the photometric magnitudes it is clear that when
combining several sets of photometric observations (even using the
same instrument) one may expect a variation in the magnitudes larger
than the quoted uncertainties. Indeed, a value of 0.2 for the
approximate uncertainty across the sample is not inconsistent with the
spread of magnitudes shown in Table \ref{irac_phot_naked} and
\ref{irac_phot_disc}.

\begin{table*}
\begin{tabular}{l|cccc|l}
\hline
Name&[3.6]&[4.5]&[5.8]&[8.0]&Sources\\
&\multicolumn{4}{c}{Magnitudes}&\\
\hline
\multirow{4}{*}{KPNO-Tau 4}&12.49$\pm$0.04&12.34$\pm$0.06&12.20$\pm$0.06&12.11$\pm$0.06&\citet{luhman_2006}\\
&12.52&12.34&12.14&12.10&\citet{guieu_2007}\\
&12.57&12.37&12.21&12.08&\citet{monin_2010}\\
&12.56$\pm$0.02&12.35$\pm$0.02&12.23$\pm$0.04&12.11$\pm$0.04&\citet{luhman_2010}\\
$\sim\Delta$&0.08&0.03&0.09&0.03\\
\hline
\multirow{4}{*}{CFHT-Tau 15}&13.17$\pm$0.03&13.06$\pm$0.07&12.95$\pm$0.06&12.86$\pm$0.10&\citet{luhman_2006}\\
&13.19&13.16&13.04&13.04&\citet{guieu_2007}\\
&13.24&13.15&13.25&13.06&\citet{monin_2010}\\
&13.20$\pm$0.02&13.05$\pm$0.02&12.98$\pm$0.04&13.00$\pm$0.07&\citet{luhman_2010}\\
$\sim\Delta$&0.07&0.11&0.30&0.20\\
\hline
\multirow{4}{*}{KPNO-Tau 5}&11.00$\pm$0.06&10.90$\pm$0.05&10.84$\pm$0.04&10.79$\pm$0.04&\citet{luhman_2006}\\
&11.03&10.99&10.90&10.84&\citet{guieu_2007}\\
&11.05&11.02&10.94&10.83&\citet{monin_2010}\\
&11.05$\pm$0.02&10.92$\pm$0.02&10.81$\pm$0.03&10.85$\pm$0.03&\citet{luhman_2010}\\
$\sim\Delta$&0.05&0.10&0.13&0.06\\
\hline
\multirow{4}{*}{CFHT-Tau 16}&13.15$\pm$0.06&13.10$\pm$0.06&12.95$\pm$0.09&12.83$\pm$0.09&\citet{luhman_2006}\\
&13.26&13.16&13.01&12.97&\citet{guieu_2007}\\
&12.23&13.16&13.04&12.99&\citet{monin_2010}\\
&13.21$\pm$0.02&13.11$\pm$0.02&13.03$\pm$0.05&13.02$\pm$0.06&\citet{luhman_2010}\\
$\sim\Delta$&0.11&0.06&0.09&0.19\\
\hline
\multirow{4}{*}{CFHT-Tau 13}&12.79$\pm$0.05&12.70$\pm$0.04&12.58$\pm$0.06&12.66$\pm$0.05&\citet{luhman_2006}\\
&12.74&12.73&12.84&12.60&\citet{guieu_2007}\\
&12.90&12.75&12.72&12.67&\citet{monin_2010}\\
&12.85$\pm$0.02&&12.73$\pm$0.04&\\
$\sim\Delta$&0.16&0.05&0.26&0.07\\
\hline
\multirow{4}{*}{CFHT-Tau 7}&9.92$\pm$0.04&9.77$\pm$0.05&9.70$\pm$0.04&9.69$\pm$0.03&\citet{luhman_2006}\\
&9.83&9.91&9.70&9.69&\citet{guieu_2007}\\
&9.98&9.87&9.76&9.72&\citet{monin_2010}\\
&9.91$\pm$0.02&9.79$\pm$0.02&9.74$\pm$0.03&9.72$\pm$0.03&\citet{luhman_2010}\\
$\sim\Delta$\&0.15&0.14&0.06&0.03\\
\hline
\multirow{4}{*}{CFHT-Tau 5}&10.37$\pm$0.04&10.21$\pm$0.05&10.06$\pm$0.04&10.02$\pm$0.03&\citet{luhman_2006}\\
&10.48&10.25&10.03&10.01&\citet{guieu_2007}\\
&10.46&10.27&10.10&10.07&\citet{monin_2010}\\
&10.42$\pm$0.02&10.19$\pm$0.02&10.08$\pm$0.03&10.05$\pm$0.03&\citet{luhman_2010}\\
$\sim\Delta$&0.11&0.08&0.07&0.06\\
\hline
\multirow{4}{*}{CFHT-Tau 11}&11.12$\pm$0.07&11.03$\pm$0.06&10.98$\pm$0.04&10.94$\pm$0.04&\citet{luhman_2006}\\
&11.13&11.09&10.96&10.96&\citet{guieu_2007}\\
&11.19&11.12&11.04&10.99&\citet{monin_2010}\\
&11.19$\pm$0.02&11.06$\pm$0.02&11.02$\pm$0.03&10.98$\pm$0.03&\citet{luhman_2010}\\
$\sim\Delta$&0.07&0.09&0.08&0.05\\
\hline
\multirow{4}{*}{KPNO-Tau 9}&13.51&&&&\citet{guieu_2007}\\
&13.63&13.52&13.67&13.41&\citet{monin_2010}\\
&13.61$\pm$0.02&13.53$\pm$0.02&13.44$\pm$0.05&13.48$\pm$0.08&\citet{luhman_2010}\\
$\sim\Delta$&0.12&0.01&0.23&0.07\\
\hline
\multirow{4}{*}{CFHT-Tau 2}&11.62$\pm$0.05&11.38$\pm$0.05&11.35$\pm$0.04&11.29$\pm$0.03&\citet{luhman_2006}\\
&11.54&11.41&11.32&11.33&\citet{guieu_2007}\\
&11.63&11.43&11.34&11.33&\citet{monin_2010}\\
&11.57$\pm$0.02&11.38$\pm$0.02&11.37$\pm$0.03&11.36$\pm$0.03&\citet{luhman_2010}\\
$\sim\Delta$&0.09&0.05&0.05&0.07\\
\hline
\multirow{4}{*}{CFHT-Tau 3}&11.71$\pm$0.05&11.62$\pm$0.05&11.54$\pm$0.04&11.56$\pm$0.04&\citet{luhman_2006}\\
&11.78&11.69&11.61&11.59&\citet{guieu_2007}\\
&11.79&11.69&11.60&11.57&\citet{monin_2010}\\
&11.77$\pm$0.02&11.62$\pm$0.02&11.56$\pm$0.03&11.54$\pm$0.03&\citet{luhman_2010}\\
$\sim\Delta$&0.08&0.07&0.07&0.05\\
\hline
\multirow{4}{*}{ITG 2}&9.53$\pm$0.04&9.41$\pm$0.04&9.31$\pm$0.03&9.29$\pm$0.03&\citet{luhman_2006}\\
&9.56&9.45&9.31&9.28&\citet{guieu_2007}\\
&9.60&9.47&9.37&9.31&\citet{monin_2010}\\
&9.57$\pm$0.02&9.42$\pm$0.02&9.33$\pm$0.03&9.34$\pm$0.03&\citet{luhman_2010}\\
$\sim\Delta$&0.07&0.06&0.06&0.05\\
\hline
\end{tabular}
\caption{IRAC photometry for our target objects without discs as reported by several studies. $\Delta$ here is the approximate range of the quoted magnitudes.\label{irac_phot_naked}}
\end{table*}

\begin{table*}
\begin{tabular}{l|cccc|l}
\hline
Name&[3.6]&[4.5]&[5.8]&[8.0]&Sources\\
&\multicolumn{4}{c}{Magnitudes}&\\
\hline
\multirow{4}{*}{CFHT-Tau 9}&11.13$\pm$0.04&10.85$\pm$0.05&10.49$\pm$0.03&9.80$\pm$0.03&\citet{luhman_2006}\\
&11.14&10.86&10.45&9.80&\citet{guieu_2007}\\
&11.16&10.88&10.51&9.83&\citet{monin_2010}\\
&11.11$\pm$0.02&10.80$\pm$0.02&10.48$\pm$0.03&9.85$\pm$0.03&\citet{luhman_2010}\\
$\sim\Delta$&0.05&0.08&0.06&0.05&\\
\hline
\multirow{4}{*}{KPNO-Tau 6}&13.07$\pm$0.06&12.74$\pm$0.06&12.36$\pm$0.08&11.59$\pm$0.07&\citet{luhman_2006}\\
&13.08&12.79&12.47&11.68&\citet{guieu_2007}\\
&13.12&12.77&12.42&11.58&\citet{monin_2010}\\
&13.08$\pm$0.02&12.77$\pm$0.02&12.41$\pm$0.04&11.82$\pm$0.04&\citet{luhman_2010}\\
$\sim\Delta$&0.05&0.05&0.11&0.23&\\
\hline
\multirow{4}{*}{KPNO-Tau 7}&12.54$\pm$0.04&12.23$\pm$0.03&11.89$\pm$0.05&11.24$\pm$0.02&\citet{luhman_2006}\\
&12.59&12.27&11.90&11.22&\citet{guieu_2007}\\
&12.62&12.29&11.99&11.25&\citet{monin_2010}\\
&12.60$\pm$0.02&12.27$\pm$0.02&11.93$\pm$0.03&11.26$\pm$0.03&\citet{luhman_2010}\\
$\sim\Delta$&0.08&0.06&0.10&0.04&\\
\hline
\multirow{2}{*}{CFHT-Tau 12}&10.86&10.63&10.34&9.95&\citet{monin_2010}\\
&10.77$\pm$0.02&10.54$\pm$0.02&10.31$\pm$0.03&9.93$\pm$0.03&\citet{luhman_2010}\\
$\sim\Delta$&0.09&0.07&0.03&0.02&\\
\hline
\multirow{4}{*}{BDD399}&10.71$\pm$0.03&10.12$\pm$0.05&9.55$\pm$0.03&8.86$\pm$0.03&\citet{luhman_2006}\\
&10.77&10.19&9.66&8.91&\citet{guieu_2007}\\
&10.80&10.21&9.64&8.92&\citet{monin_2010}\\
&10.79$\pm$0.02&10.14$\pm$0.02&9.61$\pm$0.03&8.91$\pm$0.03&\citet{luhman_2010}\\
$\sim\Delta$&0.09&0.09&0.11&0.06&\\
\hline
\multirow{4}{*}{GM Tau}&9.16$\pm$0.05&8.70$\pm$0.07&8.38$\pm$0.03&7.79$\pm$0.02&\citet{luhman_2006}\\
&9.25&8.76&8.38&7.80&\citet{guieu_2007}\\
&9.27&8.77&8.43&7.81&\citet{monin_2010}\\
&9.44$\pm$0.02&8.95$\pm$0.02&8.64$\pm$0.03&7.97$\pm$0.03&\citet{luhman_2010}\\
$\sim\Delta$&0.28&0.25&0.26&0.18&\\
\hline
\multirow{2}{*}{CFHT-Tau 6}&10.66$\pm$0.05&10.37$\pm$0.06&9.93$\pm$0.03&9.10$\pm$0.03& \citet{luhman_2006}\\
&10.78$\pm$0.02&10.37$\pm$0.02&10.03$\pm$0.03&9.16$\pm$0.03& \citet{luhman_2010}\\
$\sim\Delta$&0.12&0.00&0.10&0.06&\\
\hline
\multirow{4}{*}{CFHT-Tau 4}&9.38$\pm$0.05&8.97$\pm$0.06&8.54$\pm$0.02&7.79$\pm$0.03&\citet{luhman_2006}\\
&9.48&9.06&8.58&7.79&\citet{guieu_2007}\\
&9.54&9.07&8.60&7.79&\citet{monin_2010}\\
&9.51$\pm$0.02&9.08$\pm$0.02&8.62$\pm$0.03&7.85$\pm$0.03&\citet{luhman_2010}\\
$\sim\Delta$&0.16&0.11&0.08&0.06&\\
\hline
\multirow{3}{*}{CFHT-Tau 8}&10.85$\pm$0.06&10.29$\pm$0.03&9.86$\pm$0.04&9.19$\pm$0.02&\ \citet{luhman_2006}\\
&10.83&10.31&9.86&9.18&\citet{guieu_2007}\\
&&10.23$\pm$0.02&&9.14$\pm$0.03&\citet{luhman_2010}\\
$\sim\Delta$&0.02&0.08&0.10&0.07&\\
\hline
\multirow{4}{*}{BDD304}&11.38$\pm$0.05&10.85$\pm$0.04&10.40$\pm$0.04&9.52$\pm$0.03&\citet{luhman_2006}\\
&11.37&10.88&10.43&9.53&\citet{guieu_2007}\\
&11.43&10.93&10.50&9.54&\citet{monin_2010}\\
&&10.87$\pm$0.02&&9.59$\pm$0.03&\citet{luhman_2010}\\
$\sim\Delta$&0.06&0.08&0.10&0.07&\\
\hline
\multirow{4}{*}{BDD164}&9.48$\pm$0.05&8.92$\pm$0.06&8.28$\pm$0.03&7.40$\pm$0.03&\citet{luhman_2006}\\
&9.51&8.99&8.33&7.40&\citet{guieu_2007}\\
&9.56&9.00&8.36&7.43&\citet{monin_2010}\\
&9.75$\pm$0.02&9.34$\pm$0.02&8.87$\pm$0.03&7.89$\pm$0.03&\citet{luhman_2010}\\
$\sim\Delta$&0.27&0.32&0.59&0.49&\\
\hline
\end{tabular}
\caption{IRAC photometry for our target objects with discs as reported by several studies. $\Delta$ here is the approximate range of the quoted magnitudes.\label{irac_phot_disc}}
\end{table*}

\subsubsection{JHK Photometry}
\label{jhk_phot}

For the JHK photometry we found an uncertainty of 0.2 mags to best
represent the combined uncertainties of the models. However, in almost
all the sources quoted in the caption of Table \ref{target_phot} the
JHK magnitudes were consistent, this is as they are all taken from the
2MASS catalogue. However, one would expect temporal variation in the
JHK photometry of $\sim$0.2 mag over the timescale of a few days, as
shown by \cite{carpenter_2001}. Therefore, when combining the JHK
observations with non-contemporaneous observations, reflecting the
star-and disc properties at a different time, this variation must be
incorporated into the uncertainties.

\subsubsection{Optical Photometry}
\label{opt_phot}

The final regime where we must model our uncertainties is in the
optical. Photometry in bands shortward of $R$ and $I$ proved
impossible to include in the fitting. The associated uncertainties
required to create a sensible weighting between such observations and
longer wavelength magnitudes, resulted in these observations providing
no real constraint to the fit. For observations in the $R$ and $I$
bands uncertainties, again for the entire sample, of 0.8 and 0.3 mags
respectively were found to result in the most appropriately weighted
fits. 

Observations in this regime are the most uncertain due to the
relatively cool effective temperatures of BDs and LMS (compared to MS
stars). These objects, therefore, appear faint at these wavelengths
observations in optical filters are difficult to obtain with any
precision. Additionally, cool atmospheres contain many more molecular
species and current atmospheric models are known to include incomplete
lists of the spectral lines, on which the opacity depends
\cite{stauffer_2007}. However, the most significant problem with
optical photometry is uncertainty about the natural system of the
observations. Small changes in the filter responses, especially for
asymmetric filters such as the $R$ filter, will lead to large changes
in the derived fluxes, as the SED gradient is steep across this
wavelength range.

The $R$ and $I$ photometry presented in Table \ref{target_phot} comes
from four sources. \cite{martin_2001} used the CFHT 12K camera and
state that they observe using the $R_C, I_C$ filters (although the
CFHT 12K camera only has Mould $R$, $I$ filters). They go on to say
that the photometry will be explained in further detail in Dougados et
al (in preparation) which we assume refers to \cite{dougados_2001},
although no further explanation is provided. Therefore, it is unclear
whether these photometric observations are in the CFHT 12K-Mould $R$,
$I$ system or have, in some way, been transformed to the Cousins
system. The transformation between photometric systems is usually done
using observation of Landolt standards \citep{landolt_1992}. However,
for BDs and LMS such transforms are likely to be extremely unreliable
as the SED has a steep gradient across the filter, meaning
uncertainties in the filter shape will lead to large inaccuracies in
the photometry. \cite{briceno_2002} observed using the KPNO $I$
filters and calibrated this to the Cousins system using
\cite{landolt_1992} standards. \cite{luhman_2004} observes using an
$I$ filter on the four-shooter camera at the Fred Lawrence Whipple
Observatory, and employs conversions using \cite{landolt_1992}
standards. The final source is the series of papers \cite{guieu_2005},
\cite{guieu_2006} and \cite{guieu_2007}. \cite{guieu_2005} observe
using the CFHT 12K $R$ and $I$, and the Mega Prime $r'$ and $i'$
filters, and the calibration is not explained, suggesting that these
measurements are presented in the natural systems in which they are
taken. These data are extended by observations presented in
\cite{guieu_2006} where observations, again using the CFHT 12K and
Mega Prime systems, were combined using overlapping fields, resulting
in a photometric magnitudes in the Mega Prime $r'$ and $i'$ system,
however the magnitudes are not explicitly presented. Finally, $R$ and
$I$ magnitudes appear in \cite{guieu_2007} where they state: ``All $R$
and $I$ data have been transformed to the CFHT12k camera's Cousins
system \citep[see][]{guieu_2006,briceno_2002} before conversion to
absolute fluxes.'' However, the magnitudes presented in
\cite{guieu_2005} match those presented (for common objects) in
\cite{guieu_2007}. Therefore, as no transformation or calibration to
Cousins is explained in either \cite{guieu_2005} or \cite{guieu_2006}
we assume that the magnitudes presented in \cite{guieu_2007} are
actually in the Mega Prime $r'$, $i'$ system.
 
Table \ref{dup_phot} shows sources and photometry for the targets
where we have duplicate $I$ band measurements (all stated to be in the
Cousins $I$ photometric system). It is clear that for stars both with
and without discs large variation, as much as 0.23 mags, is apparent
in the $I$ band photometry. It is reasonable to assume that the
predictive power of our models will decrease towards shorter
wavelengths as the dust opacity becomes more important and the slope
of the SED across the filter responses leads to an exaggeration of
slight errors in the adopted response. Therefore, we conclude that a
minimum of 0.2 mags could be contributed to the uncertainty budget due
to variation in the photometry.

\begin{table*}
\begin{tabular}{l|c|l}
\hline
Name&I&Sources\\
&Magnitudes&\\
\hline
\multicolumn{3}{c}{Sources without infrared excess}\\
\hline
\multirow{3}{*}{CFHT-Tau 2}&16.81&\citet{briceno_2002}, \citet{guieu_2005}, \citet{guieu_2007}\\
&16.69&\citet{martin_2001}\\
$\sim$$\Delta$&0.12&\\
\hline
\multirow{3}{*}{CFHT-Tau 3}&16.88&\citet{briceno_2002}, \citet{guieu_2007}\\
&16.77&\citet{martin_2001}\\
$\sim\Delta$&0.11&\\
\hline
\multicolumn{3}{c}{Sources with infrared excess}\\
\hline
\multirow{3}{*}{KPNO-Tau 6}&17.90&\citet{guieu_2005}, \citet{guieu_2007}\\
&17.80&\citet{briceno_2002}\\
$\sim\Delta$&0.10&\\
\hline
\multirow{2}{*}{CFHT-Tau 6}&15.40&\citet{guieu_2005}\\
&15.63&\citet{luhman_2004}\\
$\sim\Delta$&0.23&\\
\hline
\multirow{2}{*}{CFHT-Tau 4}&15.78&\citet{luhman_2004}\\
&15.64&\citet{martin_2001}\\
$\sim\Delta$&0.14&\\
\hline
\multirow{2}{*}{CFHT-Tau 8}&16.43&\citet{guieu_2005}, \citet{guieu_2007}\\
&16.24&\citet{luhman_2004}\\
$\sim\Delta$&0.19&\\
\hline
\end{tabular}
\caption{Duplicate photometry for which differences are above 0.05 mag.\label{dup_phot}}
\end{table*}

\begin{figure}
  \includegraphics[scale=0.3,angle=90]{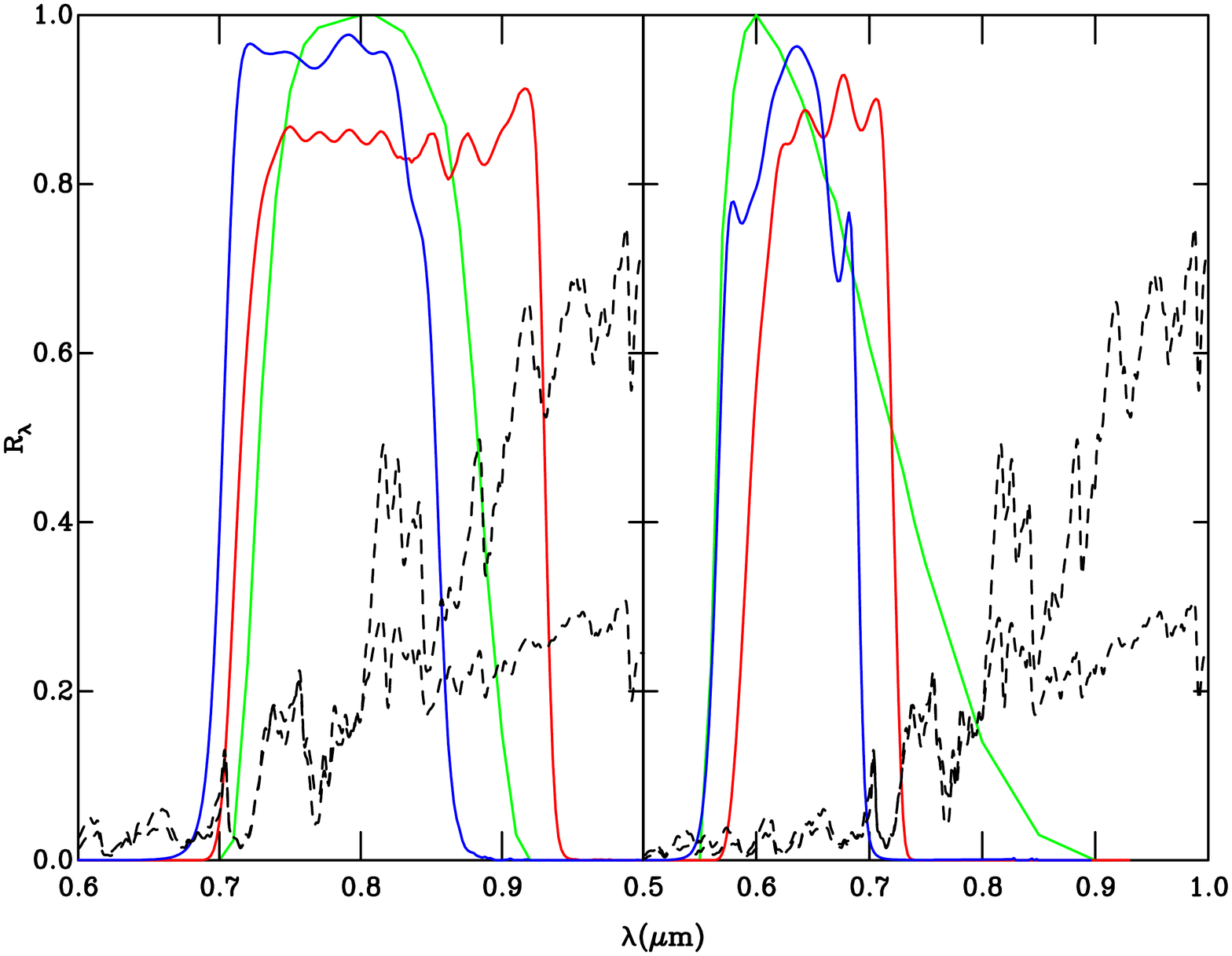}
  \caption{The normalised filters responses as a function of
    wavelength for the $I$ (left panel) and $R$ (right panel) filters of
    the Johnson-Cousins system \citep[green solid
    line][]{bessell_2005}, Mega Prime (blue solid lines, $r'$, $i'$ of
    \cfht{}) and CFHT 12K (red solid line, Mould $R$ and $I$,
    \cfht{}). Overlaid, normalised to arbitrary units, are the SEDs
    (dashed lines) of a MS (unreddenned) and pre-MS ($A_V=$0) star at
    a colour typical of a Taurus object ($R-I$=2.3).\label{filters}}
\end{figure}

Table \ref{r_mag} lists the sources of the $R$ band photometry.  For
the $R$ band photometry there are no duplicates (where the photometry
is from different underlying sources), so we cannot assess the
uncertainty of the transforms applied (as all of the photometry is
stated as having been transformed to the Cousins system). However, we
can assume that the uncertainty will be significantly larger than that
found for the $I$ band photometry due to the shapes of the $R$ band
filters. Figure \ref{filters} shows the filter responses (normalised
to the same scale) for the Mega Prime, CFHT and Cousins $R$ and $I$
filters. The shapes of the $I$ band filters are roughly consistent
with the assumption of a flat response. For the $R$ band however there
exists a longer tail (extending to 0.0 $\mu$m) in the Cousins filter
\citep{bessell_2005}. This will exacerbate problems with
transformations between these filters as the slope of the SED is a
sharp function of wavelength in this region (see Figure
\ref{filters}).

\begin{table}
\begin{tabular}{l|c|l}
\hline
Name&$R$&Source\\
\hline
\multicolumn{3}{c}{Sources without infrared excess}\\
\hline
KPNO-Tau 4&20.54&\citet{guieu_2006}, \citet{guieu_2007}\\
KPNO-Tau 5&19.10&\citet{guieu_2007}\\
CFHT-Tau 7&16.63&\citet{guieu_2005}, \citet{guieu_2007}\\
CFHT-Tau 5&23.37&\citet{guieu_2005}\\
CFHT-Tau 2&20.21&\citet{martin_2001}\\
CFHT-Tau 3&20.33&\citet{martin_2001}\\
ITG 2&20.21&\citet{guieu_2007}\\
\hline
\multicolumn{3}{c}{Sources without infrared excess}\\
\hline
KPNO-Tau 6&20.56&\citet{guieu_2005}, \citet{guieu_2007}\\
BDD399&20.33&\citet{guieu_2007}\\
CFHT-Tau 6&18.39&\citet{guieu_2005}\\
CFHT-Tau 4&19.10&\citet{martin_2001}\\
CFHT-Tau 8&19.27&\citet{guieu_2005}, \citet{guieu_2007}\\
\hline
\end{tabular}
\caption{A table showing the $R$ band photometry and sources.\label{r_mag}}
\end{table}

Transformations are generally performed using \cite{landolt_1992}
standards, which are unreddened main-sequence (MS) stars in a limited
colour range. To demonstrate why such transforms are unreliable for BD
stars we have performed a simple experiment to construct a simulated
version of a transformation between the relevant photometric
systems. In practice transformations are made using a conversion
equation as a function of colour. Table \ref{transforms} shows the
simulated $R$ and $I$ band photometry in the Cousins, CFHT 12K and
Mega Prime systems for three unreddened\footnote{at 140\,pc, although
  distance is unimportant this has been set to the distance of Taurus
  for illustrative purposes} MS stars at colours $R_{\rm c}-I_{\rm
  c}\sim$0, 1.2 and a pre-MS BD at 2.3 behind 2 magnitudes of
extinction (typical for the targets in Taurus). The simulated
photometry was created using the models of \cite{siess_2000},
\cite{chabrier_2000} and \cite{hauschildt_1999}. \cite{briceno_2002},
for example, used the standard fields SA 95 and SA 101 to calibrate
their photometry to the Cousins system. These standard fields contain
MS stars covering a range of colours from $\sim$0.0 to 0 1.2. Whereas,
the pre-MS stars of colour 2.3 is typical of the bluest BD stars in
Taurus.

\begin{table}
\begin{tabular}{lcccc}
\hline
$R_{\rm C}-I_{\rm c}$&Band&Cousins&CFHT 12K&Mega Prime\\
\hline
\multicolumn{5}{c}{pre-MS, $A_V=$2}\\
\hline
\multirow{2}{*}{$\sim2.3$}&$R$&19.63&20.31&20.87\\
&$I$&17.31&17.08&17.71\\
&$R-I$&2.32&3.23&3.16\\
\hline
\multicolumn{5}{c}{MS, $A_V=$0}\\
\hline
\multirow{2}{*}{$\sim0$}&$R$&6.35&6.36&6.37\\
&$I$&6.37&6.38&6.37\\
&$R-I$&$-$0.02&$-$0.02&0.00\\
\hline
\multirow{2}{*}{$\sim1.2$}&$R$&10.93&11.08&11.26\\
&$I$&9.73&9.68&9.95\\
&$R-I$&1.20&1.40&1.31\\
\hline
\end{tabular}
\caption{Table showing the simulated photometry for MS stars across a range of colours and a pre-MS BD typical of the Taurus targets.\label{transforms}}
\end{table}

Table \ref{transforms} shows that if we construct transformation
equations (assuming a linear relationship in colour between the
instrumental and calibrated magnitude) using the MS stars in the
colour range usually found in the \cite{landolt_1992} standard fields
we derive (assuming a negligible airmass component)
\begin{eqnarray}
I_{\rm c}=I_{\rm CFHT} - 0.009 + 0.042\times (R_{\rm CFHT}-I_{\rm CFHT})\\
I_{\rm c}=I_{\rm Mega} + 0.000 - 0.168\times (R_{\rm Mega}-I_{\rm Mega})\\
R_{\rm c}=R_{\rm CFHT} - 0.012 - 0.099\times (R_{\rm CFHT}-I_{\rm CFHT})\\
R_{\rm c}=R_{\rm Mega} - 0.020 - 0.252\times (R_{\rm Mega}-I_{\rm Mega}).
\label{convert_one}
\end{eqnarray}
These relationships would then be used to extrapolate out to the
required colour for transformation of Taurus BDs. The results of
evaluating the transform equations and comparing to the values derived
from the simulated observations are shown in Table \ref{trans_comp}.

\begin{table}
\begin{tabular}{cccc}
\hline
Band&CFHT 12K&Mega Prime\\
\hline
$I_{\rm transforms}$&+0.127&$-$0.531\\
$I_{\rm simulated}$&+0.230&$-$0.400\\
\hline
$R_{\rm transforms}$&$-$0.332&$-$0.816\\
$R_{\rm simulated}$&$-$0.680&$-$1.240\\
\hline
\end{tabular}
\caption{Table showing the $\Delta$ magnitudes predicted from the transforms equations which simulate the use of \citet{landolt_1992} standards, and the actual $\Delta$ magnitudes for a typical simulated Taurus object.\label{trans_comp}}
\end{table}

Table \ref{trans_comp} shows that if \cite{landolt_1992} transforms
are used we would expect the transforms to be incorrect by around
$\sim$0.1 for the $I$ band and $\sim$0.4--0.5 for the $R$ band for
pre-MS BDs. This is caused by the shape of the SED changing across the
filter responses as one moves from MS stars to BDs. The change in SED
between BDs and MS is driven by a change in the surface gravity of
these objects, $\sim$5.25 for a MS star and 3.79 for a pre-MS star at
similar colours. Figure \ref{filters} shows the normalised filter
responses for the $I$ and $R$ bands of the Cousins, CFHT 12K and Mega
Prime filter systems, overlaid with SEDs of a MS BD and pre-MS BD
(arbitrarily scaled) for illustrative purposes.

Figure \ref{filters} shows that the area under the filter response is
similar for the $I$ band filters but the $R$ band Cousins filters
shows a longer wavelength tail. The overlaid SEDs then show that for
the MS (greater flux, or higher line) and the pre-MS the long
wavelength tail of the Cousins filter becomes significant. This leads
to the problems with transforms illustrated by Table
\ref{trans_comp}. Therefore, for the $I$ band photometry the
combination of variations in the photometry and transformation between
photometric systems, leads to an uncertainty budget of $\sim$ 0.3,
matching the value we have adopted for the sample. For the $R$ band
photometry, uncertainties caused by transformations, combined with
variation in the photometric magnitudes, lead to uncertainties of
$\sim$0.7. However, during our fitting process we found uncertainties
of 0.8 were required, suggesting that some further source of
uncertainty may be required, for instance the missing opacity sources
in the BD atmosphere models \citep{stauffer_2007}.

As discussed previously other authors have also tried to perform
fitting using the NOMAD catalogue \citep{zacharias_2005}, which is
derived from photographic plate measurements leading to extreme
uncertainty in their conversion to Johnson-Cousins magnitudes,
therefore as we are unsure of how to transform this photometry
correctly we have not included these measurements.
 
Therefore, for BD objects, which have low effective temperatures
(compared to MS stars), photometric transformations, within the
optical regime, are likely to produce significant uncertainties in the
resulting magnitudes, which are exacerbated due to the shape of the
$R$ filter.

\section{Calibration and Simulated Photometry}
\label{calibration_grid}

Initially, our simulated stellar photospheres were diluted by the
factor $(R_*/{\rm distance})^2$ to a distance of 10\,pc to produce
SEDs in absolute flux. These SEDs then represented simulations of
`naked' stars, which, as the emission is isotropic are inclination
independent. For those systems where the disc has been modelled similar
absolute flux SEDs were also constructed at the inclinations detailed
in Table \ref{grid_tab}.

\subsection{Photometric Systems and Calibration}
\label{phot_calib}

To allow comparison and fitting to real observations these SEDs must
be adjusted to account for the distance and extinction to a given
target. We chose an extinction system consistent with
\cite{robitaille_2007}, using the same extinction curve
\citep{indebetouw_2005} interpolated onto our wavelength
sampling. This allowed the construction of SEDs for stars with and
without discs at any distance and extinction (using $F_{\rm
  app}=F_{\rm abs}\times10^{(-0.4A_\lambda)}\times(\frac{10}{distance
  (pc)})^2$, where $F_{\rm app}$ is apparent flux, $F_{\rm abs}$
absolute flux and $A_\lambda$ the extinction in the required band),
for comparison to observed spectra. However, most observations of such
systems are photometric and presented as either monochromatic fluxes
or in a specified magnitude system. Therefore, to enable fitting we
derive these photometric quantities.

To derive magnitudes one requires the underlying filter responses of
the instrument, the method by which the instrument sums flux
(i.e. counting photons or energy) and the assumed zero points. 

The magnitudes are defined as
\begin{equation}
M_{\lambda}=-2.5\log _{10}(N^{\rm counts})-{\rm Zeropoint},
\label{magnitude}
\end{equation}
where counts is usually the number of electrons, or in the case of
IRAC where counts=$e^{-}/{\rm Gain}$, as the gain is significant
($\sim$3--4). Subsequent integration across the filter is performed as
follows \citep[as described in][]{bessell_1998}. For photon counting,
i.e. responses in electrons/photon,
\begin{equation}
N^{\rm counts} = \int \! \frac{\lambda}{hc}R_{\lambda}(\lambda)F_{\lambda}(\lambda) \, {\rm d} \lambda
\label{nrg_int}
\end{equation}
and for energy summing, i.e. responses electrons/unit energy,
\begin{equation}
N^{\rm counts} = \int \! R_{\lambda}(\lambda)F_{\lambda}(\lambda) \, {\rm d} \lambda
\label{phot_int}
\end{equation}
The derivation of monochromatic fluxes is explained in Appendix A of
\cite{robitaille_2007}. The summed flux is still calculated, as with
magnitudes, but this flux is then redistributed across the filter
using an underlying (well chosen) assumption as to the SEDs shape (for
instance 10\,000\,K Black body for the \textit{Spitzer} MIPS
fluxes). Finally, the ``monochromatic'' flux is then quoted as the
value of this redistributed flux curve at the central wavelength of
the filter. After integration using equations \ref{phot_int} or
\ref{nrg_int} an additional normalisation factor is required. For a
monochromatic flux,
\begin{equation}
\frac{\int \! U(\lambda)F_{\lambda}[{\rm actual}]R_{\lambda}(\lambda) \, {\rm d}\lambda}{\int \! R_{\lambda}(\lambda) \, {\rm d}\lambda} = \frac{\int \! U(\lambda)F_{\lambda}[{\rm assumed}]R_{\lambda}(\lambda) \, {\rm d}\lambda}{\int \! R_{\lambda}(\lambda) \, {\rm d}\lambda},
\label{mono_int}
\end{equation}
where $U(\lambda)$ is 1 if the response is in $e^-/{\rm unit\,energy}$
or $\lambda/(h\times c)$ if the response is in $e^-/{\rm
  photon}$. Then, as the assumed shape is known,
\begin{equation}
\lambda F_{\lambda}[assumed] \equiv \lambda_0 F_{\lambda_0}[quoted].
\label{mono_int_2}
\end{equation}
Therefore, to correctly calibrate magnitude and monochromatic flux
derivations one require the filter responses, assumed SED shape and
photometric zero point.  A summary of all of the calibration
information for our derived photometry is available
online\footnote{(\mysitecalib{})} and is presented in Table
\ref{calibration}. Presenting this information is vital as it means
that researchers can recreate what we have done and compare quantities
derived from the models with the same observational quantities
regardless of the system chosen.

\subsection{Fitting Photometric Measurements}
\label{phot_detail}

The derivation of magnitudes and monochromatic fluxes is performed
during fitting on the apparent SEDs. At this stage the reader may ask
the question why do we spend the extra computational time to derive
the magnitudes and fluxes directly (at run time) from the apparent
SEDs, as oppose to simply adjusting the photometry from the absolute
model by the extinction and distance. Additionally, the reader might
question why we derive the monochromatic fluxes so rigorously, as
oppose to simply interpolating for a flux point at the assumed
wavelength on the SED.

Previous fitting, in general, has been performed using these two
shortcuts to provide a much faster fitting routine. However, these
sacrifices imply major assumptions. Either, the filters are assumed to
be infinitesimally thin, or the SED is assumed to be a smooth curve
which matches the assumed SED shape across the filter as well as the
filter response curves being assumed to be flat across the
bandpass. In most, but not all, cases the uncertainties in the
observations and data are large enough to blur any errors made due to
the breakdown of these assumptions.

This breakdown is apparent for objects under heavy extinction but is
perhaps best demonstrated by the use of optical photometry for BDs. In
this regime the SED is a steep function of wavelength.

Most studies will convert optical magnitudes to so-called
monochromatic fluxes to provide better constraint of the stellar
parameters \citep[e.g.][]{guieu_2007,luhman_2007,bouy_2008}. This
conversion is usually performed by calculating a summed flux from the
magnitude (i.e. reversing equation \ref{magnitude}), then assuming
some underlying SED shape (usually those presented in Table
\ref{calibration}) and quoting the flux at the filters central
wavelength. These fluxes are then fit directly to interpolated values
of the flux at the same wavelength.

\begin{figure}
  \includegraphics[scale=0.3,angle=90]{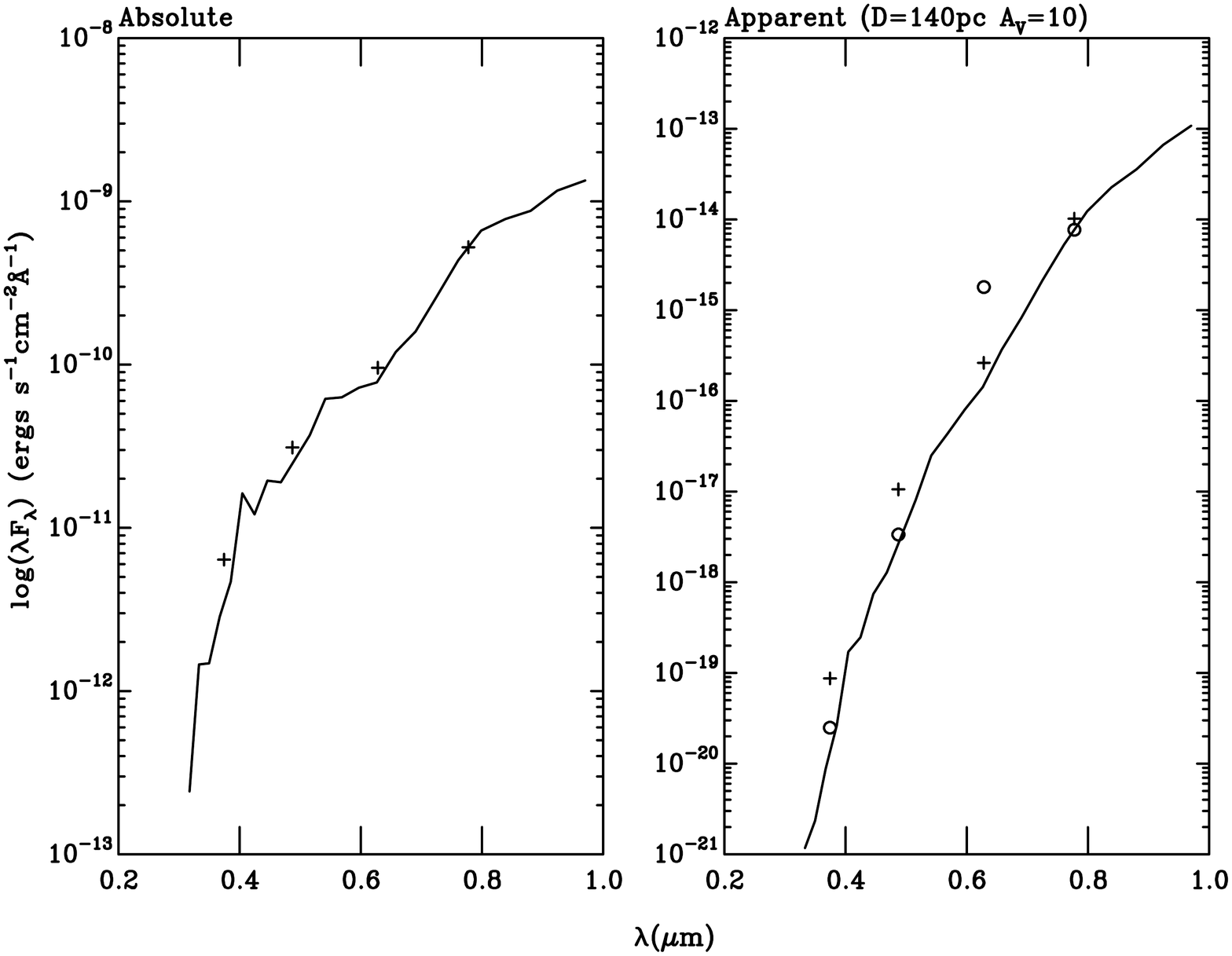}
  \caption{SEDs (solid lines) in $\lambda$ ($\mu$m), $\lambda
    F_{\lambda}$ ($ergs\,s^{-1} cm^{-2} \AA^{-1}$), at 10\,pc (left
    panel) and 140\,pc behind an extinction of $A_V$=10 (right
    panel). The crosses denote the derived monochromatic fluxes for
    each SED and the circles on the right panel show the fluxes
    resulting from simply reddening the absolute monochromatic
    fluxes.\label{mono_test}}
\end{figure}

Figure \ref{mono_test} shows as solid lines the SEDs for a typical
brown dwarf and disc (BDD, hereafter) model from our model grid (A
model from the ad-hoc grid: $M_*=0.04M_{\odot}$, Age=1\,Myr,
negligible accretion, a rotation period of 0.5 days, a disc mass of
0.01$M_*$ a disc outer radius of 100 AU and $\alpha$=2.15), at 10\,pc
(left panel) and 140\,pc behind an extinction of 10$A_V$ (right
panel). The monochromatic fluxes derived rigorously for each SED are
shown for the CFHT Mega Prime u*, g', r' and i' filter bands (as
crosses). If one applies the extinction to the monochromatic fluxes
directly there results a significant difference in flux from those
derived from the reddened SED. Additionally, the derived monochromatic
fluxes do not lie directly on the SED, in either case. Whilst, the
deviation is small given the gradient of the SED in these region the
affect on the fitting statistic is significant.  Essentially, if we do
not derive monochromatic fluxes properly from the underlying SEDs, and
furthermore, simply treat them as points along the SED we are not
comparing two equal quantities. Therefore, for our fitting we only
compare observed monochromatic fluxes with those derived from the
underlying SEDs in the same way.

This problem is exacerbated in the optical regime as, due to the steep
gradient across the wavelength space, the monochromatic fluxes in this
region are very powerful constraints. In addition the photometric
systems and conversions of these fluxes from magnitude space are often
unclear or unreliable. For instance several studies have utilised the
photometry of \cite{zacharias_2005} or the NOMAD survey and converted
the optical magnitudes to monochromatic fluxes. However, these
magnitudes are often derived from photographic plates and it is not
obvious how to convert these correctly to comparable monochromatic
fluxes. In fact transformations between magnitude systems are usually
derived using observations of main-sequence stars and then applied,
without adjustment, to LMS and BDs (this is discussed more in Section
\ref{data_problems}).

A further complication for overlaying monochromatic fluxes on spectra
(if one does not derive the monochromatic fluxes rigorously) is the
quoted central wavelength. In reality the effective wavelength of a
filter shifts position with the input spectrum as shown by
\cite{bessell_1998} and is defined as,
\begin{equation}
\lambda_{\rm eff}=\frac{\int \! \lambda F_{\lambda}[actual]R_{\lambda}(\lambda) \, {\rm d}\lambda}{\int \! F_{\lambda}[actual]R_{\lambda}(\lambda) \, {\rm d}\lambda}.
\label{eff_lambda}
\end{equation}
For a typical spectrum of a BD the shift in wavelength is small for
filters at longer wavelengths (typically $\sim$0.01$\mu$m).  However,
for shorter wavelength filters, where the photospheric flux changes
rapidly across the filter, the shift can be as large as
$\sim$0.1$\mu$m. If one fits observed monochromatic fluxes to
monochromatic fluxes derived for an SED in the same fashion then the
effective wavelength of the filter does not affect the difference (it
is just a direct comparison of fluxes). However, if the observed
monochromatic flux is fitted to a simple point lying on the SED, for
optical colours, then the shift in effective wavelength may well be
significant, again exacerbated by the steep rise in stellar flux
across the optical filters. Essentially, this leads to flux points in
the optical being shifted away from the expected position.  Therefore,
one must be careful when overlaying optical ``monochromatic fluxes''
derived from optical magnitudes on model SEDs. This is discussed and
an example presented in Section \ref{results}.

As the uncertainties in the models are large most previous studies
have not required this level of sophistication in the derivation of
photometric values. However, as this effect becomes particularly
prominent in the optical the SED curve and monochromatic flux points
are often dropped in this regime \citep[see, for example,the
incomplete SEDs at lower wavelengths of][]{guieu_2007}. As, using this
fitting tool, we are now able to precisely model observations in the
optical we no longer have this limitation and, additionally, as models
improve have the software to test their predictions. Simply stated to
improve the models we must ensure we are comparing them, on an equal
footing, to observations.

\subsection{Other Tools}
\label{tools}

In addition to the fitting tool several other tools are available at
\mysite{} enabling users to explore the model grids. A browsing and
downloading tool is available to select, view and retrieve, SEDs and
flux and magnitude information for any model within any grid
(\mysiteplots{}). Additionally, we have constructed model isochrones
and mass tracks using our grids and these can be retrieved from
\mysiteiso{}. Finally, we have also made available an interpolation
tool allowing users to produce SEDs, monochromatic fluxes and
magnitudes for models within, but not covered by, our parameter space
at a user entered distance and extinction.

\onecolumn
\begin{longtable}{llllllll}
\hline
\multirow{2}{*}{Filter Set}&Response&SED Shape&Zero Magnitude&Filter&Zero&Central $\lambda$&Width\\
&Units (e$^-$/?)&(Assumed)&Flux, $F_{\nu}(M_0)$ (Jy)&&points& ($\mu$m)&($\mu$m)\\
\hline
\multirow{8}{*}{Johnson-Cousins (1) \& (2)}&\multirow{8}{*}{Photons}&\multirow{8}{*}{Flat$^{(1)}$}&\multirow{8}{*}{Vega$^{(2)}$}&U&$-$14.18&0.36&0.10\\
&&&&B&$-$15.30&0.46&0.18\\
&&&&V&$-$14.86&0.585&0.21\\
&&&&R&$-$15.11&0.71&0.30\\
&&&&I&$-$14.58&0.81&0.20\\
&&&&J&$-$14.26&1.23&0.36\\
&&&&H&$-$13.49&1.62&0.44\\
&&&&K&$-$13.09&2.18&0.56\\
\hline
\multirow{2}{*}{Tycho (3) \& (2)}&\multirow{2}{*}{Photons}&\multirow{2}{*}{Flat}&\multirow{2}{*}{Vega}&V$_t$&$-$15.09&0.527&0.21\\
&&&&B$_t$&$-$14.94&0.422&0.145\\
\hline
\multirow{9}{*}{Bessell (4) \& (5)}&\multirow{9}{*}{Photons}&\multirow{9}{*}{Flat}&\multirow{9}{*}{Vega}&U&$-$14.23&0.37&0.11\\
&&&&B&$-$15.32&0.44&0.18\\
&&&&V&$-$14.86&0.55&0.21\\
&&&&R&$-$15.09&0.64&0.29\\
&&&&I&$-$14.57&0.80&0.20\\
&&&&J&$-$13.92&1.22&0.36\\
&&&&H&$-$13.65&1.63&0.96\\
&&&&K&$-$12.71&2.19&0.36\\
&&&&L&$-$11.74&3.45&0.72\\
\hline
\multirow{5}{*}{SDSS (6) \& (7)}&\multirow{5}{*}{Photons}&\multirow{5}{*}{Flat}&\multirow{5}{*}{$F_{\nu}(M_0)$=3631}&u&$-$12.43&0.3555&0.0599\\
&&&&g&$-$14.22&0.473&0.1379\\
&&&&r&$-$14.22&0.6261&0.1382\\
&&&&i&$-$13.78&0.7672&0.1535\\
&&&&z&$-$11.84&0.9097&0.137\\
\hline
\multirow{3}{*}{2MASS (8), (9) \& (10)}&\multirow{3}{*}{Photons}&\multirow{3}{*}{Flat}&$F_{\nu}(M_0)$=1594&J&15.73&1.235&0.162\\
&&&$F_{\nu}(M_0)$=1024&H&16.37&1.662&0.251\\
&&&$F_{\nu}(M_0)$=666.7&K$_S$&17.38&2.159&0.262\\
\hline
\multirow{3}{*}{MKO (11), (12) \& (13)}&\multirow{3}{*}{Photons}&\multirow{3}{*}{Flat}&\multirow{3}{*}{Vega}&J&$-$13.48&1.24&0.16\\
&&&&H&$-$13.46&1.65&0.29\\
&&&&K&$-$12.64&2.2&0.34\\
\hline
\multirow{5}{*}{UKIRT (14), (12) \& (13)}&\multirow{5}{*}{Photons}&\multirow{5}{*}{Flat}&\multirow{5}{*}{Vega}&Z&$-$13.84&0.8775&0.095\\
&&&&Y&$-$13.46&1.02&0.1\\
&&&&J&$-$14.01&1.28&0.42\\
&&&&H&$-$13.29&1.65&0.45\\
&&&&K&$-$12.82&2.2&0.66\\
\hline
\multirow{4}{*}{IRAS (15), (16) \& (17)}&\multirow{4}{*}{Energy}&\multirow{4}{*}{Flat}&$F_{\nu}(M_0)$=28.3&12&21.00&12.0&6.5\\
&&&$F_{\nu}(M_0)$=6.73&25&23.55&25.0&11.0\\
&&&$F_{\nu}(M_0)$=1.19&60&26.24&60.0&40.0\\
&&&$F_{\nu}(M_0)$=0.43&100&28.41&100.0&37.0\\
\hline
\multirow{2}{*}{SCUBA$^{(4)}$ (18) \& (19)}&\multirow{2}{*}{Energy}&10\,000K&\multirow{2}{*}{N/A$^{(3)}$}&450WB&0&443.0&138.92\\
&&Black Body&&850WB&0&863.0&230.636\\
\hline
\multirow{4}{*}{IRAC (20) \& (21)}&\multirow{4}{*}{Photons}&\multirow{4}{*}{Flat}&$F_{\nu}(M_0)$=280.9&3.6&$-$10.15&3.55&1.2\\
&&&$F_{\nu}(M_0)$=179.7&4.5&$-$9.70&4.493&1.56\\
&&&$F_{\nu}(M_0)$=115.0&5.8&$-$8.07&5.731&2.27\\
&&&$F_{\nu}(M_0)$=64.9&8.0&$-$8.55&7.872&4.69\\
\hline
\multirow{3}{*}{MIPS (22)}&\multirow{3}{*}{Energy}&10\,000K&$F_{\nu}(M_0)$=7.17&24&25.97&23.68&14.20\\
&&Black body&$F_{\nu}(M_0)$=0.778&70&29.62&71.42&61.06\\
&&&$F_{\nu}(M_0)$=0.160&160&32.48&155.9&99.83\\
\hline
\multirow{3}{*}{CIT (23) \& (24)}&\multirow{3}{*}{Photons}&\multirow{3}{*}{Flat}&\multirow{3}{*}{Vega}&J&$-$13.73&1.25&0.24\\
&&&&H&$-$13.44&1.65&0.30\\
&&&&K$_S$&$-$12.97&2.215&0.41\\
\hline
\multirow{3}{*}{PACS (25), (26) \& (27)}&\multirow{3}{*}{Photons}&\multirow{3}{*}{Flat}&\multirow{3}{*}{Vega}&blue&$-$6.00&70&90.5\\
&&&&green&$-$5.26&100&109\\
&&&&red&$-$4.27&160&391.25\\
\hline
\multirow{3}{*}{SPIRE (28), (29) \& (30)}&\multirow{3}{*}{Photons}&\multirow{3}{*}{Flat}&\multirow{3}{*}{Vega}&250&$-$3.16&250&176.30\\
&&&&350&1.87&350&281.26\\
&&&&500&0.0$^{(5)}$&500&290.24\\
\hline
\multirow{3}{*}{CFHT 12K (Mould) (31)}&\multirow{3}{*}{Photons}&\multirow{3}{*}{Flat}&\multirow{3}{*}{Vega}&V&$-$14.86&0.5374&0.0974\\
&&&&R&$-$14.69&0.6581&0.1251\\
&&&&I&$-$14.78&0.8223&0.2164\\
\hline
\multirow{5}{*}{CFHT Mega Prime (31)}&\multirow{5}{*}{Photons}&\multirow{5}{*}{Flat}&\multirow{5}{*}{Vega}&u*&$-$14.33&0.3743&0.0758\\
&&&&g'&$-$15.49&0.4872&0.1455\\
&&&&r'&$-$14.76&0.6282&0.1219\\
&&&&i'&$-$14.61&0.7776&0.1508\\
&&&&z'&$-$15.07&1.1702&0.6868\\
\hline
\caption{A table describing the calibration information for each
  photometric band used. Sources are given as numbers in brackets: (1)
  \citet{johnson_1966}, (2) \citet{bessell_2005}, (3)
  \citet{bessell_2000}, (4) \citet{bessell_1988}, (5)
  \citet{bessell_1998}, (6) \citet{fukugita_1996}, (7) \sdss{}, (8)
  \citet{cohen_2003}, (9) \citet{skrutskie_2006}, (10) \twomass{},
  (11) \citet{tokunaga_2002}, (12) \citet{simons_2002}, (13) \ukirt{},
  (14) \citet{hawarden_2001}, (15) \citet{neugebauer_1984}, (16)
  \irasresp{}, (17) \iraszero{}, (18) \citet{holland_1999}, (19)
  \jcmt{}, (20) \citet{reach_2005}, (21) \irac{}, (22) \mips{}, (23)
  \citet{elias_1982}, (24) \citet{stephens_2004}, (25)
  \citet{poglitsch_2008}, (26) \citet{poglitsch_2010}, (27) \pacs{},
  (28) \citet{griffin_2008}, (29) \citet{griffin_2010}, (30) \spire{}
  and (31) \cfht{}. Notes are superscript numbers in bracket: (1) A
  Constant flux across the filter. (2) The Vega spectrum was a NextGen
  atmosphere model with $T_{\rm eff}$=9950K and $\log (g)$=3.95, scaled to
  the Vega flux of \citet{hayes_1985}. (4) SCUBA 2 filters
  approximately equal to SCUBA filters. (5) Wavelength longer than the
  maximum available in the Vega spectrum. \label{calibration}}
\end{longtable}
\twocolumn